\documentclass[12pt]{article}
\pdfoutput=1
\usepackage{amsfonts}
\usepackage{amsmath}
\usepackage{url}
\usepackage{hyperref}
\usepackage{bbold}
\usepackage{slashed}
\usepackage{tikz}
\usepackage{graphicx}
\usepackage{epstopdf}
\usepackage{subfigure}
\usepackage{pgfplots}
\usepackage{amssymb}
\usepackage{color}
\usepackage{mathrsfs}
\usepackage{fancybox}
\usepackage{lipsum}
\usepackage{eurosym}
\usepackage{tcolorbox}

\numberwithin{equation}{section}
\def\Deltal{\ell}
\usepackage{environ}
\NewEnviron{myequation}{%
\begin{eqnarray}
\scalebox{1.1}{$\BODY$}
\end{eqnarray}
}

\newcommand{\beq}{\begin{equation}}
\newcommand{\eeq}{\end{equation}}
\newcommand{\nn}{\nonumber\\} 
\newcommand{\bea}{\begin{eqnarray}}
\newcommand{\ea}{\end{eqnarray}}
\newcommand{\barr}{\!\begin{array}}
\newcommand{\earr}{\end{array}\!}
\newcommand{\lb}{{\langle}}
\newcommand{\rb}{{\rangle}}

\newcommand{\sixj}[6]{\Bigl\{\mbox{\small$\!\begin{array}{ccc} #1 \! & \!\! #3 \! & \!\! #5 \nspc \\[-1mm]  #2 \!  & \!\! #4 \!  & \!\! #6 \nspc \end{array}\!$}\!\Bigr\}}

\newcommand{\sixjAN}[6]{\Bigl\{\mbox{\small$\!\begin{array}{ccc} #1 \! & \!\! #3 \! & \!\! #5 \nspc \\[-1mm]  #2 \!  & \!\! #4 \!  & \!\! #6 \nspc \end{array}\!$}\!\Bigr\}_b^{\rm an}}

\def\d{{\partial}}
\def\n{{\bf \widehat n}}
\def\k{{\bf k}}

\begin{document}
\begin{titlepage}

\setcounter{page}{1} \baselineskip=15.5pt \thispagestyle{empty}

\vfil

${}$
\vspace{1cm}

\begin{center}

\def\thefootnote{\fnsymbol{footnote}}
\begin{changemargin}{0.05cm}{0.05cm} 
\begin{center}
{\Large \bf Solving the Schwarzian \\
~\\
via the Conformal Bootstrap}
\end{center} 
\end{changemargin}

~\\[1cm]
{Thomas G. Mertens${}^{\rm a,c}$\footnote{\href{mailto:thomas.mertens@ugent.be}{\protect\path{thomas.mertens@ugent.be}}, \href{mailto:tmertens@princeton.edu}{\protect\path{tmertens@princeton.edu}}}, Gustavo J. Turiaci${}^{\rm a}$\footnote{\href{mailto:turiaci@princeton.edu}{\protect\path{turiaci@princeton.edu}}} and Herman L. Verlinde${}^{\rm a,b}$\footnote{\href{mailto:verlinde@princeton.edu}{\protect\path{verlinde@princeton.edu}}}}
\\[0.3cm]

{\normalsize { \sl ${}^{\rm a}$Physics Department and ${}^{\rm b}$Princeton Center for Theoretical Science 
\\[1.0mm]
Princeton University, Princeton, NJ 08544, USA}} \\[3mm]
{\normalsize { \sl ${}^{\rm c}$Department of Physics and Astronomy
\\[1.0mm]
Ghent University, Krijgslaan, 281-S9, 9000 Gent, Belgium}}

\end{center}


 \vspace{0.2cm}
\begin{changemargin}{01cm}{1cm} 
{\small  \noindent 
\begin{center} 
\textbf{Abstract}
\end{center} }
We obtain exact expressions for a general class of correlation functions in the 1D quantum mechanical model described by the Schwarzian action, that arises as the low energy limit of the SYK model. The answer takes the form of an integral of a momentum space amplitude obtained via a simple set of diagrammatic rules. The derivation relies on the precise equivalence between the 1D Schwarzian theory and a suitable large $c$ limit of 2D Virasoro CFT. The mapping from the 1D to the 2D theory is similar to the construction of kinematic space. We also compute the out-of-time ordered four point function. The momentum space amplitude in this case contains an extra factor in the form of a crossing kernel, or R-matrix, given by a 6j-symbol of SU(1,1). We argue that the R-matrix describes the gravitational scattering amplitude near the horizon of an AdS${}_2$ black hole. Finally, we discuss the generalization of some of our results to ${\cal N}=1$ and ${\cal N}=2$ supersymmetric Schwarzian~QM.
\end{changemargin}
 \vspace{0.3cm}
\vfil
\begin{flushleft}
\today
\end{flushleft}

\end{titlepage}

\newpage
\tableofcontents
\newpage

\addtolength{\abovedisplayskip}{.5mm}
\addtolength{\belowdisplayskip}{.5mm}

\def\plus{\raisebox{.5pt}{\tiny$+$\smpc}}

\addtolength{\parskip}{.6mm}
\def\spc{\hspace{1pt}}

\def\nspc{{\hspace{-2pt}}}
\def\ff{\rm\smpc f\smpc} 
\def\fff{\mbox{Y}}
\def\ww{{\rm w}}
\def\smpc{{\hspace{.5pt}}}

\def\zz{{\spc \rm z}}
\def\xx{{\rm x\smpc}}
\def\xxi{\mbox{\footnotesize \spc $\xi$}}
\def\jj{{\rm j}}
 \addtolength{\baselineskip}{-.1mm}

\renewcommand{\Large}{\large}

\def\calO{{b}}
\def\be{\begin{equation}}
\def\ee{\end{equation}}




\def\mathbi#1{\textbf{\em #1}} 
\def\som{{ \textit{\textbf s}}} 
\def\tom{{ \textit{\textbf t}}} 
\def\nom{n} 
\def\mom{m} 
\def\la{\langle}
\def\bea{\begin{eqnarray}}
\def\eea{\end{eqnarray}}
\def\is{\!  & \!  = \!  &  \!}
\def\ra{\rangle}
\def\half{{\textstyle{\frac 12}}}
\def\cL{{\cal L}}
\def\halfi{{\textstyle{\frac i 2}}}
\def\ba{\bea}
\def\ea{\eea}
\def\lb{\langle}
\def\rb{\rangle}
\newcommand{\rep}[1]{\mathbf{#1}}

\def\uU{\bf U}
\def\be{\bea}
\def\ee{\eea}
\def\delbar{\overline{\partial}}
\def\ra{\bigr\rangle}
\def\la{\bigl\langle}
\def\ccdot{\!\spc\cdot\!\spc}
\def\nspc{\!\spc\smpc}
\def\tr{{\rm tr}}
\def\ra{\bigr\rangle}
\def\la{\bigl\langle}
\def\li{\bigl|\spc}
\def\ri{\bigr |\spc}

\def\hf{\textstyle \frac 1 2}

\def\bfcdot{\raisebox{-1.5pt}{\bf \LARGE $\spc \cdot\spc $}}
\def\spc{\hspace{1pt}}
\def\is{\! &  \! = \! & \!}
\def\d{{\partial}}
\def\n{{\bf \widehat n}}
\def\k{{\bf k}}
\def\GO{{\cal O}}

\def\pp{{\mbox{\tiny$+$}}}
\def\mm{{\mbox{\tiny$-$}}}

\setcounter{tocdepth}{2}
\addtolength{\baselineskip}{0mm}
\addtolength{\parskip}{.4mm}
\addtolength{\abovedisplayskip}{1mm}
\addtolength{\belowdisplayskip}{1mm}

\def\fff{e}

\section{Introduction: Schwarzian QM}\label{sec:Sch}

In the past few years it has been recognized that holographic CFTs at finite temperature exhibit characteristics of many body quantum chaos \cite{SS, KitaevTalks, MSS, Polchinski:2015cea, Jensen:2016pah}.  The SYK model is a soluble many body quantum system with a well-controlled large $N$ limit that exhibits maximal chaos and other characteristics that indicate it has a holographic dual given by a 2D gravity theory on AdS${}_2$ \cite{KitaevTalks, Sachdev:1992fk, Polchinski:2016xgd, Jevicki:2016bwu, Maldacena:2016hyu,Jevicki:2016ito, Witten:2016iux}. The Schwarzian theory describes the quantum dynamics of a single 1D degree of freedom $f(\tau)$ and forms the theoretical gateway between the microscopic SYK model and the dual 2D dilaton gravity theory \cite{Jackiw:1984je, Almheiri:2014cka, Maldacena:2016upp, Engelsoy:2016xyb, Cvetic:2016eiv}. 

In this paper we will study the finite temperature correlation functions in the 1D quantum mechanical theory described by the action
\bea
\label{schwaction}
\qquad S[f] \is 
-C\int_0^{\beta}\!\!\! d\tau\spc  \left(\bigl\{\spc f,\spc \tau\spc \bigr\} + \frac{2\pi^2}{\beta^2} f'^2\right) \\[3mm]
\is  - C\int_0^{\beta}\!\!\! d\tau \spc \bigl\{\spc F,\spc \tau\spc \bigr\}, \quad\qquad F\, \equiv \, \tan\left(\frac{\pi f(\tau)}{\beta}\right),
\eea
where $C$ is the coupling constant of the zero-temperature theory. We will set $C=1/2$ from here on out, unless explicitly stated.
Here $f(\tau+\beta) = f(\tau) +\beta$ runs over the space Diff$(S^1)$ of diffeomorphisms on the thermal circle,  and
\bea
\bigl\{\spc f,\spc \tau\spc \bigr\} \is \frac {f'''}{f'} - \frac 3 2 \left(\frac{f''}{f'}\right)^2 \,
\eea
denotes the Schwarzian derivative. 

The action $S[f]$ is invariant under ${SL(2,\mathbb{R})}$ M\"obius transformations that act on $F$ via 
\bea
\label{mobiusF}
F \to \frac{ a F + b}{c F+ d}.
\eea 
The model possesses a corresponding set of conserved charges $\ell_a$ that generate the $\mathfrak{sl}(2,\mathbb{R})$ algebra $[\ell_a,\ell_b] = i\epsilon_{abc} \ell_c$ and commute with the Hamiltonian $H$. In fact, as reviewed in section \ref{sect:schro}, the Hamiltonian $H$ is found to be equal to the $SL(2,\mathbb{R})$ Casimir, $H = \frac {1} 2 \ell_a \ell_a.$ The energy spectrum and dynamics are thus uniquely determined by the $SL(2,\mathbb{R})$ symmetry. 
 The Schwarzian theory is integrable and expected to be exactly soluble at any value of the inverse temperature $\beta$.
In the following, we will label the energy eigenvalues $E$ in terms of the $SL(2,\mathbb{R})$ spin $j= -\frac 1 2 + i k$ via
\bea
\label{eka}
E(k) \is - \spc j(j+1) \, = \, \frac 1 4 + k^2 .
\eea
The constant $\frac{1}{4}$ can be removed by choosing appropriate normal ordering in the quantum theory, and we will drop it throughout most of this work. 
If we mod out by the overall
$SL(2,\mathbb{R})$ symmetry, the partition sum 
\bea
Z (\beta)\is \int_{{\!}_{\cal M}}\!
\! {{\cal D}\! \spc f}\, e^{-S[f] } 
\eea
reduces to an integral over the infinite dimensional quotient space
\bea
\label{coad}
{\cal M} \is {\rm Diff}(S^1)/SL(2,\mathbb{R}).
\eea
This space ${\cal M}$ equals the coadjoint orbit of the identity element $\mathbb{1} \in$ Diff$(S^1)$, which is known to be a symplectic manifold that upon quantization gives rise to the identity representation of the Virasoro group Diff$(S^1)$, i.e. the identity module of the Virasoro algebra \cite{Kirillov, LazutkinSegal, Witten:1987ty}.
We choose the functional measure $d\mu(f)$ to be the one derived from the symplectic form on ${\cal M}$, which as shown in \cite{AS, altland, wittenstanford} takes the form ${\cal D}f = \prod_\tau {df}/{f'}$. 

The fact that the space ${\cal M}$  is a symplectic manifold was exploited in \cite{wittenstanford} to show that  the partition function $Z$ is one-loop exact and given by
\bea
Z(\beta) \is \Bigl(\frac{\pi}{\beta}\Bigr)^{3/2}\, e^{{\pi^2}/{ \beta}} \, = \, \int_0^\infty\! \!\! d\mu(k) \, e^{-\beta E(k)}
\eea
with $E(k)$ as in (\ref{eka}) and where  the integration measure is given in terms of $k$ by
\bea
d\mu(k) \is dk^2 \sinh(2\pi k). 
\eea
This exact result for the spectral density 
\bea
\label{smeasure}
\rho(E) = \sinh\bigl(2\pi \sqrt{E}\bigr)
\eea
is further indication that the Schwarzian theory is completely soluble. In this paper we will show that this is indeed the case.

For our analysis we will make use of the more detailed property that the space ${\cal M}$ in (\ref{coad}) is not just any phase space, but forms the quantizable coadjoint orbit space that gives rise to the identity module of the Virasoro algebra. As we will show in section \ref{sect3}, this observation implies that the correlation functions of the Schwarzian theory 
\bea
\la \spc {\cal O}_1 \, ...\, {\cal O}_n\spc \ra \is \frac 1 Z \int_{{\!}_{\cal M}}\!
\! {{\cal D}\!\spc f}\, e^{-S[f] }\,{\cal O}_1\, ...\, {\cal O}_n = \frac{1}{Z}\text{Tr}\bigl(e^{-\beta H} \,{\cal O}_1\, ...\, {\cal O}_n\bigr)\, 
\eea
can be obtained by taking a suitable large $c$ limit of well-studied correlation functions of an exactly soluble 2D CFT with Virasoro symmetry. In subsequent sections, we will then use this relation to explicitly compute the correlation functions of a natural class of $SL(2,\mathbb{R})$ invariant observables ${\cal O}_i$. 
We will now first summarize our main results.

\pagebreak

\subsection{Overview of results}
\label{sect:overview}

We will study the correlation functions of the following bi-local operators
\bea
\label{bilocal}
\mathcal{O}_\Deltal(\tau_1,\tau_2) \equiv 
\left( \frac{\sqrt{f'(\tau_1)f'(\tau_2)}}{\frac{\beta}{\pi}\sin \frac{\pi}{\beta}[f(\tau_1)-f(\tau_2)]} \right)^{2\Deltal}.
\eea
We can think of this expression as the two-point function $
\GO_{\Deltal}(\tau_1,\tau_2) = \langle \mathcal{O}(\tau_1) \mathcal{O}(\tau_2) \rangle_{\text{CFT}}$ of some 1D `matter CFT' at finite temperature coupled to the Schwarzian theory, or equivalently, as the boundary-to-boundary propagator of a bulk matter field coupled to the 2D dilaton-gravity theory in a classical black hole background. 

The bi-local operator (\ref{bilocal}) is invariant under the $SL(2,\mathbb{R})$ transformations  (\ref{mobiusF}). This in particular implies that $\GO_\ell$ commutes with the Hamiltonian $H$ of the Schwarzian theory
\bea
\label{hcom}
[H, \GO_{\Deltal}(\tau_1,\tau_2) ] \is 0.
\eea
So the bi-local operators are diagonal between energy eigenstates. We will see that the time-ordered correlation functions of $\GO_{\Deltal}(\tau_1,\tau_2)$ indeed only depend on the time-difference~$\tau_2-\tau_1$.

Below we will give the explicit formulas for the correlation function with one and two insertions of the bi-local operator $\GO_{\Deltal}$. We will call these the two-point and four-point functions, since they depend on  two and four different times $\tau_i$, respectively. In
 the holographic dual theory they correspond to the AdS${}_2$ gravity amplitude with one and two boundary-to-boundary propagators. Our eventual interest is to compute the out-of-time ordered (OTO) four point function, which exhibits maximal Lyapunov behavior and contains the gravitational scattering amplitudes of the bulk theory as an identifiable subfactor.
 
\medskip

\begin{center}
{\bf Two-point function}
\end{center}

The two-point function at finite temperature is defined by the functional integral with a single insertion of the bi-local operator 
\bea
\la\, \GO_{\Deltal}(\tau_1,\tau_2)\, \ra \; \is 
\frac 1 Z  \int \! {\cal D}\!\spc f \, e^{-S[f]} \, \GO_{\Deltal}(\tau_1,\tau_2) \    =  \ 
 \raisebox{1mm}{ \begin{tikzpicture}[scale=0.6, baseline={([yshift=0cm]current bounding box.center)}]
\draw[thick] (0,0) circle (1.5);
\draw[thick] (-1.5,0) -- (1.5,0);
\draw[fill,black] (-1.5,0) circle (0.1);
\draw (-2,0) node {\small $\tau_2$};
\draw[fill,black] (1.5,0) circle (0.1);
\draw (2,0) node {\small $\tau_1$};
\draw (-0,.4) node {\small $\ell$};
\end{tikzpicture}}
\eea
Here we introduced a diagrammatic notation that will be useful below. 

The two-point function of the Schwarzian theory at zero temperature was obtained in \cite{altland}. As we will show in section \ref{sec:ZZbosonic}, the generalization of their result to finite temperature is given by a double integral over intermediate $SL(2,\mathbb{R})$ representation labels $k_1$ and $k_2$
\bea
\la \spc \GO_{\Deltal}(\tau_1,\tau_2)\spc
\ra \, \is\,  \int\spc \prod_{i=1}^2 d\mu(k_{i})\,  \, {\cal A}_2(k_i, \ell, \tau_i).
\eea
We will call the integrand the `momentum space amplitude'. In section \ref{sec:ZZbosonic} we will obtain the following explicit formula for ${\cal A}_2(k_i, \ell, \tau_i)$
\bea
{\cal A}_2(k_i, \ell, \tau_i)
\is e^{-(\tau_2-\tau_1)k_1^2 - (\beta -\tau_2+\tau_1)k_2^2} \,\;  {\frac{ \Gamma(\Deltal\pm i k_1\pm i k_2)
\!}{\Gamma(2\Deltal)} } ,
\eea
where $\Gamma(x\pm y\pm z)$ is short-hand for the product of four gamma functions with all four choices of signs. In the following sections, we will derive the above result from the relation between the Schwarzian theory and 2D Virasoro CFT, by taking a suitable large $c$ limit of known results in the latter. We will also perform a number of non-trivial checks on the result. In particular, it reduces to the zero-temperature result of \cite{altland} in the limit $\beta \to \infty$.

\medskip

\begin{center}
{\bf Propagators and vertices}
\end{center}

From the above expression for the two-point function, we can extract the following combinatoric algorithm, analogous to the Feynman rules, for computing time-ordered correlation functions of bi-local operators in the Schwarzian theory. We remark that these rules are still non-perturbative in the Schwarzian theory and merely represent a convenient packaging of the exact amplitudes.

We represent the momentum space amplitude ${\cal A}_2(k_i, \ell, \tau_i)$ diagrammatically as
\bea
{\cal A}_2(k_i,\ell, \tau_i)\  \is \  \ \ \begin{tikzpicture}[scale=0.61, baseline={([yshift=0cm]current bounding box.center)}]
\draw[thick] (0,0) circle (1.5);
\draw[thick] (-1.5,0) -- (1.5,0);
\draw[fill,black] (-1.5,0) circle (0.1);
\draw (0,2) node {\small $k_1$};
\draw (-2,0) node {\small $\tau_2$};
\draw[fill,black] (1.5,0) circle (0.1);
\draw (2,0) node {\small $\tau_1$};
\draw (0,-2) node {\small $k_2$};
\draw (-0,.4) node {\small $\ell$};
\end{tikzpicture}
\eea
The thermal circle factorizes into two propagators, one with `momentum' $k_1$ and one with `momentum' $k_2$. The Feynman rule for the propagator and vertices read
\bea 
\label{frules}
\boxed{\ \begin{tikzpicture}[scale=0.57, baseline={([yshift=-0.1cm]current bounding box.center)}]
\draw[thick] (-0.2,0) arc (170:10:1.53);
\draw[fill,black] (-0.2,0.0375) circle (0.1);
\draw[fill,black] (2.8,0.0375) circle (0.1);
\draw (3.4, 0) node {\footnotesize $\tau_1$};
\draw (-0.7,0) node {\footnotesize $\tau_2$};
\draw (1.25, 1.6) node {\footnotesize $k$};
\draw (6.5, 0) node {$\raisebox{6mm}{$\ \ =\ \ e^{-\spc \spc k^2 \spc (\tau_2-\tau_1)}$}$};
\end{tikzpicture}, ~~~~~~~~~~\ \ \begin{tikzpicture}[scale=0.7, baseline={([yshift=-0.1cm]current bounding box.center)}]
\draw[thick] (-.2,.9) arc (25:-25:2.2);
\draw[fill,black] (0,0) circle (0.08);
 \draw[thick](-1.5,0) -- (0,0);
\draw (.3,-0.95) node {\footnotesize $\textcolor{black}{k_2}$};
\draw (.3,0.95) node {\footnotesize$\textcolor{black}{k_1}$};
\draw (-1,.3) node {\footnotesize$\Deltal$};
\draw (2.5,0.1) node {$\mbox{$\ =\  \, \gamma_\ell(k_1,k_2)\spc .$}$}; \end{tikzpicture}\ }
\eea
The propagator with momentum $k$ represents the phase factor between $\tau_1$ and $\tau_2$ of an energy eigenstate with energy $E = k^2$.
Each vertex corresponds to a factor
\beq
\label{gammaell}
\boxed{\ \gamma_\ell(k_1,k_2)\; = \; \sqrt{\frac{ \Gamma(\Deltal\pm i k_1\pm i k_2)\!}{\Gamma(2\Deltal)}}. \ }
\eeq
This vertex factor represents the matrix element of each endpoint of the bi-local operator between the corresponding two energy eigenstates.

\medskip

\begin{center}
{\bf Time ordered 4-point function}
\end{center}

The time-ordered 4-point function comes in different types, depending on the ordering of the four different times. The simplest ordering is
\beq
\la \spc \GO_{\Deltal_1}(\tau_1,\tau_2)\, \GO_{\Deltal_2}(\tau_3,\tau_4)\spc
\ra \ = \   \begin{tikzpicture}[scale=0.65, baseline={([yshift=0cm]current bounding box.center)}]
\draw[thick] (0,0) circle (1.5);
\draw[thick] (1.3,.7) arc (300:240:2.6);
\draw[thick] (-1.3,-.7) arc (120:60:2.6);
\draw[fill,black] (-1.3,-.68) circle (0.1);
\draw (-1.8,-.7) node {\small $\tau_3$};
\draw (-1.8,.7) node {\small $\tau_2$};
\draw (1.8,-.7) node {\small $\tau_4$};
\draw (1.8,.7) node {\small $\tau_1$};
\draw[fill,black] (1.3,-.68) circle (0.1);
\draw[fill,black] (-1.3,0.68) circle (0.1);
\draw[fill,black] (1.3,0.68) circle (0.1);
\draw (0,.8) node {\small $\ell_1$};
\draw (0,-.8) node {\small $\ell_2$};
\end{tikzpicture}
\eeq
where we assume that the four times are cyclically ordered via $\tau_1 < \tau_2 < \tau_3 < \tau_4.$
This ordering ensures that the legs of the two bi-local operators do not cross each other.
This time-ordered 4-point function is given by a triple integral over intermediate momenta
\bea
\la \spc \GO_{\Deltal_1}(\tau_1,\tau_2)\, \GO_{\Deltal_2}(\tau_3,\tau_4)\spc
\ra \is \int  \prod_{i=1}^3 d\mu(k_i) \; {\cal A}_4\bigl(k_i,\ell_i,\tau_i\bigr).
\eea
The momentum amplitude is represented by the diagram
\beq
{\cal A}_4\bigl(k_i,\ell_i,\tau_i\bigr)\ =  \ \,\begin{tikzpicture}[scale=0.65, baseline={([yshift=0cm]current bounding box.center)}]
\draw[thick] (0,0) circle (1.5);
\draw[thick] (1.3,.7) arc (300:240:2.6);
\draw[thick] (-1.3,-.7) arc (120:60:2.6);
\draw[fill,black] (-1.3,-.68) circle (0.1);
\draw[fill,black] (1.3,-.68) circle (0.1);
\draw[fill,black] (-1.3,0.68) circle (0.1);
\draw[fill,black] (1.3,0.68) circle (0.1);
\draw (2,0) node {\footnotesize $k_s$};
\draw (-2,0) node {\footnotesize  $k_s$};
\draw (0,.8) node {\footnotesize $\ell_1$};
\draw (0,-.8) node {\footnotesize $\ell_2$};
\draw (0,1.88) node {\footnotesize  $k_1$};
\draw (0,-1.88) node {\footnotesize  $k_4$};
\end{tikzpicture}\eeq
Here we took into account the aforementioned result (\ref{hcom}) that the bi-local operators commute with the Hamiltonian, so that the same energy eigenstate (labeled by the momentum variable $k_s$) appears on both sides of each bi-local operator.

Applying the Feynman rules formulated above, we find that the momentum amplitude of the time-ordered four point function reads
\bea
\label{4pt}
{\cal A}_4\bigl(k_i,\ell_i,\tau_i\bigr)\is  e^{- k_1^2(\tau_2-\tau_1) - k_4^2(\tau_4-\tau_3) - k_s^2(\beta -\tau_2+ \tau_3 -\tau_4 +\tau_1)}\, \gamma_{\Deltal_1}(k_1,k_s)^2 \gamma_{\Deltal_2}(k_s,k_4)^2.\qquad
\eea
In section \ref{sec:ZZbosonic}, we will explicitly compute the four-point function from the relationship between the Schwarzian and 2D CFT and confirm that this is indeed the correct result.\footnote{Note that the amplitude (\ref{4pt}) factorizes into a product of two 2-point amplitudes
\bea
{\cal A}_4\bigl(k_i,\ell_i,\tau_i\bigr) \is e^{\beta k_s^2}\, {\cal A}_2\bigl(k_1,k_s,\ell_1,\tau_{21}\bigr) \;  {\cal A}_2\bigl(k_4,k_s,\ell_2,\tau_{43}\bigr) 
\eea
and thus  indeed only depends on the two time differences $\tau_{21} = \tau_2-\tau_1$ and $\tau_{43} = \tau_4-\tau_3$, as dictated by equation (\ref{hcom}).}

\bigskip

\begin{center}
{\bf OTO 4-point function}
\end{center}

Finally we will turn to our main interest, the out-of-time-ordered 4-point function \cite{KitaevTalks}. We will diagrammatically represent the OTO 4-point function as
\bea
\label{otodiagram}
\la \spc \GO_{\Deltal_1}(\tau_1,\tau_2)\, \GO_{\Deltal_2}(\tau_3,\tau_4)\spc
\ra_{\rm OTO} \is\ 
\begin{tikzpicture}[scale=0.65, baseline={([yshift=0cm]current bounding box.center)}]
\draw (-1.5,-1.3) node {\small $\tau_2$};
\draw (-1.5,1.3) node {\small $\tau_3$};
\draw[thick] (-1.05,1.05) -- (-.15,.15);
\draw[thick] (.15,-.15) -- (1.05,-1.05);
\draw[thick] (-1.05,-1.05) -- (1.05,1.05);
\draw[thick] (0,0) circle (1.5);
\draw[fill,black] (-1.05,-1.05) circle (0.1);
\draw[fill,black] (1.05,-1.05) circle (0.1);
\draw[fill,black] (-1.05,1.05) circle (0.1);
\draw[fill,black] (1.05,1.05) circle (0.1);
\draw (-.8,.35) node {\footnotesize $\ell_2$};
\draw (.88,.35) node {\footnotesize $\ell_1$};
\draw (1.5,-1.3) node {\small $\tau_4$};
\draw (1.5,1.3) node {\small $\tau_1$};
\end{tikzpicture}
\eea
where in spite of their new geometric ordering along the circle, we in fact assume that the four time instances continue to be ordered according to $\tau_1 < \tau_2 < \tau_3 < \tau_4.$  Operationally, we define the OTO correlation function via analytic continuation starting from the time ordered correlation function with the ordering $\tau_1 < \tau_3 < \tau_2 < \tau_4$ as indicated by the above diagram. Since for this configuration, the legs of the bi-local operators do in fact cross, the resulting time ordered 4-point function differs from the analytic continuation of the uncrossed 4-point function (\ref{4pt}).

In section \ref{sect5}, we will show that the OTO correlation function can be expressed as an integral over four momentum variables
\bea
\la \spc \GO_{\Deltal_1}(\tau_1,\tau_2)\, \GO_{\Deltal_2}(\tau_3,\tau_4)\spc
\ra_{\rm OTO} \is \int \! \prod_{i=1}^4 d\mu(k_i) \spc 
\, {\cal A}^{\rm OTO}_4\bigl(k_i,\ell_i,\tau_i\bigr),
\eea
where the momentum space amplitude is represented by the following diagram (to avoid clutter, we again suppressed the times $\tau_i$ labeling the end points of the bi-local operators)
\bea
\label{otoamp}
{\cal A}^{\rm OTO}_4\bigl(k_i,\ell_i,\tau_i\bigr)\, \is \ \ \begin{tikzpicture}[scale=0.65, baseline={([yshift=0cm]current bounding box.center)}]
\draw[thick] (-1.05,1.05) -- (-.15,.15);
\draw[thick] (.15,-.15) -- (1.05,-1.05);
\draw[thick] (-1.05,-1.05) -- (1.05,1.05);
\draw[thick] (0,0) circle (1.5);
\draw[fill,black] (-1.05,-1.05) circle (0.1);
\draw[fill,black] (1.05,-1.05) circle (0.1);
\draw[fill,black] (-1.05,1.05) circle (0.1);
\draw[fill,black] (1.05,1.05) circle (0.1);
\draw (2,0) node {\footnotesize $k_s$};
\draw (-2,0) node {\footnotesize  $k_t$};
\draw (-.8,.35) node {\footnotesize  $\ell_2$};
\draw (.88,.35) node {\footnotesize  $\ell_1$};
\draw (0,1.88) node {\footnotesize  $k_1$};
\draw (0,-1.88) node {\footnotesize $k_4$};
\end{tikzpicture}
\eea
Note that we now have four different momentum variables $k_i$. The correlation function will indeed depend on all four time differences $\tau_{i+1} - \tau_i$.

The final answer for the momentum amplitude of the OTO 4-point function reads 
\bea
{\cal A}^{\rm OTO}_4\bigl(k_i,\ell_i,\tau_i\bigr)\is  e^{ -   k_1^2(\tau_3-\tau_1) - k_t^2(\tau_3 -\tau_2) -k_4^2(\tau_4-\tau_2)- k_s^2(\beta-\tau_4 +\tau_1) }
\, 
\\[3mm]
\! & \! \! & \! \!\!\!\!\!\!\!\!\!\!\!\!\!\!\!\!\!\!\! \times  \, \gamma_{\Deltal_1}(k_1,k_s) \gamma_{\Deltal_2}(k_s,k_4)\spc \gamma_{\Deltal_1}(k_4,k_t) \gamma_{\Deltal_2}(k_t,k_1)\times R_{k_sk_t}\! \left[\; {}^{k_4}_{k_1} \;{}^{\Deltal_2}_{\Deltal_1} \right]. \nonumber
\eea
Comparing with the diagram \eqref{otoamp}, we recognize the same propagators and vertex factors as before.  However, the momentum amplitude now also contains an additional factor $R_{k_sk_t}\! \left[\; {}^{k_4}_{k_1} \;{}^{\Deltal_2}_{\Deltal_1} \right]$, which takes into account the effect of the two crossing legs in the diagram \eqref{otoamp}. From the holographic dual perspective, it represents the scattering amplitude of particles in the AdS$_2$ black hole background \cite{SS,Jackson:2014nla}. Computing this crossing kernel is one of the main goals of this paper. We will describe this computation in section \ref{sect5}.

\begin{center}
{\bf The crossing kernel}
\end{center}
The crossing kernel enters as a new entry in the Feynman rules for the Schwarzian correlation function. It relates the crossed diagram to the uncrossed diagram via
\bea
\label{crossing}
\boxed{ \ \begin{tikzpicture}[scale=.6, baseline={([yshift=0cm]current bounding box.center)}]
\draw[thick] (-1.05,1.05) -- (-.15,.15);
\draw[thick] (.15,-.15) -- (1.05,-1.05);
\draw[thick] (-1.05,-1.05) -- (1.05,1.05);
\draw[thick] (0,0) circle (1.5);
\draw[fill,black] (-1.05,-1.05) circle (0.08);
\draw[fill,black] (1.05,-1.05) circle (0.08);
\draw[fill,black] (-1.05,1.05) circle (0.08);
\draw[fill,black] (1.05,1.05) circle (0.08);
\draw (1.85,0) node {\scriptsize $k_s$};
\draw (-1.85,0) node {\scriptsize $k_t$};
\draw (-.75,.33) node {\scriptsize $\ell_2$};
\draw (.78,.33) node {\scriptsize $\ell_1$};
\draw (0,1.75) node {\scriptsize  $k_1$};
\draw (0,-1.75) node {\scriptsize $k_4$};
\end{tikzpicture}~~\raisebox{-3pt}{$\ \ \  = \ \ R_{k_sk_t}\! 
 \left[\, {}^{k_4}_{k_1} \,{}^{\Deltal_2}_{\Deltal_1}\right] $}~~~
\begin{tikzpicture}[scale=.6, baseline={([yshift=0cm]current bounding box.center)}]
\draw[thick] (0,0) circle (1.5);
\draw[thick] (1.3,.7) arc (300:240:2.6);
\draw[thick] (-1.3,-.7) arc (120:60:2.6);
\draw[fill,black] (-1.3,-.68) circle (0.08);
\draw[fill,black] (1.3,-.68) circle (0.08);
\draw[fill,black] (-1.3,0.68) circle (0.08);
\draw[fill,black] (1.3,0.68) circle (0.08);
\draw (1.85,0) node {\scriptsize $k_s$};
\draw (-1.85,0) node {\scriptsize $k_t$};
\draw (0,.75) node {\scriptsize $\ell_1$};
\draw (0,-.75) node {\scriptsize $\ell_2$};
\draw (0,1.75) node {\scriptsize $k_1$};
\draw (0,-1.75) node {\scriptsize $k_4$};
\end{tikzpicture}\ }
\eea

\noindent
An alternative name for the crossing kernel is the $R$-matrix.
The matrix $R_{k_sk_t}$ in fact depends on six numbers, $k_1, k_4, k_s,k_t, \ell_1$ and $\ell_2$,  that all label the spin of a corresponding sextuplet of representations of $SL(2,\mathbb{R})$.   It satisfies the unitarity property
\bea
\int\! d\mu(k)\; R_{k_s k} R^{\dag}_{kk_{t}} \is \frac{1}{\rho(k_s)}\, \delta(k_s - k_{t}), \qquad \qquad \rho(k) = 2k \sinh(2\pi k).
\eea

The explicit form of the R-matrix can be found in several different ways. The most convenient method uses the relation between the Schwarzian QM and 2D CFT. In section \ref{sect5} we will compute $R_{k_sk_t}
$ by taking a large $c$ limit of the CFT R-matrix that expresses the monodromy of 2D conformal blocks under analytic continuation over the lightcone. This 2D crossing kernel is explicitly known, thanks to the work of Ponsot and Teschner \cite{PT}, see also \cite{Teschner:2001rv, Teschner:2012em}. As shown in \cite{PT}, the 2D kernel can be expressed as a quantum 6j-symbol of the non-compact quantum group $U_q (\mathfrak{sl}(2,\mathbb{R}))$.
 Taking the large $c$ limit of their formulas, we obtain that
\bea
R_{k_sk_t}\! \left[\; {}^{k_4}_{k_1} \;{}^{\Deltal_2}_{\Deltal_1}\right] \! & \! =\! & \! \mathbb{W}(k_s, k_t ; \Deltal_1 + i k_4,\Deltal_1 - i k_4, \Deltal_2 - i k_1,\Deltal_2 + i k_1) \\[3.5mm]
\! & \!\! & \!\!\!\!\!\!\!\!\!\!\!\!\!\!\!\!\!\!\!\!\!\!\!\!\!\!\!\!\!\!\!\!\!\!\!  \times\,\sqrt{\Gamma(\Deltal_1 \pm i k_1 \pm ik_s)\Gamma(\Deltal_2 \pm i k_4 \pm ik_s)\Gamma(\Deltal_2 \pm ik_1\pm ik_t)\Gamma(\Deltal_1 \pm i k_4 \pm i k_t)}\nonumber\eea
where $ \mathbb{W}(a,b,c,d,e,f)$ denotes a so-called Wilson function, defined as a particular linear combination of two generalized hypergeometric functions ${}_4F_{3}$. The explicit formula is given in Appendix \ref{app:ZZ}. The Wilson function was introduced in \cite{groenevelt}, where it was shown that the above expression in fact coincides with the classical 6j-symbol of the Lie group $SU(1,1)$. 

The appearance of the 6j-symbols in OTO correlation functions should not come as a surprise. States and operators in the Schwarzian theory are specified by a representation label of ${SL}(2,\mathbb{R})$. The crossing kernel relates the OTO 4-point function with the corresponding time-ordered amplitude. It thus applies an isomorphism between two different orderings of taking a triple tensor product. 
The 6j-symbols satisfy some remarkable identities known as the pentagon and hexagon identities. From the point of view of the Schwarzian theory, these identities are consistency requirements that follow from locality, analyticity and associativity of the operator algebra. 

In a future paper, we intend to elaborate on the relation between the R-matrix of the Schwarzian theory and the gravitational scattering amplitudes of the 2D Jackiw-Teitelbom model. We will make some preliminary comments on this relation in section \ref{sect53}, where we outline how our results can be used to exhibit the expected maximal Lyapunov growth of the OTO correlation functions. 

This concludes our overview of the explicit expressions of the correlation functions of the Schwarzian theory. 
In the following sections, we will explain the method by which we obtained these results. We also present a few more details of the derivations and perform some non-trivial checks.

\section{Schr\"odinger formulation}
\label{sect:schro}
In this section, we outline the Hamiltonian formulation of the Schwarzian theory, and how it is related to other 1D systems with $SL(2,\mathbb{R})$ symmetry.  We temporarily set $\beta  = 2\pi$. The reader familiar with the basic properties of Schwarzian QM can choose to skip this section.

\subsection{Zero temperature}
We first consider the Schwarzian theory at zero temperature. In this limit, the $\dot{f}^2$-term is dropped in the action (\ref{schwaction}), reducing it to the pure Schwarzian action $S =  \int d\tau \left\{f,\tau\right\}$.\footnote{Here, in this section only, we will write $\dot{f}(\tau)$ instead of $f'(\tau)$.} To transit to a Hamiltonian description, it is useful to recast the Lagrangian into a first order form as
\bea
L \is \pi_\phi \dot{\phi} + \pi_f \dot{f} - (\pi_\phi^2 + \pi_f e^{\phi}).
\eea
This first-order form makes clear that the Schwarzian theory has a four dimensional phase space, labeled by two pairs of canonical variables $(\phi, \pi_\phi)$ and $(f,\pi_f)$. Alternatively, we may view the quantity $\pi_f$ as a Lagrange multiplier, enforcing the constraint $\dot{f} = e^{\phi}$. Setting $\phi = \log \dot f$ and integrating out $\pi_\phi$, it is readily seen that the above first-order Lagrangian indeed reduces to the Schwarzian theory. Upon quantization, the variables satisfy canonical commutation relations $
[f,\pi_f] \, = \, i $ and $ [\phi, \pi_\phi] = i$.

\begin{figure}[t]
\centering
\includegraphics[width=.96\textwidth]{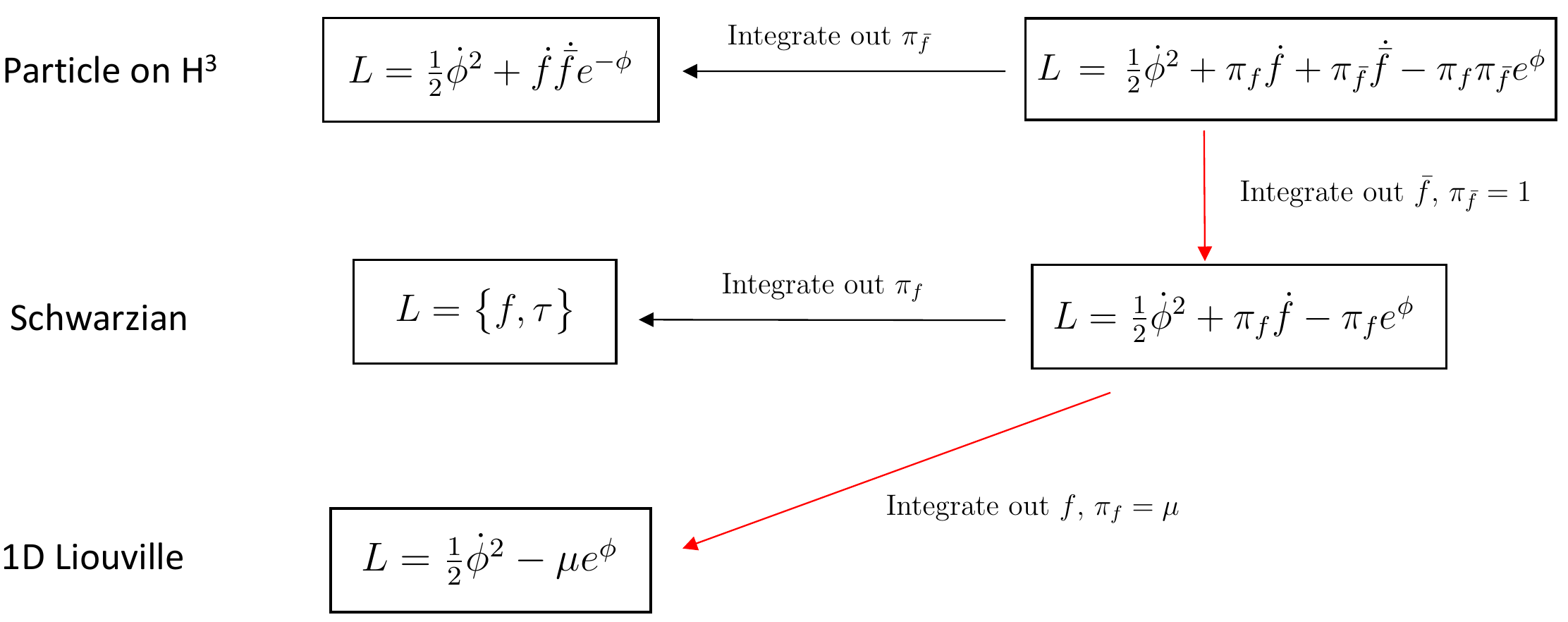}
\caption{Overview of different models with  underlying $SL(2,\mathbb{R})$ symmetry. Red lines indicate one-way lines: they are projections that reduce the dimension of the phase space.}
\label{schemeofmodels}
\end{figure}

The invariance of the Schwarzian action under M\"obius transformations 
\bea
\label{mobius}
f \to \frac{af+b}{cf+d}
\eea
implies the presence of a set of conserved charges
\bea
\label{lmin}
\ell_{-1} \is \pi_f, \qquad \nonumber \ \
\ell_0 \, = \, f \pi_f  + \pi_\phi, \qquad \ \
\ell_1 \, = \, f^2\pi_f  + 2 f \pi_\phi + e^{\phi}, \nonumber
\eea
that satisfy an $\mathfrak{sl}(2,\mathbb{R})$ algebra. The Hamiltonian $H$ is equal to the quadratic Casimir 
\bea
\label{hamus}
H \is \pi_\phi^2 + \pi_f e^{\phi} \, = \ \ell_0^2  -\textstyle  \frac 1 2  \{ \ell_{-1}, \ell_1\} 
\eea 
and thus manifestly commutes with the $SL(2,\mathbb{R})$ symmetry generators. In particular, we can define a mutual eigenbasis of $H$ and $\pi_f = \ell_{-1}$
\bea
\label{eigenbasis}
\pi_f \bigl| \lambda, k\ra \! \is\! \lambda \bigl|  \lambda, k\ra, \qquad \quad 
H \bigl| \lambda, k\ra = E(k)\bigl| \lambda, k\ra, \qquad  E(k) \equiv \textstyle \frac 1 4 + k^2,
\eea
which spans the complete Hilbert space of the theory.

The 1D Schwarzian theory is closely related to the free particle on the 3D Euclidean AdS space $H_3^+$ with coordinates $(\phi,f, \bar f)$ and metric $ds^2 = d\phi^2 + 2 e^{-\phi} df d\bar{f}$, and to 1D Liouville theory. The different 1D models and their connections are summarized in Figure \ref{schemeofmodels}. The ${H}^3_+$ model has $SL(2,\mathbb{R}) \times SL(2,\mathbb{R})$ symmetry, which is broken to $SL(2,\mathbb{R})$ by setting the momentum variable $\bar\pi_f$ equal to a constant. Similarly, the reduction to the 1D Liouville theory proceeds by setting $\pi_f = \mu$, which breaks all symmetry.

\subsection{Finite temperature}
Putting the theory at finite temperature (we continue to set $\beta = 2\pi$ for convenience) reintroduces the extra $\dot{f}^2$-term in the action (\ref{schwaction}). The effect of this term in the first order formulation is taken into account by changing the Hamiltonian to
\bea
H \is \pi_\phi^2 + \pi_f e^{\phi} + e^{2\phi}.
\eea
Upon solving the constraint $\dot{f} = e^\phi$, the added term reduces to $e^{2\phi} = \dot{f}^2$.
This Hamiltonian still has $SL(2,\mathbb{R})$ symmetry generated by the charges
\begin{align}
\ell_{-1} \! & \!=  \cos^2(f)\, \pi_f - \sin(2f)\pi_\phi + \cos(2f)e^{\phi},\nonumber \\[1mm]
\ell_{0} \! & \!= \textstyle \frac 1 2 \sin(2f) \, \pi_f + \cos(2f)\pi_\phi + \sin(2f) e^{\phi}, \\[1mm]
\ell_{1} \! & \!= \sin^2(f)\, \pi_f + \sin(2f)\pi_\phi - \cos(2f)e^{\phi}.\nonumber
\end{align}
These charges satisfy $\left[\ell_{0},\ell_{\pm 1}\right] = \mp \ell_{\pm 1}$ and $\left[\ell_{1},\ell_{- 1}\right] = 2\ell_0$ and all commute with the Hamiltonian, which can again be identified with the quadratic Casimir operator $H = \ell_0^2 - \frac 1 2 \{ \ell_1, \ell_{-1}\}$. The $SL(2,\mathbb{R})$ symmetry generated by these charges acts via broken linear transformations on the uniformizing variable $F$
\bea
\label{gmobius}
F \to \frac{aF + b}{cF + d}, \qquad \quad F = \tan(f/2), \qquad ad-bc=1.
\eea 

Since $\pi_f = \ell_1 + \ell_{-1}$ commutes with $H$, we can again define a mutual eigenbasis (\ref{eigenbasis})
that span the full Hilbert space of the model. The Schr\"odinger wavefunctions of the eigenstates take the form $
\Psi_{\lambda,k}(f,\phi) = e^{i \lambda f} \psi_{\lambda, k} (\phi)$
where $\psi_{\lambda, k} (\phi)$ solves the Schr\"odinger equation
\bea
\left(-\partial_\phi^2 + \lambda e^{\phi} + e^{2\phi}\right) \psi_{\lambda,k}(\phi) \is k^2 \psi_{\lambda,k}(\phi),
\eea
given by a 1D particle in a Morse potential $V(\phi) = \lambda e^{\phi} + e^{2\phi}$. The solutions are given in terms of Whittaker $W$-functions. The full eigenmode functions normalized in the flat measure $df\, d\phi$ are given by
\bea
\label{eigenf}
\Psi_{\lambda,k}(f, \phi) \is \sqrt{\frac{k \sinh(2\pi k)}{4\pi^3}}\; \bigl|\Gamma\bigl(ik+\textstyle \lambda/2 + 1/2\bigr)\bigr|\, e^{i\lambda f} e^{-\phi/2}\, W_{-{\lambda}/2, ik}\left(2e^{\phi}\right)_{\strut{}}.
\eea

\subsection{Particle in a magnetic field}
There exists an interesting and useful connection between the Schwarzian model and a particle on the hyperbolic plane $H_2^+$ in a constant magnetic field \cite{Kitaev16}. The Landau problem on $H_2^+$ was first analyzed by A.~Comtet and P.~J.~Houston in \cite{Comtet}. A main result of \cite{Comtet}, which also turns out to be useful for our problem, is an explicit formula for the spectral density of states.

Writing the $H^2_+$ metric as $ds^2 = d\phi^2+e^{-2\phi}df^2,$
the Lagrangian of the particle is given by
\bea
S \is  \int \!  dt \, \Bigl( \frac 1 4 \dot{\phi}^2+ \frac 14 \spc e^{-2\phi}\spc {\dot{f}^2}  + B \dot{f}e^{-\phi}\Bigr),
\eea
which identifies the magnetic vector potential as $q A_f = B e^{-\phi}$ with $q$ the charge of the particle. 
The Hamiltonian of this system, for fixed constant $B$, is
\bea
H_B \is p_\phi^2 + \left(p_f e^{\phi} - B\right)^2,
\eea
where we denoted the canonical conjugate variables by $p_\phi$ and $p_f$.
The model is again invariant under M\"obius transformations \eqref{mobius} and possesses a corresponding set of $SL(2,\mathbb{R})$ symmetry generators 
\bea
\ell_{-1}\! \is\! p_f \qquad,\ \ 
\ell_{0} \spc = \spc f p_f + p_\phi \qquad,\ \ 
\ell_1 \spc = \spc f^2 p_f + 2f p_\phi - p_f e^{2\phi} + 2Be^\phi.
\eea
Once again, the Hamiltonian is equal to the quadratic Casimir. The normalized simultaneous eigenmodes of $p_f$ (with eigenvalue $\nu$) and $H_B$ (with eigenvalue $E(k) = \frac 1 4 + k^2 + B^2$) take the form \cite{Comtet}
\beq
\Psi_{\nu,k}(f,\phi) = \sqrt{\frac{k \sinh(2\pi k)}{4\pi^3|\nu |}}\; \bigl|\Gamma\bigl(ik-B+ 1/2\bigr)\bigr|\, e^{i \nu f}\, e^{-\phi/2}\, W_{B, ik}\left(2|\nu| e^{\phi}\right).
\eeq
This should be compared with formula (\ref{eigenf}) for the eigenmodes of the Schwarzian model.

Using the above formula for the eigenmodes, it is straightforward to compute the density of states for the Landau problem on $H^2_+$. The result for spectral measure reads
\bea
\label{comtet}
d\mu_B(k) \is \rho_B(k) dk \, = \, dk^2 \frac{\sinh(2\pi k )}{\cosh(2\pi k ) + \cos(2\pi B)}.
\eea

We can use this result to compute the spectral measure of the Schwarzian theory via the following observation \cite{Kitaev16}.  Upon shifting $\phi \to \phi - \log(-2B)$ with $B\to i\infty$, the Hamiltonian $H_B$ reduces to
\bea
H_B \is p_\phi^2 + p_f e^{\phi} + B^2,
\eea
which, up to the irrelevant constant $B^2$-contribution,  coincides with the Hamiltonian (\ref{hamus}) for the Schwarzian model at zero temperature. We can use this correspondence to derive the exact formula for the spectral measure (\ref{smeasure}) of the Schwarzian theory quoted in the introduction. Starting from Comtet's result (\ref{comtet}) and using that $\cos(2\pi B)$ diverges as $B \to i\infty$, we deduce that (up to an irrelevant overall normalization)
\bea
d\mu(k) \is dk^2 \sinh(2\pi k ).
\eea

\section{Partition function: a 2D Perspective}
\label{sect3}
In this section we will study the path integral formulation of the Schwarzian theory at finite temperature. In particular, we will use its relationship to the group Diff($S^1$) to reformulate 1D Schwarzian QM as a suitable large $c$ limit of 2D Virasoro CFT.\footnote{Related ideas are formulated in \cite{Mandal:2017thl}.}

The partition function of the Schwarzian theory \eqref{schwaction} is defined as the integral
\bea
\label{zpart}
Z (\beta)\is \int
\! \frac{{\cal D}\! \spc f}{SL(2,\mathbb{R})} \, e^{- \spc S[f] } 
\eea
over invertible functions $f$, satisfying the periodicity and monotonicity constraints $f(\tau+\beta) = f(\tau) + \beta$  and $f'(\tau) > 0.$
The space of functions with these properties specifies the group Diff($S^1$) of diffeomorphisms of the circle, also known as the Virasoro group.  

The $SL(2,\mathbb{R})$ quotient in (\ref{zpart}) indicates that the functional integral runs over the infinite dimensional quotient space \bea
\label{mspace}
{\cal M} \, = \, {\rm Diff}(S^1)/SL(2,\mathbb{R})
\eea
of diffeomorphisms modulo the group of M\"obius transformations (\ref{gmobius}) acting on $F\!=\!\tan(\frac{\pi f}\beta)$.
This space ${\cal M}$ is called the coadjoint orbit of the identity element $\mathbb{1} \in$ Diff$(S^1)$, which is known to be a symplectic manifold \cite{Kirillov, LazutkinSegal}. Its symplectic form takes the following form
\beq
\label{symp}
\omega \, = \, \int_0^{2\pi}\!\!\! dx \, \left[ \frac{df' \wedge df''}{f'^2} - df\wedge df' \right].
\eeq
This observation was used by Stanford and Witten \cite{wittenstanford} to evaluate the functional integral with the help of the Duistermaat-Heckman (DH) formula \cite{Duistermaat:1982vw}. 

The DH formula applies to any integral over a symplectic manifold of the schematic form
\bea
\label{phasint}
{I} \is  \int\! {\rm dp d q} \, e^{-\spc {\rm H}({\rm p,q})}
\eea
where H(p,q) generates, via the Poisson bracket $\{ q, p\}=1$,  a U(1) symmetry of the manifold.  In this paper we will apply a somewhat different argument: 
instead of the DH theorem, we will use the general fact that 
the phase space integral of the form (\ref{phasint}) is equal to the $\hbar \to 0$ limit of the trace of the quantum operator $e^{-{\rm H(p,q)}}$ over the Hilbert space obtained by quantizing the phase space:
\bea
\label{ilimit}
I \is\, \lim_{\hbar \to 0}  \; {\rm Tr}\bigl(\spc e^{-\spc {\rm H(p,q})}\spc \bigr).
\eea 
The physical intuition that underlies this equality is that for small $\hbar$, the Hilbert space admits an orthogonal basis of states each localized within a Planck cell in phase space. The trace then takes the form of a sum over all Planck cells, which in the $\hbar \to 0$ limit reduces to the phase space integral defined via the symplectic measure. 

The strategy  that we plan to follow is to exploit the fact that, if there exists a precise way to quantize the phase space ${\cal M}$ and construct the corresponding Hilbert space, then the formula (\ref{ilimit}) provides an exact and efficient way of computing the integral~$I$.

\subsection{Spectral density from modular bootstrap} 

In our problem, the phase space $({\cal M}, \omega)$ specified by equations (\ref{mspace}) and (\ref{symp}) can be quantized through the standard methods of co-adjoint orbit quantization. The details of this quantization step are explained in detail in \cite{Kirillov, LazutkinSegal, Witten:1987ty, AS}. It is customary to label the quantization parameter $\hbar$ via
\bea
\hbar \is \frac{24\pi}{c}
\eea
and introduce the following basis of $SL(2,\mathbb{R})$ invariant functions on ${\cal M}$
\bea
L_n \is \frac{\beta\spc c }{48\pi^2} \int_0^{\beta}\!\! d\tau\, e^{2\pi i n\tau/\beta}\, \bigl\{F,\tau\bigr\}.
\eea
The main statement that we will need for our purpose is that in the quantum theory, these functions $L_n$ become identified with the generators of the Virasoro algebra
\bea
[L_n,L_m] \is (n-m) L_{n+m} + \frac{c}{12} (n^3-n)\delta_{n+m}
\eea
at central charge $c$. The classical limit $\hbar \to 0$ corresponds to the large central charge limit $c\to \infty$.  The Hilbert space of the quantum theory is given by the identity module of the Virasoro algebra, i.e the linear space spanned by all states obtained by acting with $L_{-n}$'s with $n>2$ on the $SL(2,\mathbb{R})$ invariant vacuum state $|0\ra$. 

The Schwarzian action 
\bea
S[f] \is - \frac 1 2 \int_0^\beta\!\! d\tau\, \{ F, \tau\} \; = \; -\frac{24\pi^2}{\beta\spc c}\, L_0 
\eea
is the generator of a $U(1)$ symmetry $f(\tau) \to f(\tau + \delta)$. This fact was used in \cite{wittenstanford} to invoke the DH formula and conclude that the partition function $Z(\beta)$ is one-loop exact. 

For our purpose, the relevant observation is that the exponential of the Schwarzian action can be expressed as an evolution operator 
\bea
e^{-S[f]} \is q^{L_0}, \qquad  \qquad  q \equiv  e^{\mbox{\footnotesize $-{\frac{24\pi^2}{\beta c}}$}}
\eea
in the quantum theory. We are now ready to apply the above argument, that relates the phase space integral (\ref{phasint}) and the $\hbar\to 0$ limit of the trace (\ref{ilimit}), to  the Schwarzian partition function (\ref{zpart}). 
We obtain the following identity
\bea
\label{zbetatrace}
\quad Z (\beta) \is \, \lim_{\raisebox{-4pt}{${{}_{c\to \infty}}\atop{{}^{q \to 1}}$}}{\rm Tr}\bigl(q^{L_0} \bigr), \qquad \qquad q^{\frac{c}{24}} = e^{-\frac{\pi^2}{ \beta}} = {\rm fixed}. 
\eea
where the trace is over the identity module of the Virasoro algebra. The quantity $\chi_0(q) = {\rm Tr}(q^{L_0})$  is the identity character of the Virasoro algebra. Geometrically, it represents the torus partition function of a chiral identity sector of a 2D CFT. Taking the limit $q\to 1$ amounts to sending the modular parameter $\tau\to 0$. In this limit, the torus degenerates into an infinitesimally thin circular tube. The long direction of the circular tube is the original thermal circle of the Schwarzian theory. The short direction is a fiducial circle that we added in order to write the integral over ${\cal M}$ as a trace.
 
The identity character of a $c>1$ CFT takes the form
\beq
\label{chizero}
 {\rm Tr}\bigl(q^{L_0} \bigr)\, \equiv \, \chi_0 (q)\; =\, \frac{q^{\mbox{\footnotesize $\frac{1-c}{24}$}} (1-q)}{\eta(\tau)},
 \eeq
 where $\eta(\tau)$ denotes the Dedekind eta function $\eta(\tau) = q^{\frac 1 {24}}
 \prod_{n=1}^\infty (1-q^n)$ with $q = e^{2\pi i\tau}$.\footnote{We apologize to the reader for temporarily also using the symbol $\tau$ for the modular parameter $q=e^{2\pi i \tau}$ of the torus. Using equation (\ref{zbetatrace}), we can express the modular parameter $\tau$ in terms of the temperature $\beta$  of the Schwarzian and the central charge $c$ of the auxiliary 2D CFT via 
 \bea
 \tau \is \frac{12\pi i}{\beta c}.
 \eea}
 The factor $(1-q)$ in the above formula for the identity character accounts for the presence of the null state $L_{-1} | 0 \rangle = 0$.
 
It is now straightforward to combine equations (\ref{zbetatrace})-(\ref{chizero}) and extract an exact expression for the Schwarzian partition function. This can be done in two ways. First, from the identity $\eta(-\frac{1}{\tau})=\sqrt{\tau_2}\spc \eta(\tau)$ we derive that for $q \sim 1$, we can replace $\eta(\tau) \sim (\tau_2)^{-1/2} e^{i \pi \tau/6}$. Using this result, we can directly take the large $c$ limit of equation (\ref{zbetatrace}) and deduce that $Z(\beta)$ takes the following form  
\bea
Z(\beta) \is e^{S_0 + \beta E_0}\, \left(\frac{\pi}{\beta} \right)^{3/2}\! \exp{\Bigl(\spc \frac{\pi^2}{\beta}\Bigr)}.
\eea
Here we absorbed a (divergent) zero-point entropy $S_0$ and a zero-point energy $E_0$ contribution in the prefactor. This formula matches with the exact result found in \cite{wittenstanford, Cotler:2016fpe}.

Alternatively, we can apply the modular transformation $\tau\to-{1}/{\tau}$ directly to the identity character $\chi_0(q)$ as a whole, and use the known formula for the modular $S$-matrix for $c>1$ Virasoro CFT to decompose the result in terms of Virasoro characters in the dual channel. For this it is convenient to parametrize the highest weights $\Delta$ of the Virasoro representations and the central charge $c$ as follows\footnote{This parametrization is familiar from Liouville CFT. We emphasize, however, that in this section we are using completely general properties of genus one Virasoro characters.}
\bea
\label{lxparam}
\Delta(P) \is \frac{Q^2}{4} + P^2, \qquad \qquad 
c \, = \,  1 + 6 Q^2 \, =\, 1 + 6\bigl(b + b^{-1}\bigr)^2.
\eea
The modular transformation rule of the Virasoro characters then reads
\bea
\label{mod}
\chi_0 \left(q\right)\! \is\! \int_0^\infty\!\!\! dP~S^P_0~\chi_P(\tilde{q})
, \qquad \  \tilde{q} \spc = \spc e^{-\frac{\beta c}{6}},\qquad \ 
\chi_P(\tilde{q}) \spc = \spc \frac{ \tilde{q}^{P^2}}{\eta(\tilde\tau)},
\eea
 where the modular S-matrix is given by 
\bea
S^P_0 \is 4\sqrt{2} \sinh \bigl( 2 \pi b P\bigr) \sinh\Bigl( \mbox{\Large $\frac{\mbox{\footnotesize $2$} \pi P}{b}$} \Bigr).
\eea
We now set 
\bea
 k \is \frac{P}{b} , \qquad \qquad  E\spc\, = \, \frac{\Delta - \frac{c-1}{24}}{b^2}\, =\,  k^2_{\strut{}}
\eea
and take the limit $b\to 0$ (which sends $c\to \infty$) while keeping $k$, $E$ fixed. In this limit
\bea
\label{blimit}
S_0^P\!  &  \!\sim  \! & \! 2k \sinh(2\pi k), \qquad \qquad \chi_P(\tilde{q})\, \sim \, e^{-\beta k^2}.
\eea
The second formula has a clear physical significance. The large $c$ limit sends $\tilde{q} \to 0$, which turns the operator $\tilde{q}^{L_0}$ into a projection operator on the lowest energy state in the given channel. 
Combining (\ref{zbetatrace}), (\ref{mod}) and (\ref{blimit}) we obtain that
\bea
\label{zspectral}
Z(\beta) \is  \int_0^\infty\! \!\! d\mu(k) \, e^{- \beta E(k)}, \qquad \quad d\mu(k) = dk^2 \sinh(2\pi k),
\eea
reproducing the result obtained in \cite{wittenstanford}.

\begin{figure}[t]
\centering
\includegraphics[width=.62\textwidth]{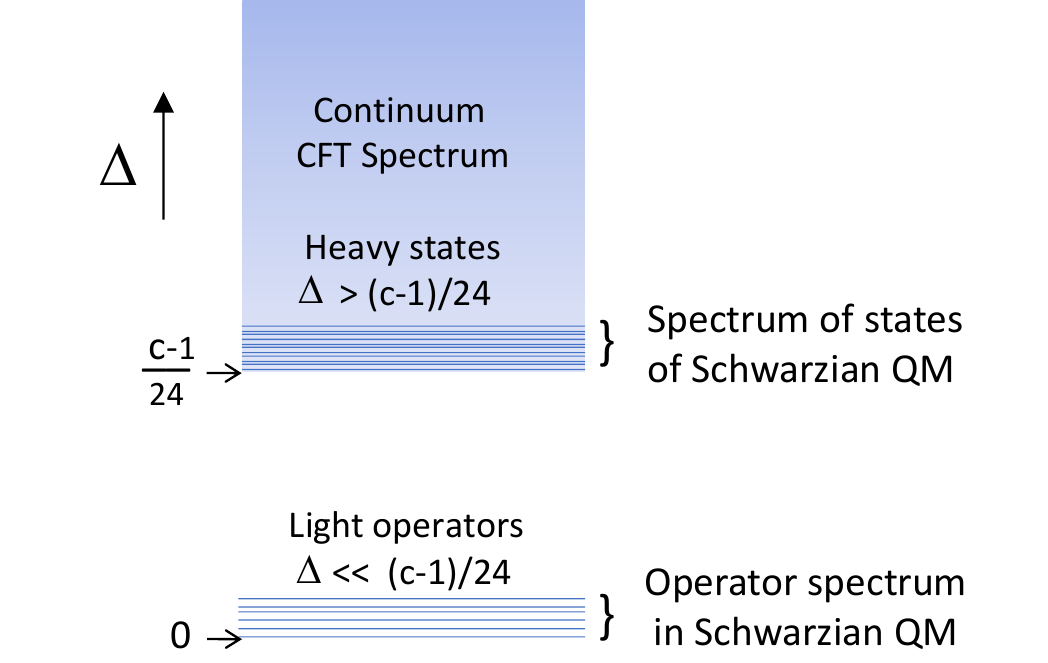}
\caption{\small The spectrum of states in the Schwarzian theory arises from the CFT spectrum of states with conformal dimension $\Delta = \frac{c-1}{24} + b^2 E$, in the limit $b\to 0$. The operators in the Schwarzian are all light CFT operators with conformal dimension $\Delta = \ell$.}
\label{schpectrum}
\end{figure}

While the explicit formula (\ref{zspectral}) for the spectral density is not a new result, our derivation provides a new and useful perspective on the Schwarzian theory. Specifically, it indicates that the 1D model arises as a special $c\to \infty$ limit of 2D Virasoro CFT, in which we only keep the states with conformal dimensions $\Delta$ close to the threshold $\Delta_c = \frac{c}{24}$ (Figure \ref{schpectrum}). 

The above modular bootstrap argument identifies a natural spectral density on the space of Virasoro representations, given by the modular S-matrix element $S_0^P$ \cite{LaurenHV}. This spectral density is not a specific property of a particular 2D CFT, but a universal measure analogous to the Plancherel measure on the space of continuous series representations of $SL(2,\mathbb{R})$. This measure is defined for any value of the central charge $c$. We have shown that, after taking the large $c$ limit while zooming in close to  $\Delta_c = \frac{c-1}{24}$, it coincides with the exact spectral density of the Schwarzian theory.
In the following sections we will generalize this observation with the aim of studying correlation functions. 

\subsection{Spectral density from ZZ branes}

As further preparation for the study correlation functions, it is useful to derive the formula for the spectral density from yet another slightly different perspective. As mentioned above, the identity character $\chi_0(q)$ represents the chiral genus one partition function of the identity sector of the Virasoro CFT. Alternatively, we can identify $\chi_0(q)$ with the partition function of the Virasoro CFT on the annulus. This annulus partition function is equal a trace over an open string sector of the Virasoro CFT, or by using channel duality, as the transition amplitude between two ZZ boundary states \cite{Cardy, Fateev:2000ik}.
\bea
\chi_0(q) \is \lb ZZ | \tilde{q}^{L_{0}} | ZZ \rb.
\eea
The Schwarzian theory arises in the limit $q\to 1$, which in the dual closed string channel corresponds to the limit $\tilde{q} \to 0$, as shown in Figure \ref{fig:open-close}. Note that the insertion of the ZZ branes cuts the thermal circle of the Schwarzian theory into two halves.\footnote{This approach to the geometric quantization of the Virasoro group seems related to the one put forward in \cite{Gukov:2008ve} for compact groups, but using a topological theory instead of a CFT.}

\begin{figure}[t!]
\begin{center}
 \begin{tikzpicture}[scale=0.62]
\draw (-7,0) node {${\rm Tr}\bigl(q^{L_0}\bigr) \ = $};
\draw[thick] (-3-1,-0.5) -- (3-1,-0.5);
\draw[thick] (3-1,0.5) -- (-3-1,0.5);
\draw[thick] (-1-3,0) ellipse (0.17 and 0.5);
\draw[thick] (-1+3,0) ellipse (0.17 and 0.5);
\draw[dashed] (-1,0) ellipse (0.17 and 0.5);
\draw[thick,->] (5,0) --(6,0);
\draw[thick] (9,-3) -- (9,3);
\draw[thick] (10,-3) -- (10,3);
\draw[thick] (9.5,-3) ellipse (0.5 and 0.17);
\draw[thick] (9.5,3) ellipse (0.5 and 0.17);
\draw[dashed] (9.5,0) ellipse (0.5 and 0.17);
\draw (9.5,-3.8) node {\small $|{\rm ZZ}\rb$};
\draw (9.5,3.8) node {\small $\lb{\rm ZZ}|$};
\draw (14,0) node {= \ $\langle ZZ |\, \tilde{q}^{L_0}\spc |ZZ\rangle$};
\end{tikzpicture}
\end{center}
\vspace{-0.5cm}
\caption{\small  The identity character can be represented as the annulus partition sum of the Virasoro CFT, or by using channel duality, as the transition amplitude between two ZZ boundary states.}
\label{fig:open-close}
\end{figure}

The ZZ boundary state is given as an integral over Ishibashi boundary states \cite{Fateev:2000ik}\footnote{Here and in the following, we drop irrelevant overall constant factors.}
\bea
\label{zzwave}
|ZZ\rb\! \is \! \int_0^\infty\!\! dP \, \Psi_{\rm ZZ}(P)\, |\hspace{-0.3mm}| P\rb\!\rb,~~~~\Psi_{\rm ZZ}(P)=\frac{2 \pi i P }{\Gamma(1-2 i b P)\Gamma(1+\frac{2iP}{b})}. 
\qquad
\eea
In the limit we are considering the boundary states are associated to a circle with a radius that goes to zero (if we map the cylinder to the complex plane) and this allows us to approximate $|\hspace{-0.3mm}| P\rb\!\rb \to |P\rb$. This is the main feature that will allow us later to compute correlation functions since it can be used to turn a correlation function between ZZ-branes into an integral of a correlation function on the sphere. Using this and taking $\tilde{q} = e^{-\beta /b^{2}}$, where $\beta$ is the temperature of the Schwarzian theory, the partition function becomes 
\bea
Z \! \is\! \int_0^\infty\!\!\! dP ~|\Psi_{\rm ZZ}(P)|^2 ~e^{- \beta \frac{P^2}{b^2}},~~~~~|\Psi_{\rm ZZ}(P)|^2= \sinh\bigl( 2 \pi b P\bigr) \sinh\Bigl(\mbox{\Large $\frac{\mbox{\small 2} \pi P}{b}$}\Bigr).
\eea
For small $b$ this integral is dominated by states with $P$ of order $b $. Therefore we define $P=k b$ and take the  $b\to 0$ limit; we recover the result (\ref{zspectral}).

\section{Schwarzian correlators from ZZ branes}\label{sec:ZZbosonic}
In this section we will exploit the relationship between the Schwarzian theory and Virasoro CFT to compute finite temperature correlation functions of $SL(2,\mathbb{R})$ invariant operators in the Schwarzian theory. The simplest such operator is the Schwarzian itself. Its correlation functions are completely fixed by symmetries and are described in Appendix \ref{sect:stress}.

A more interesting class of correlation functions are those involving the bi-local operators
\beq
\label{bilocalo}
\mathcal{O}_\Deltal(\tau_1,\tau_2) \equiv 
\left( \frac{\sqrt{f'(\tau_1)f'(\tau_2)}}{\frac{\beta}{\pi}\sin \frac{\pi}{\beta}[f(\tau_1)-f(\tau_2)]} \right)^{2\Deltal}.
\eeq
These operators naturally live on the 2D space ${\cal K}$ parametrized by pairs of points $(\tau_1,\tau_2)$ on the thermal circle. We will call ${\cal K}$ kinematic space, since it plays an analogous geometrical role as the kinematic space associated with 2D holographic CFTs ~\cite{Czech:2015qta, Czech:2016xec}.

To exhibit the geometry of kinematic space ${\cal K}$, let us  -- motivated by the form (\ref{bilocalo}) of the bi-local operators -- associate to any point $(u,v) \in {\cal K}$ a classical field $\phi_{cl}(u,v)$ via
\beq\label{ZZcl}
e^{\phi_{cl}(u,v)} = \frac{\sqrt{f'(u)f'({v})}}{\frac{\beta}{\pi}\sin \frac{\pi}{\beta}[f(u)-f({v})]}.
\eeq
This field satisfies the Liouville equation
\bea
\partial_u \partial_v \phi_{cl}(u,v) = e^{2\phi_{cl}(u,v)}.
\eea
Hence kinematic space ${\cal K}$ naturally comes with a constant curvature metric $ds^2 = e^{2\phi(u,v)} du dv$, and looks like a hyperbolic cylinder with an asymptotic boundary located at $u=v$. Note, however, that the metric on kinematic space is now a dynamical quantity that depends on the dynamical diffeomorphism $f(\tau)$.

\begin{figure}[t!]
\begin{center}
 \begin{tikzpicture}[xscale=1.3,yscale=0.8,rotate=90]
 \fill [draw=none, fill=blue, opacity=0.1] (-2,-1.985) to [bend right=60] (-2,2) -- (-1.8,2.10)-- (-1.7, 2.12) -- (-0.2,2.2) --  (0,2.2)--  (0.2,2.2) -- (1.7,2.12) -- (1.8,2.10)-- (2,2) to [bend right=60] (2,-1.985) -- (2, -2) arc (0:-180:2 and 0.2)-- cycle ;
\draw[thick] (2, -2) arc (0:-180:2 and 0.2);
\draw[thick,dashed] (2, -2) arc (0:180:2 and 0.2);
\draw[thick, fill= blue, opacity= 0.13] (0,2) ellipse (2 and 0.2);
\draw[draw=none, fill= blue,opacity= 0.07] (0,-2) ellipse (2 and 0.2);
\draw[thick] (0,2) ellipse (2 and 0.2);
\draw (0,-2.6) node {$ZZ$};
\draw (0,2.6) node {$ZZ$};
\end{tikzpicture}
\vspace{-3mm}
\end{center}
\caption{\small  Geometry of the classical Liouville background between two ZZ branes.}
\label{figZZgeo}
\end{figure}
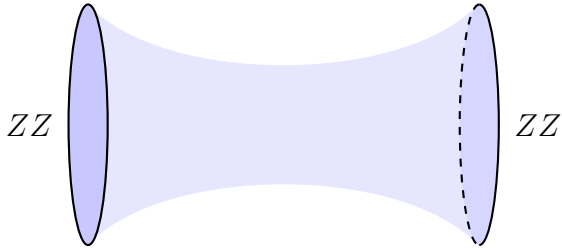

\subsection{ZZ branes and kinematic space}

Given the similarity between the two geometric structures, it is tempting to look for a direct identification between the kinematic space ${\cal K}$ and the geometry of 
Liouville CFT bounded by two ZZ-branes. To make this idea more explicit, let us consider Liouville CFT with ZZ branes placed at the spatial positions $\sigma=0$ and $\sigma=\pi$. The time direction is parametrized by $\tau$. The action describing this system is 
\bea\label{eq:LiouvilleAction}
S  \is \frac{c}{192\pi} \int \! d\tau \int_0^\pi\!\! d\sigma\, \left[ (\partial \phi)^2 +  4 \mu e^{2 \phi} \right] 
\eea
For our application, the only role of the Liouville CFT is to provide a convenient geometrical description of the Virasoro partition function and conformal blocks. Indeed, Liouville theory is known to be equivalent to the geometric Lagrangian associated with the symplectic form $\omega$ on Diff$(S^1)$ quoted in the previous section.\footnote{The parameters $Q = b + b^{-1}$ and  $P$ used in the expressions (\ref{lxparam}) of the central charge $c$ and the conformal dimension $\Delta$  are naturally identified with the background charge of the Liouville CFT and  the `Liouville momenta' of the  vertex operators $V_P$ with conformal dimension $\Delta$.}

We introduce the light-cone coordinates $u=\tau+\sigma$ and $v=\tau-\sigma$. 
We are interested in the limit $c\to \infty$. In this limit, the functional integral localizes on the space of classical solutions to the Liouville equation of motion.
The boundary conditions of $\phi$ are that the regions near $\sigma = 0$  and $\sigma = \pi$ corresponds to the asymptotic  regions of a hyperbolic cylinder. It is shown in \cite{Dorn:2006ys} that the lowest energy solution is $4\mu^4 e^{2\phi} = \sin^{-2} \sigma$. Written in the form $ds^2=e^{2\phi}dudv$ this describes a hyperbolic geometry of the form shown in Figure \ref{figZZgeo}. 

\begin{figure}[t]
\begin{center}
 \begin{tikzpicture}[scale=1,rotate=0]
 \fill [draw=none, fill=blue, opacity=0.1] (-2,-1.985) to [bend right=60] (-2,2) -- (-1.8,2.10)-- (-1.7, 2.12) -- (-0.2,2.2) --  (0,2.2)--  (0.2,2.2) -- (1.7,2.12) -- (1.8,2.10)-- (2,2) to [bend right=60] (2,-1.985) -- (2, -2) arc (0:-180:2 and 0.2)-- cycle ;
\draw[thick] (2, -2) arc (0:-180:2 and 0.2);
\draw[thick,dashed] (2, -2) arc (0:180:2 and 0.2);
\draw[thick, fill= blue, opacity= 0.13] (0,2) ellipse (2 and 0.2);
\draw[draw=none, fill= blue,opacity= 0.07] (0,-2) ellipse (2 and 0.2);
\draw[thick] (0,2) ellipse (2 and 0.2);
\draw (0,-.5) node {\small $e^{\ell\phi(u,v)}$};
\draw[thick] (-1,-2.17) to [bend right=10] (0,-1) to [bend right=10](1,-2.17);
\draw[thick,fill] (0,-1) circle (0.1);
\draw (-1,-2.5) node {\small $u$};
\draw (1,-2.5) node {\small $v$};
\end{tikzpicture}
\vspace{-0.7cm}
\end{center}
\caption{\small The kinematic space of the Schwarzian theory. The bi-local operator (\ref{bilocalo}) in the 1D QM is represented by a local Liouville CFT vertex operator in the 2D bulk. The boundary of the kinematic space corresponds to the limit where the two end-points of the bi-local operator coincide.}
\label{figZZkin}
\end{figure}
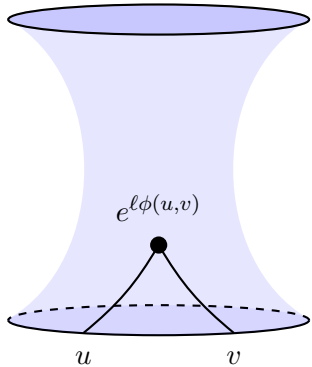

As explained e.g. in \cite{Balog:1997zz}, the most general classical solution of Liouville theory can be obtained by starting with a representative $\phi(u,v)$ for a given conformal class and then apply a general conformal transformation $e^{2\phi(u,v)}\to {f'(u)f'(v)} e^{2\phi(f(u),f(v))}$. 
The most general solution thus takes the form given in equation (\ref{ZZcl}), after performing a rescaling that maps the distance between the ZZ-branes from $\pi$ to $\beta/2$. These solutions are all isomorphic to the geometry shown in Figure \ref{figZZkin}. We can interpret this 2d space as a kinematic space of the Schwarzian theory. Note, however, that in our case, the kinematic space is in fact dynamical.

Finally, we remark that the equivalence between the Schwarzian and the large $c$ limit of Liouville CFT is of course not surprising. It is well-known that the Liouville stress tensor $T = \frac 1 2 (\phi')^2 + \phi''$ reduces to the Schwarzian derivative when evaluated on a general classical solution of the form (\ref{ZZcl}). 
This observation can be used to show that the Liouville lagrangian in a combined large $c$ and DLCQ limit reduces to the Schwarzian action.

\subsection{Two-point function}\label{sec:2ptZZ}

In this section we will study the correlation function of a single bi-local operator 
\bea
G_{\Deltal}(\tau_1,\tau_2)=\lb \mathcal{O}_{\Deltal}(\tau_1,\tau_2)\rb.
\eea
It is here that our method really gets extra mileage compared to other approaches. From the saddle-point solution (\ref{ZZcl}) for the field $\phi$ we see that the Liouville vertex operators $e^{2\Deltal \phi(u,v) }$ and the bi-local operators $\mathcal{O}_\Deltal(\tau_1,\tau_2)$ placed between two ZZ branes become identical, if we identify $u=\tau_1$ and $v=\tau_2$.
Motivated by this, we will propose the following identification between the correlation functions of both theories 
\beq
\boxed{\ \text{Insertion${}^{\strut{}}$ of $\GO_{\Deltal}(\tau_1,\tau_2)$ in Schwarzian}~~\leftrightarrow~~\text{Insertion of $V_\ell = e^{2\ell \phi(\tau_1,\tau_2)}$ in Liouville CFT${}_{\strut{}}$\ }}\nonumber
\eeq
In the rest of this section we will present detailed evidence in support of this proposal.

Now we will compute the one-point function of the operator $V_\ell = e^{2\ell \phi(\tau_1,\tau_2)}$ between ZZ-branes. Via the method of images, we can map this one-point function to the chiral two-point function on a torus. There is no known closed expression for this two-point function for finite $c$. Nevertheless we will be able to compute it in the limit relevant for the comparison with the Schwarzian theory. 
Using the known wavefunction (\ref{zzwave}) of the ZZ-branes and approximating the Ishibashi states by primary states we can write (dropping overall constant prefactors)
\bea
\lb ZZ | V_{\ell} (z,\bar{z}) | ZZ\rb = \int dPdQ~\Psi_{\rm ZZ}^\dagger(P)\Psi_{\rm ZZ}(Q) \lb P| V_{\Deltal} (z=e^{\frac{\tau_1}{b^{2}} },\bar{z}=e^{-\frac{\tau_2}{b^{2}}}) | Q\rb.
\ea
We are taking the limit $b\to 0$ in which the spatial circle in the closed string channel is going to zero. This enables us to use the minisuperspace approximation for the Liouville CFT wavefunctions, which in effect amounts to a truncation of the full Liouville CFT to the Liouville zero-mode. Thus we can compute the correlation function in the following way 
\bea
\lb ZZ| V_\ell |ZZ\rb \is \int dPdQ ~\Psi_{ZZ}^\dagger(P) \Psi_{ZZ}(Q) \lb P | e^{ - (\beta-\tau) H } e^{2\ell \phi} e^{-\tau H} |Q\rb,
\eea
where the external states are associated to a wavefunction of the Liouville zero mode
\bea
\Psi_P(\phi)=\lb \phi | P\rb \is \frac{2}{\Gamma\bigl(-\frac{2 i P}{b}\bigr)} \; K_{2i P/b} \left(e^{\phi} \right).
\eea
This state has energy $E =  P^2/b^2$. If we call $P=b k_2$ and $Q=b k_1$, the integral that gives the amplitude between states $|P\rb$ and $|Q\rb$ matches exactly with the $b\to 0$ limit of the DOZZ formula \cite{DOZZ}
\bea
 \lb P= bk_2 | e^{2\Deltal \phi}  |Q=bk_1\rb\! \is \! \int\!\! d\phi~ \Psi^\dagger_{P}(\phi)\Psi_Q(\phi) e^{2\Deltal \phi}
\, = \, \frac{ \Gamma(\Deltal \pm i (k_1\pm k_2))}{\Gamma(2 i k_2)\Gamma(2 i k_1)\Gamma(2\Deltal)}.
\ea

Combining this result with the known exact form (\ref{zzwave}) of the wavefunction of the ZZ-branes and dividing by the partition function, we obtain the following formula for the Euclidean two-point function of the Schwarzian theory 
\bea\label{eq:2ptexact}
~~\boxed{\ G^\beta_\Deltal(\tau_1,\tau_2)\, = \,   \frac{(\frac{\beta}{\pi^2})^{3/2}}{2\sqrt{\pi} e^{\frac{ \pi^2}{\beta}}}\int^{\strut{}}_{\strut{}} \! d\mu(k_1) d\mu(k_2) \spc e^{-|\tau| k_1^2 - (\beta-|\tau|) k_2^2}\,  \frac{\Gamma\bigl(\Deltal\pm i(k_1\pm k_2)\bigr)}{\Gamma(2\Deltal)}\ }
\ea
where $\tau=\tau_{12}$, $d\mu(k) = dk^2 \sinh(2\pi k)$ and the $\pm$ signs mean that one multiplies the $\Gamma$-functions with all combinations of signs.\footnote{$\Gamma\bigl(\Deltal\pm i(k_1\pm k_2)\bigr) = \Gamma\bigl(\Deltal + i(k_1+ k_2)\bigr)\Gamma\bigl(\Deltal + i(k_1- k_2)\bigr)\Gamma\bigl(\Deltal - i(k_1+ k_2)\bigr)\Gamma\bigl(\Deltal - i(k_1- k_2)\bigr)$.} This is valid for $-\beta < \tau < \beta$ and needs to be periodically continued beyond this interval. This was done for Euclidean time, to obtain the Lorenzian two-point function we need to Wick rotate $\tau$ to imaginary values which will be discussed shortly. If the Schwarzian is thought of as a low energy limit of the SYK model then $\Deltal=1/q$ is the conformal dimension of the fermions in the theory, if the interaction involves $q$ fermions.   

The formula (\ref{eq:2ptexact}) is the result quoted in the introduction.\footnote{There and in the following we omit the overall constant in front of the integral expression (\ref{eq:2ptexact}).} As explained in section \ref{sect:overview}, one can decompose the integrand in (\ref{eq:2ptexact}) into elementary factors analogous to propagators and vertices in the Feynman rules of QFT. We will not repeat these rules here.

In the remainder of this subsection we will check this result in special limiting regimes. First we can check that \eqref{eq:2ptexact} behaves for $\tau_{12}\to 0$ as $G^\beta_\ell (\tau_1,\tau_2) = {\tau_{12}^{-2\Deltal}}+ ..$.
This is the expected behavior. 
We can also take the zero-temperature limit $\beta\to\infty$.  In this limit, our two-point function reduces to 
\bea
G^\infty_\Deltal(\tau_1,\tau_2) \is \int \!\! {dk^2} \spc \sinh{(2 \pi  k)}\, e^{-|\tau| \spc k^2}\, \frac{ \Gamma^2( \Deltal+ i k)\Gamma^2(\Deltal- i k)}{{2\pi^2}\, \Gamma(2\Deltal)},
\eea
which coincides with the result found in \cite{altland}.\footnote{ In particular this implies that in the zero-temperature limit the two-point function at large times $\tau\to\infty$ behaves as a power-law independently of $\Deltal$ as 
\beq
G^{\infty}_\Deltal(\tau_1,\tau_2) = \frac{\Gamma(\Deltal)^4}{2^{2\ell}\Gamma(2\Deltal)} ~\frac{1}{\tau_{12}^{3/2}}+\ldots~,~~~\tau\to\infty.
\eeq
 This behavior matches with the numerical computation done in \cite{altland} using the SYK model.}

\begin{figure}[t!]
\begin{center}
\begin{tikzpicture}[scale=0.85]
    \begin{axis}[
        xlabel=$\beta$,
        ylabel=$\footnotesize 6(G_2^\beta-G_2^\infty)\big|_{\tau\to0}$,
        domain=3:60
        ]
          \addplot[smooth, thick]{19.7392*x^(-2)+3*x^(-1)};
    \addplot[only marks,blue] coordinates {(4,1.98566)(5,1.39039)(6.5,0.931549)(10,0.500246)(20,0.199814)(30,0.123356)(40,0.0884273)(50,0.0691018)};
    \end{axis}
    \end{tikzpicture}~~~~~~
\begin{tikzpicture}[scale=0.85]
    \begin{axis}[
        xlabel=$\tau/\beta$,
        ylabel=$\footnotesize G^\beta_{\Deltal}(\tau)$
        ]
        \addplot[smooth,thick, mark=none,blue] plot coordinates {(0.042,4.6)(0.05,3.96)(0.1,2.59)(0.15,2.013)(0.2,1.68873)(0.25,1.48592)(0.3,1.35)(0.35,1.26347)(0.4,1.21)(0.45,1.17)(0.5, 1.1644)(0.55,1.17)(0.6,1.21)(0.65,1.26347)(0.7,1.35)(0.75,1.48592)(0.8,1.68873)(0.85,2.013)(0.9,2.59)(0.95,3.96)(0.968,4.7)};
    \addlegendentry{\tiny $g^{-2}=0.06$}
    \addplot[smooth,thick, mark=none,brown] plot coordinates {(0.042,4.7)(0.05,4.36)(0.1,3.04)(0.15,2.46)(0.2,2.137)(0.25,1.93)(0.3,1.789)(0.35, 1.69)(0.4,1.63)(0.45,1.599)(0.5, 1.5881)(0.55,1.599)(0.6,1.63)(0.65,1.69)(0.7,1.789)(0.75,1.93)(0.8,2.137)(0.85,2.46)(0.9,3.04)(0.95,4.36)(0.968,4.7)};
    \addlegendentry{\tiny $g^{-2}=0.3$}
    \addplot[smooth,color=violet,mark=none]
        plot coordinates {(0.042, 4.88)(0.05,4.47)(0.1,3.18)(0.15,2.62)(0.2,2.29)(0.25,2.093)(0.3, 1.95)(0.35, 1.862)(0.4,1.8)(0.45,1.76)(0.5, 1.757)(0.6,1.8)(0.65,1.862)(0.7,1.95)(0.75,2.093)(0.8,2.301)(0.85,2.62)(0.9,3.18)(0.95,4.475)(0.968,4.88)};
     \addlegendentry{\tiny $g^{-2}=4$}     
    \addplot[smooth,dashed,color=black,mark=none,thick]
        plot coordinates {(0.042, 4.88659)(0.05,4.48135)(0.1,3.18848)(0.15,2.63058)(0.2,2.31188)(0.25,2.10781)(0.3,1.97059)(0.35,1.87774)(0.4,1.81749)(0.45,1.78347)(0.5,1.77245)(0.55,1.78347)(0.6,1.81749)(0.65,1.87774)(0.7,1.97059)(0.75,2.10781)(0.8,2.31188)(0.85,2.63058)(0.9,3.18848)(0.95,4.48135)(0.968,4.88659)};
          \addlegendentry{\tiny saddle-point}
    \end{axis}
    \end{tikzpicture}
    \end{center}
    \vspace{-0.85cm}    
    \caption{\small  Left: On the y-axis we plot the numerical evaluation of the $\tau\to0$ limit of $6(G^\beta(\tau)-G^\infty(\tau))$ (blue dots), which coincides with the expectation value of the Schwarzian (full black line).
    Right: Exact two-point function for $\Deltal=1/4$, and different values of $\beta$. We indicate the parameter $g^{-2}=\frac{2 \pi}{ \beta}$. The dashed black line indicates the saddle-point result (\ref{sp2pt}). }
    \label{fig:temp-check}
    \end{figure}
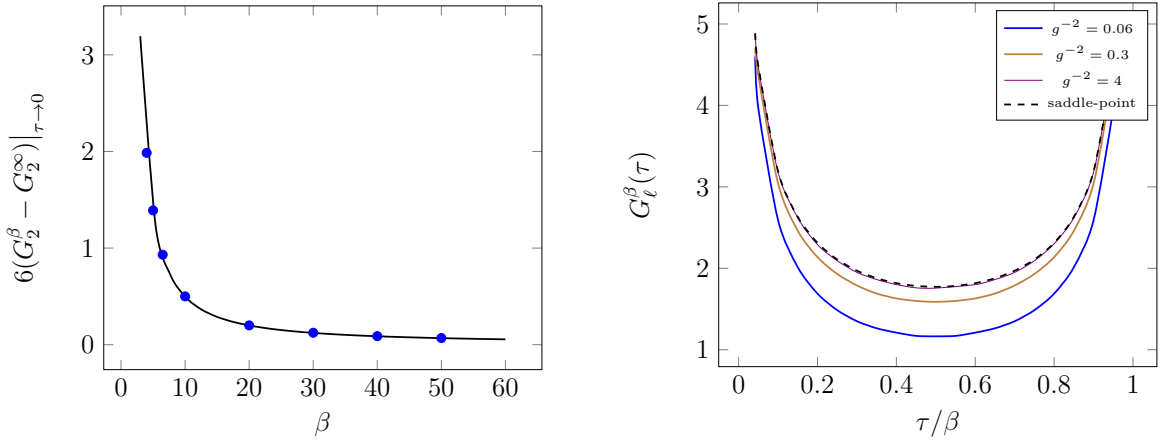

For the case $\Deltal=1$ we can go further, since we know that the next-to-leading order term for $\tau_{12} \to 0$ is given by the Schwarzian itself: 
\bea
\left(\frac{\pi \sqrt{f'(\tau_1)f'(\tau_2)}}{\beta \sin \frac{\pi}{\beta}[f(\tau_1)-f(\tau_2)]} \right)^2 \is \frac{1}{\tau_{12}^2} \, + \, \frac{1}{6}\, \bigl\{ \tan \frac{\pi f(\tau)}{\beta} , \tau\bigr\} 
\,  +\, \ldots
\eea
This will give us a non-trivial check on the temperature dependence of the exact two-point function. We can obtain the expectation value of the Schwarzian by taking derivatives of the partition function
\bea
\Bigl\langle \bigl\{ \tan \frac{\pi f(\tau)}{\beta} , \tau\bigr\} \Bigr\rangle \is \frac{2\pi^2}{\beta^2} + \frac{3}{\beta} + {\rm const.}
\eea
The constant factor depends in part on the zero-point energy. We can eliminate this factor by substracting the zero-temperature limit of the correlation function
\beq
\frac{G_1^\beta(\tau)-G_1^\infty(\tau)}{1/6} \bigg|_{\tau\to0} =\frac{2\pi^2}{\beta^2} + \frac{3}{\beta}.
\eeq
We checked numerically that our formula matches this expectation, as shown in Figure \ref{fig:temp-check}.

We can also take the weakly coupled (large $C$) limit. In our conventions, since we are keeping $C$ fixed, this limit is equivalent to taking $\beta\to 0$ with $\tau/\beta$ fixed. In this regime, quantum corrections are suppressed and correlation functions should be well-approximated by replacing the saddle-point solution
\beq\label{sp2pt}
\GO_{\Deltal}^{\rm cl}(\tau_1,\tau_2)  = \Bigl(\frac{\pi}{\beta\sin \frac{\pi}{\beta}\tau } \Bigr)^{2\Deltal}.
\eeq
We have checked that our exact result \eqref{eq:2ptexact} indeed has this property, see Figure \ref{fig:temp-check}. 

\begin{center}
{\bf Real-time two-point function and the thermofield double}
\end{center}
To conclude this section, we will make some remarks on the real-time continuation and physical application of these results. 

\begin{figure}[h]
\centering
\includegraphics[width=0.8\textwidth]{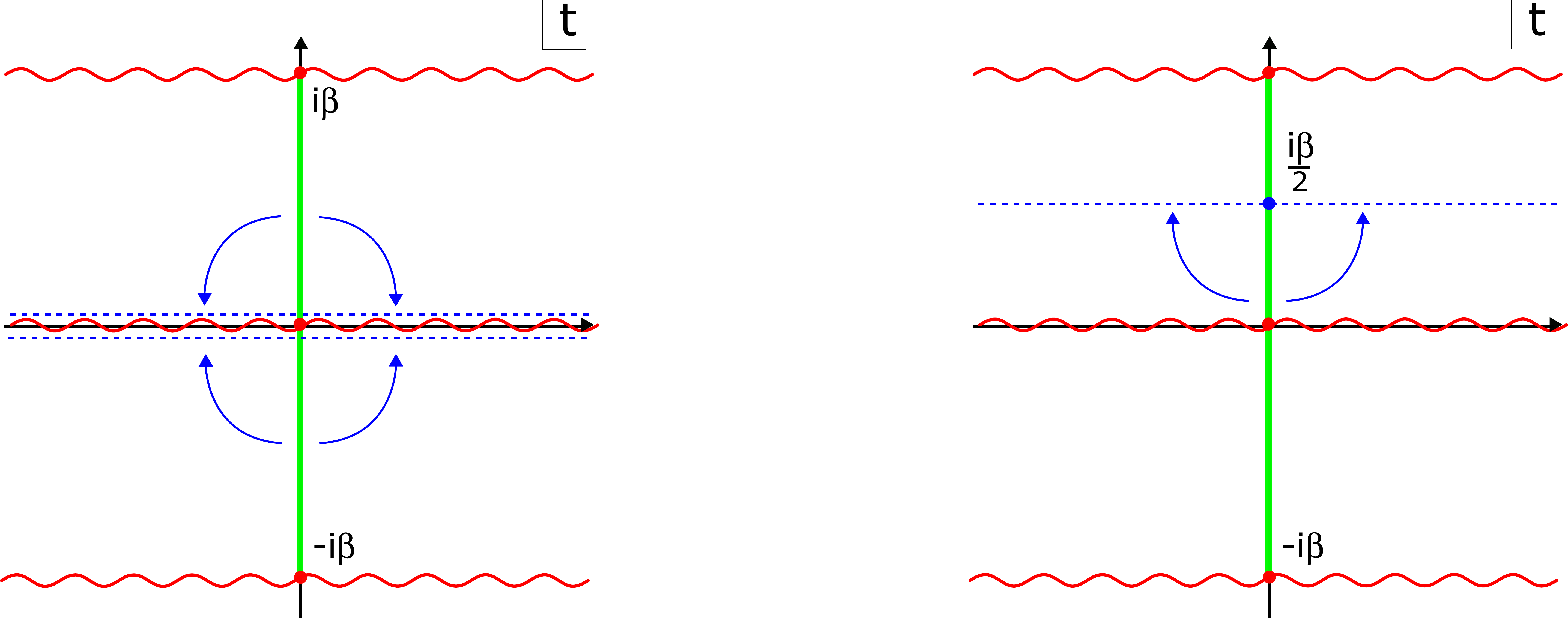}
\caption{Left: Analytic structure of the two-point function. The green line represents the Euclidean regime. Time-ordered and anti-time-ordered Lorentzian two-point functions can be found by analytically continuing these expressions to respectively $t \pm i\epsilon$ (blue lines). Right: Relevant analytic continuation for the thermofield double two-point function.}
\label{anstructure}
\end{figure}

The two-point function (\ref{eq:2ptexact}) has a branch point at $\tau=0$ and two branch cuts running on both sides along the real axis. This is because no spacelike separated points exist in 0+1D. 
This can then be periodically continued along the entire $\tau$-axis (Figure \ref{anstructure}), with periodic branch cuts.

In real time, two possible continuations exist by setting $i\tau \to t \pm i\epsilon$, where the $+$ sign is for $\tau > 0$ and the $-$ sign for $\tau<0$. These correspond to Lorentzian time-ordered $G_\ell^{+}(t_1,t_2)$ and anti-time-ordered two-point functions $G_\ell^{-}(t_1,t_2)$ respectively, for $t_1 > t_2$ given as expectation values of the following bilocal operators:
\bea
\mathcal{O}_\ell^{+}(t_{1},t_2) \is \left\langle \mathcal{O}_\ell(t_1)\mathcal{O}_\ell(t_2)\right\rangle_{\text{CFT}}, \nonumber \\[-2mm] \\[-2mm] \nonumber
\mathcal{O}_\ell^{-}(t_{1},t_2) \is \left\langle \mathcal{O}_\ell(t_2)\mathcal{O}_\ell(t_1)\right\rangle_{\text{CFT}},
\eea
where ${\cal O}_\ell^\pm(t_1,t_2) \equiv  \Bigl(\frac{\sqrt{{f}'(t_1)\, {f}'(t_2)}}{\frac{\beta}{\pi}\sinh \frac \pi \beta [f(t_1)-f(t_2)]\pm i\epsilon}\Bigr)^{2\ell}$. The explicit expression for $G_\ell^{\pm}(t)$ reads (ignoring the constant prefactor, and with $\gamma_\ell(k_1,k_2)$ as given in equation (\ref{gammaell}) )
\begin{align}
\label{realt2pt}
G^\pm_\Deltal(t)\, = \,  
\int^{\strut{}}_{\strut{}} \! d\mu(k_1) d\mu(k_2) \, e^{\pm i t k_1^2 - (\beta \pm i t )k_2^2}e^{-\epsilon k_1^2 - \epsilon k_2^2}\,  \gamma_\ell(k_1,k_2)^2 
\end{align}
with Gaussian damping introduced by $\epsilon$. 
The $\epsilon \to 0^+$ limit is well-defined for both cases, but in general different. This means the commutator of time-separated operators:
\bea
\mathcal{O}^+_\ell(t_{1},t_2) - \mathcal{O}^-_\ell(t_{1},t_2) = \left\langle \left[\mathcal{O}_\ell(t_1),\mathcal{O}_\ell(t_2)\right]\right\rangle_{\text{CFT}},
\eea
does not vanish in expectation values:
\bea
G^+_\ell(t_{12}) - G^-_\ell(t_{12}) = \la \mathcal{O}^+_\ell(t_{1},t_2) - \mathcal{O}^-_\ell(t_{1},t_2)\ra \neq 0.
\eea
This is as expected since all points are timelike separated on the 1D line. Likewise, one can consider other real-time two-point functions of interest, such as the retarded correlator:
$G^{\text{ret}}_\ell (t_1,t_2) = \left(G^+_\ell(t_1,t_2) + G^-_\ell(t_1,t_2)\right)\theta(t_{12})$. 

The long-time behavior of these correlators is easy to determine due to destructive oscillations of the $k_i$-integrals, and gives (ignoring irrelevant prefactors)
\begin{equation}
G^\pm_\ell(t) \to \frac{1}{t^{3/2}}\frac{1}{(\beta \pm it)^{3/2}} \beta^{3/2}.
\end{equation}
At intermediate times (or zero temperature) for which $1 \ll t\ll \beta$, $G^\pm_\ell(t) \sim 1/t^{3/2}$ and at long times $t \gg \beta$, $G^\pm_\ell(t) \sim 1/t^{3}$. 
In either case, the correlator decreases monotonically to zero. Hence no Poincar\'e recurrences occur at very long times. The Schwarzian theory is rather peculiar from this perspective, as it has a continuum of states, thereby foiling the standard argument for recurrences, but its density of states or entropy do not exhibit a volume-divergence. Instead the divergence in the entropy arises due to an infinite $S_0$ (which was irrelevant for our entire discussion), signaling that one needs to go back to its UV completion (e.g. SYK) to understand the very long time behavior of the theory.

\begin{figure}[t]
\centering
\includegraphics[width=0.3\textwidth]{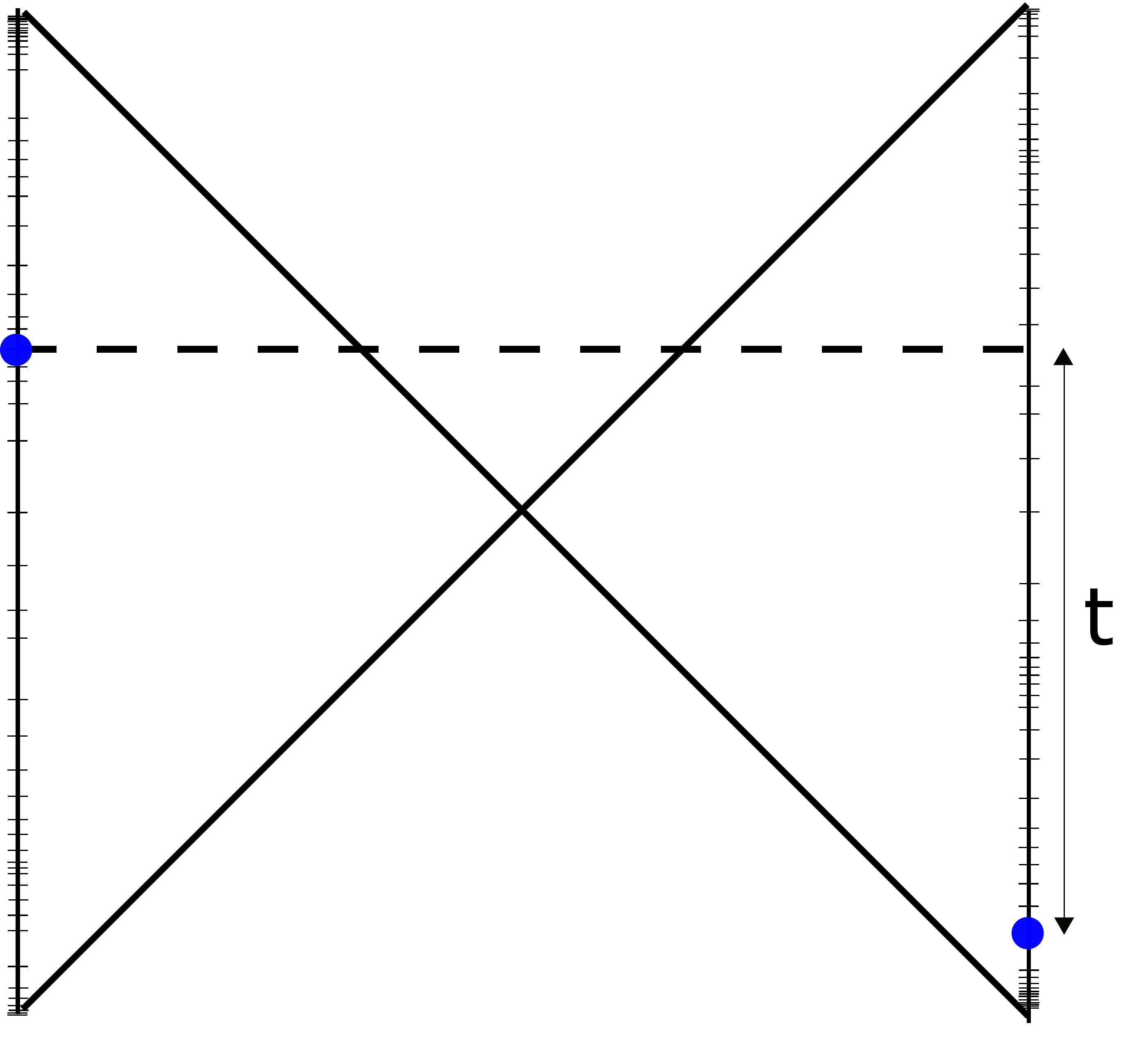}
\caption{Two-point correlator in a thermodouble system. The saddle is the black hole solution, whose Penrose diagram is drawn. One can readily compute correlators on opposite sides of the wormhole. The Schwarzian path integral contains time reparametrizations of the boundary lines that are constrained to start and end at the same points as the eternal black hole time coordinate. A sample clock-ticking configuration is drawn.}
\label{eternal}
\end{figure}

Replacing $\tau \to \tau + \frac{\beta}{2}$ has the effect of moving one operator insertion to the thermal copy of the thermofield double, which in the small coupling (small $C$) limit is just the eternal AdS${}_{2}$ black hole (Figure \ref{eternal}) \cite{Maldacena:2001kr}. We note that our discussion here does not assume a holographic bulk; in a sense we will see what can be deduced of the alleged bulk dual purely from the Schwarzian system. In real time:\footnote{The minus sign in front of $t_1$ on the right hand side corresponds to time running oppositely in the double, but can be ignored in this discussion.}
\bea
\left\langle \text{TFD}\right|\mathcal{O}^L_\ell\left(t_1\right) \mathcal{O}^R_\ell(t_2)\left|\text{TFD}\right\rangle \is \left\langle \mathcal{O}_\ell\left(\textstyle \frac{i\beta}{2}-t_1\right)\mathcal{O}_\ell(t_2)\right\rangle_{\text{CFT}} \, = \,  {\cal O}_\ell\bigl(\textstyle \frac{i\beta} 2 - t_1 ,t_2\bigr)
\eea
where the thermofield double state is
\begin{equation}
\left|\text{TFD}\right\rangle = \frac{1}{\sqrt{Z}} \sum_i e^{-\frac{\beta E_i}{2}}\left|*i\right\rangle_L \otimes \left|i\right\rangle_R.
\end{equation}
When performing the Lorentzian real-time continuation, no branch cuts are encountered and $G^+$ and $G^-$ coincide (see figure \ref{anstructure}). In particular the commutator vanishes, confirming that all points on opposite sides of the thermofield double are spacelike separated. This indicates that the bulk spacetime has a horizon, and that looks identical to the eternal black hole space-time \cite{Maldacena:2001kr}. It is interesting that the causal structure of the dual bulk can be decoded from these correlation functions. Quantum fluctuations of the time reparametrization $f(\tau)$ in themselves are not sufficient to allow communication between both sides. This is of course expected, as to make the wormhole traversable, one would need to add an explicit interaction connecting the two copies of the thermofield double \cite{Gao:2016bin}\cite{Maldacena:2017axo}.
Ignoring the prefactors, the real-time thermodouble correlator can immediately be written down:
\begin{align}
G^{LR}_\Deltal(t)\, = \, \int^{\strut{}}_{\strut{}} \! d\mu(k_1) d\mu(k_2)\, \, e^{-\left(\frac{\beta}{2} + it\right) k_1^2 - \left(\frac{\beta}{2} - it\right)k_2^2}\,  \frac{\Gamma\bigl(\Deltal\pm i(k_1\pm k_2)\bigr)}{\Gamma(2\Deltal)}.
\end{align}
Taking the small temperature limit, one obtains 
\begin{equation}
G^{LR}_\ell(t) \,\, \to \,\, \frac{\beta^{3/2}}{\left(\frac{\beta}{2}-it\right)^{3/2}\left(\frac{\beta}{2}+it\right)^{3/2}} \,\,\to \,\, 0.
\end{equation}
This behavior corresponds to the disappearance of left-right correlation in the extremal black hole limit, generalizing this statement from just the classical saddle point to the full quantum gravity regime. 

\subsection{Four-point function}\label{sec:4pt}

\begin{figure}[h!]
\begin{center}
\begin{tikzpicture}[scale=1.1, baseline={([yshift=0cm]current bounding box.center)}]
\draw[thick] (0,0) circle (1.5);
\draw[thick] (1.3,.7) arc (300:240:2.6);
\draw[thick] (-1.3,-.7) arc (120:60:2.6);
\draw[fill,black] (-1.3,-.68) circle (0.08);
\draw[fill,black] (1.3,-.68) circle (0.08);
\draw[fill,black] (-1.3,0.68) circle (0.08);
\draw[fill,black] (1.3,0.68) circle (0.08);
\draw (1.85,0) node {\small $k_s$};
\draw (-1.85,0) node {\small $k_s$};
\draw (0,.75) node {\small $\ell_1$};
\draw (0,-.75) node {\small $\ell_2$};
\draw (0,1.75) node {\small $k_1$};
\draw (0,-1.75) node {\small $k_4$};
\draw (-1.7,-.66) node {\small $\tau_3$};
\draw (-1.7,.66) node {\small $\tau_2$};
\draw (1.7,-.66) node {\small $\tau_4$};
\draw (1.7,.66) node {\small $\tau_1$};
\end{tikzpicture}~~~~~~~~~~~~~~~~~~~~~~
\begin{tikzpicture}[scale=1.1, baseline={([yshift=0cm]current bounding box.center)}]
\draw[thick] (-1.05,1.05) -- (-.15,.15);
\draw[thick] (.15,-.15) -- (1.05,-1.05);
\draw[thick] (-1.05,-1.05) -- (1.05,1.05);
\draw[thick] (0,0) circle (1.5);
\draw[fill,black] (-1.05,-1.05) circle (0.08);
\draw[fill,black] (1.05,-1.05) circle (0.08);
\draw[fill,black] (-1.05,1.05) circle (0.08);
\draw[fill,black] (1.05,1.05) circle (0.08);
\draw (1.85,0) node {\small $k_s$};
\draw (-1.85,0) node {\small $k_t$};
\draw (-.75,.33) node {\small $\ell_2$};
\draw (.78,.33) node {\small $\ell_1$};
\draw (0,1.75) node {\small $k_1$};
\draw (0,-1.75) node {\small $k_4$};
\draw (-1.4,-1.2) node {\small $\tau_2$};
\draw (-1.4,1.2) node {\small $\tau_3$};
\draw (1.4,-1.2) node {\small $\tau_4$};
\draw (1.4,1.2) node {\small $\tau_1$};
\end{tikzpicture}
\vspace{-2mm}
\end{center}
\caption{\small The diagrammatic representation of the two types of four-point functions. The left diagram depicts the time-ordered four-point function (\ref{fpt}) with $\tau_1<\tau_2< \tau_3 <\tau_4$. The diagram on the right represents the out-of-time ordered four point function: in contrast with the geometric ordering, we assume that the four time instances are still ordered as $\tau_1 < \tau_2 < \tau_3 < \tau_4.$  The OTO correlation function is defined via analytic continuation from the time-ordered correlation function.} 
\label{fig:fourptfiga}
\end{figure}
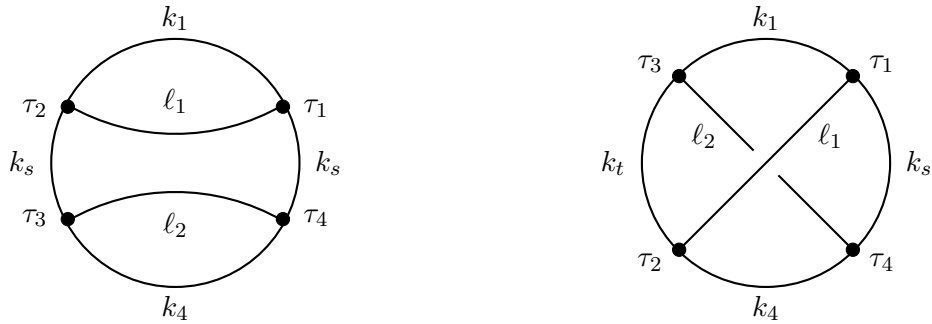

Next we consider the time-ordered four point function, given by the two-point function of two bi-local operators.
\beq
\label{fpt}
G_{\Deltal_1\Deltal_2}(\tau_1,\tau_2,\tau_3,\tau_4)\, = \, \la \mathcal{O}_{\Deltal_1}(\tau_1,\tau_2) \mathcal{O}_{\Deltal_2}(\tau_3,\tau_4)\ra.
\eeq 
There are different choices for how to order the four different times. Here we will assume that the time instances are cyclically ordered via $\tau_1 < \tau_2 < \tau_3 < \tau_4.$ In the diagrammatic representation of the amplitudes, this ordering ensures that the legs of the two bi-local operators do not cross each other, as indicated in the left-hand side diagram of Figure \ref{fig:fourptfiga}.

\begin{figure}[h!]
\begin{center}
 \begin{tikzpicture}[scale=.95,rotate=0]
 \fill [draw=none, fill=blue, opacity=0.1] (-2,-1.985) to [bend right=60] (-2,2) -- (-1.8,2.10)-- (-1.7, 2.12) -- (-0.2,2.2) --  (0,2.2)--  (0.2,2.2) -- (1.7,2.12) -- (1.8,2.10)-- (2,2) to [bend right=60] (2,-1.985) -- (2, -2) arc (0:-180:2 and 0.2)-- cycle ;
\draw[thick] (2, -2) arc (0:-180:2 and 0.2);
\draw[thick,dashed] (2, -2) arc (0:180:2 and 0.2);
\draw[thick, fill= blue, opacity= 0.13] (0,2) ellipse (2 and 0.2);
\draw[draw=none, fill= blue,opacity= 0.07] (0,-2) ellipse (2 and 0.2);
\draw[thick] (0,2) ellipse (2 and 0.2);
\draw[thick] (-2,-1.985) to [bend right=60] (-2,1.985);
\draw[thick] (2,-1.985) to [bend left=60] (2,1.985);
\draw[thick,fill] (0,-1) circle (0.08);
\draw[thick] (-1,-2.17) to [bend right=10] (0,-1) to [bend right=10](1,-2.17);
\draw[thick,fill] (0,0) circle (0.08);
\draw[thick] (-1.5-0.2,-2.11) to [bend right=10] (0,0) to [bend right=10] (1.5+0.2,-2.11);
\draw (-1,-2.5) node {\small $\tau_1$};
\draw (1,-2.5) node {\small $\tau_2$};
\draw (-1.7,-2.4) node {\small $\tau_4$};
\draw (1.7,-2.4) node {\small $\tau_3$};
\end{tikzpicture}
~~~~~~~~~~~~~~~~~~~~
 \begin{tikzpicture}[scale=.95,rotate=0]
 \fill [draw=none, fill=blue, opacity=0.1] (-2,-1.985) to [bend right=60] (-2,2) -- (-1.8,2.10)-- (-1.7, 2.12) -- (-0.2,2.2) --  (0,2.2)--  (0.2,2.2) -- (1.7,2.12) -- (1.8,2.10)-- (2,2) to [bend right=60] (2,-1.985) -- (2, -2) arc (0:-180:2 and 0.2)-- cycle ;
\draw[thick] (2, -2) arc (0:-180:2 and 0.2);
\draw[thick,dashed] (2, -2) arc (0:180:2 and 0.2);
\draw[thick, fill= blue, opacity= 0.13] (0,2) ellipse (2 and 0.2);
\draw[draw=none, fill= blue,opacity= 0.07] (0,-2) ellipse (2 and 0.2);
\draw[thick] (0,2) ellipse (2 and 0.2);
\draw[thick] (-2,-1.985) to [bend right=60] (-2,1.985);
\draw[thick] (2,-1.985) to [bend left=60] (2,1.985);
\draw[thick,fill] (-0.5-0.2,-1) circle (0.08);
\draw[thick,fill] (0.5,-1) circle (0.08);
\draw[thick] (-1.5-0.2,-2.11) to [bend right=10] (-0.5-0.2,-1) to [bend right=10] (0.5-0.2,-2.195);
\draw[thick] (-0.2,-2.195) to [bend right=10] (0.5,-1) to [bend right=10] (1.5,-2.125);
\draw (-1.5,-2.4) node {\small $\tau_1$};
\draw (1.5,-2.4) node {\small $\tau_4$};
\draw (-.3,-2.5) node {\small $\tau_3$};
\draw (.4,-2.5) node {\small $\tau_2$};
\vspace{-2mm}
\end{tikzpicture}
\end{center}
\caption{\small The four-point function in the Schwarzian theory corresponds to a two-point function of two bulk Liouville vertex operators. If the two bulk operators are timelike separated (left), the correlation function and the end-points of the two bi-local operators are time ordered. If the two bulk operators are spacelike separated (right), the legs of the bi-local operators (which follow 2D light-cone directions starting from each vertex) cross each other. Time-ordered and out-of-time ordered correlation functions are thus related by the CFT monodromy matrix that relates the timelike separated and spacelike separated two-point functions.}
\label{figZZhyp}
\end{figure}
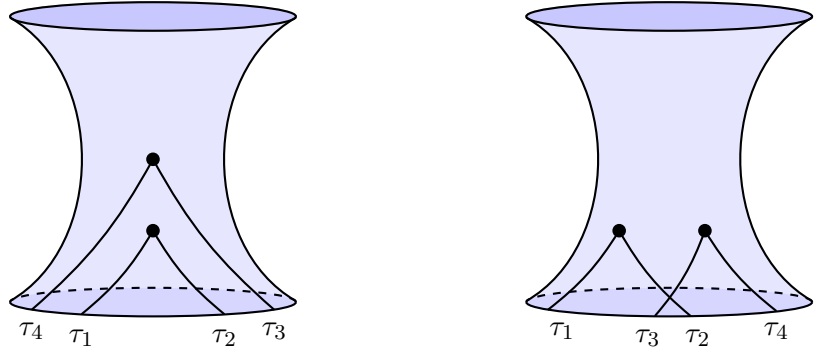

We will compute this four-point function by applying the dictionary between the Schwarzian and 2D Liouville CFT. This leads us to consider the following two-point function of primary operators between two ZZ-branes
\beq
G_{\Deltal_1\Deltal_2} \, = \, \lb ZZ| V_{\Deltal_1} (z_1,\bar{z}_1)V_{\Deltal_2} (z_2,\bar{z}_2) | ZZ\rb .
\eeq
As explained above, this can be interpreted as a four point function (\ref{fpt}) in the Schwarzian theory if we identify $z_1 \to \tau_2$, $\bar{z}_1\to \tau_1$, $z_2 \to \tau_3$ and $\bar{z}_2 \to \tau_4$. For the time-ordered operator, the locations $(z_1,\bar{z}_1)$ and $(z_2,\bar{z}_2)$ are chosen to be timelike separated, as indicated on the left-hand side in Figure \ref{figZZhyp}, so that their past lightcones do not intersect.
\bigskip

\bigskip

\bigskip

As before, we can go to the closed string channel and write the four point function as
\bea
G_{\Deltal_1\Deltal_2} \is \int dP dQ~\Psi^\dagger_{\rm ZZ}(P)\Psi_{\rm ZZ}(Q)~\lb P | V_{\Deltal_1} (z_1,\bar{z}_1)V_{\Deltal_2} (z_2,\bar{z}_2)| Q\rb.
\eea
In the Schwarzian limit we are allowed to replace the Ishibashi states $|\!\spc |P\rb\!\rb$ by the corresponding primary states $|P\rb$. The correlation function on the right-hand side is computed on the sphere and can be expanded in conformal blocks in the $V_{\Deltal_2} V_Q \to V_{\Deltal_1} V_P$ channel.

We thus arrive at the following representation for the time-ordered four point function in terms of 2D CFT data 
\bea
\lb P | V_{\Deltal_1} (z_1,\bar{z}_1)V_{\Deltal_2} (z_2,\bar{z}_2)| Q\rb \is\\[3mm]
\! & \!\! & \hspace{-4.5cm} \int dP_s~C(-P,\Deltal_1,P_s)\, C(-P_s,\Deltal_2, Q) \; \mathcal{F}_{P_s}\left[\;{}_{P}^{\Deltal_1} \;{}_{Q}^{\Deltal_2}\right](z_1,z_2)\; \mathcal{F}_{P_s}\left[\;{}_{P}^{\Deltal_1} \;{}_{Q}^{\Deltal_2}\right](\bar{z}_1, \bar{z}_2)\nonumber
\ea
where $C(1,2,3)$ is the DOZZ OPE coefficient. 
The cross-ratios in these formulae are given by $z=z_1/z_2$ and $\bar{z}=\bar{z}_1/\bar{z}_2$. The conformal blocks are defined by the following normalization 
\bea
\mathcal{F}_{P_s}\left[\;{}_{P}^{\Deltal_1} \;{}_{Q}^{\Deltal_2}\right](z_1,z_2)
\! & \!=\! & \!z_1^{\Delta_P - \Delta_s } z_2^{\Delta_s-\Delta_Q}\, \bigl(1\, +\, \ldots\bigr)\, 
\ea
where the $\ldots$ denote higher order terms in $z$ and $\bar{z}$. To take the Schwarzian limit, we set $P=bk_1$, $Q=bk_4$ and send $b\to0$, while simultaneously sending $z \to 0$. In this limit the conformal block becomes trivial.\footnote{As we will see in the next section, the 2D conformal blocks exhibit non-trivial monodromy properties under analytic continuation. These will turn out to be an essential ingredient in the computation of out-of-time-ordered correlation functions.}
 Using the same notation as in the previous section, we obtain\footnote{The zero-temperature limit factorizes as $G^\infty_{\Deltal_1\Deltal_2} = G^\infty_{\Deltal_1}G^\infty_{\Deltal_2}$. This could have been deduced immediately due to \eqref{hcom} and cluster decomposition. This zero-temperature result is in agreement with \cite{altland}. This factorization would not have happened had we computed the time-ordered correlator $\la \mathcal{O}_{\Deltal_1}(\tau_1,\tau_4) \mathcal{O}_{\Deltal_2}(\tau_2,\tau_3)\ra$.}
\bea
\label{fourpt}
G^\beta_{\Deltal_1\Deltal_2}\! \is \int\!\! dk_1^2dk_4^2dk_s^2\, \sinh 2 \pi k_1 \sinh 2 \pi k_4\sinh 2 \pi k_s~ \\[2mm]
\! & \!\! & \hspace{-2cm} \times\; e^{-k_1^2(\tau_2-\tau_1) -k_4^2(\tau_4-\tau_3) - k_s^2(\beta -\tau_2+ \tau_3 -\tau_4 +\tau_1)}\;
\frac{\Gamma(\Deltal_2\pm ik_4\pm ik_s)\, \Gamma(\Deltal_1\pm ik_1\pm i k_s)}{\Gamma(2\Deltal_1)\, \Gamma(2\Deltal_2)}.\nonumber
\ea
This integral expression is only valid in the regime $\tau_4>\tau_3>\tau_2>\tau_1$.

The  formula (\ref{fourpt}) is identical to the result (\ref{4pt}) quoted in the Introduction. Again, it is possible to disentangle the full expression (\ref{fourpt}) into propagators and vertices. Or conversely, applying the Feynman rules outlined in section \ref{sect:overview} to the diagram on the left in Figure \ref{fig:fourptfiga}, we directly obtained the full result (\ref{fourpt}) for the time-ordered four-point function.

For later reference, we summarize the above calculation of the four point function by means of the following diagram: 
\bea
G_{\Deltal_1\Deltal_2} \is \int\! dP\spc dQ \spc dP_s~\Psi_{\rm ZZ}^\dagger(P)\Psi_{\rm ZZ}(Q)~\times~
 \begin{tikzpicture}[scale=0.35, baseline={([yshift=-0.1cm]current bounding box.center)}]
 \draw[thick](0,-3.5) -- (0,3.5);
 \draw[thick](0,1.5) -- (1.9,1.5);
  \draw[thick](0,-1.5) -- (1.9,-1.5);
\draw (0.6,-3.5) node {\footnotesize $P$};
\draw (0.6,3.5) node {\footnotesize $Q$};
\draw (0.8,0) node {\footnotesize $P_s$};
\draw (2.7,1.5) node {\small $\Deltal_1$};
\draw (2.7,-1.5) node {\small $\Deltal_2$};
\end{tikzpicture}
\times 
 \begin{tikzpicture}[scale=0.35, baseline={([yshift=-0.1cm]current bounding box.center)}]
 \draw[thick](0,-3.5) -- (0,3.5);
 \draw[thick](0,1.5) -- (-1.9,1.5);
  \draw[thick](0,-1.5) -- (-1.9,-1.5);
\draw (-0.6,-3.5) node {\footnotesize $P$};
\draw (-0.6,3.5) node {\footnotesize $Q$};
\draw (-0.8,0) node {\small $P_s$};
\draw (-2.7,1.5) node {\small $\Deltal_1$};
\draw (-2.7,-1.5) node {\small $\Deltal_2$};
\end{tikzpicture}\\
\nonumber
\eea
As indicated, the insertion of the ZZ states splits the thermal circle into two halves, each given by a chiral conformal block of the 2D Virasoro CFT. Each half has the same intermediate momentum $P_s$, and all momenta are integrated over. In the figure, we have absorbed the OPE coefficients into the definition of the conformal blocks.

\section{OTO four point function}
\label{sect5}
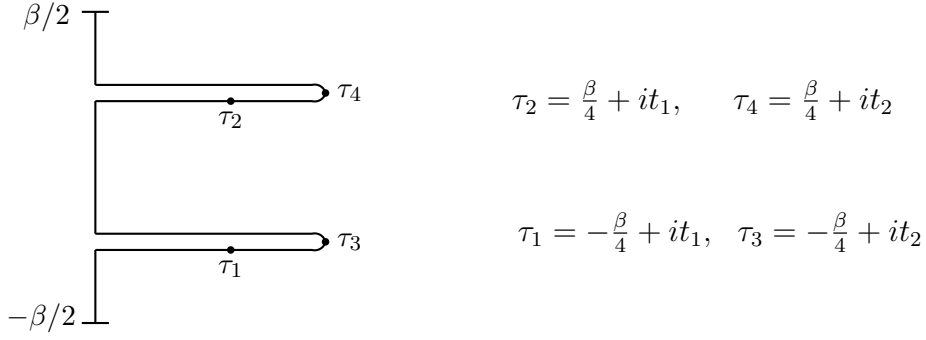
\begin{figure}[h!]
\begin{center}
 \begin{tikzpicture}[scale=0.36, baseline={([yshift=-0.1cm]current bounding box.center)}]
 \draw[color=black,thick](-.5,7.5) -- (.5,7.5);
 \draw[color=black,thick](-.5,-4) -- (.5,-4);
 \draw[color=black,thick](0,-4) -- (0,-1.3);
 \draw[color=black,thick](0,-.7) -- (0,4.2);
\draw[color=black,thick] (0,-.7) -- (8,-.7) to [bend left=45]  (8.5,-1);
\draw[color=black,thick] (0,-1.3) -- (8,-1.3) to [bend right=45]  (8.5,-1);
 \draw[color=black,thick](0,4.8) -- (0,7.5);
\draw[color=black,thick] (0,4.8) -- (8,4.8) to [bend left=45]  (8.5,4.5);
\draw[color=black,thick] (0,4.2) -- (8,4.2) to [bend right=45]  (8.5,4.5);
\draw[thick,fill] (8.5,-1) circle (0.1);
\draw[thick,fill] (8.5,4.5) circle (0.1);
\draw[thick,fill] (5,-1.3) circle (0.1);
\draw[thick,fill] (5,4.2) circle (0.1);
\draw (-2,-3.8) node {\small $-\beta/2$};
\draw (-1.8,7.3) node {\small $\beta/2$};
\draw (5,-2) node {\small $\tau_1$};
\draw (5,3.5) node {\small $\tau_2$};
\draw (9.4,-1) node {\small $\tau_3$};
\draw (9.4,4.5) node {\small $\tau_4$};
\end{tikzpicture}~~~~~~~~~~~~~~\raisebox{-25pt}{$\tau_1 = -\frac\beta 4 + i t_1, \ \  \tau_3 = -\frac \beta 4 + i t_2$} \hspace{-5.6cm}\raisebox{25pt}{$\tau_2 = \frac \beta 4 + i t_1, \ \ \ \ \, \tau_4 =\frac \beta 4 + i t_2$}
\end{center}\vspace{-2mm}
\caption{\small Time ordering prescription for the out-of-time ordered four point function at finite inverse temperature $\beta$. Note that the time operator insertion at $\tau_3$ acts before the operator insertion at $\tau_2$, even though in real time $t_1 = {\rm Im}(\tau_2)$ is earlier than  $t_2 = {\rm Im}(\tau_3)$.}
\label{otofptcontour}
\end{figure}

In this section we consider the out-of time ordered four-point function
\bea
\label{fptoto}
G^{\rm OTO}_{\Deltal_1\Deltal_2}(\tau_1,\tau_2,\tau_3,\tau_4)\is \la \mathcal{O}_{\Deltal_1}(\tau_1,\tau_2) \mathcal{O}_{\Deltal_2}(\tau_3,\tau_4)\ra_{\rm OTO}
\eea 
The OTO prescription can be implemented in different ways. One convenient choice is to complexify the time coordinates, and choose the real and imaginary parts as follows
\bea
\tau_1 \! \is\! -\frac\beta 4 + i t_1, \qquad  \tau_2 = \frac \beta 4 + i t_1, \qquad \tau_3 = -\frac \beta 4 + i t_2 \qquad \tau_4 =\frac \beta 4 + i t_2,
\eea
and consider a time contour prescription as indicated in Figure \ref{otofptcontour}.

To compute the OTO correlation functions, one could try to explicitly perform the three integrals in equation (\ref{fourpt}) for the time-ordered correlation function, write $G_{\Deltal_1\Deltal_2}$ as an analytic function of the four times $\tau_i$ and then perform the appropriate analytic continuation. 
However, at present we do not know how to perform the integrals explicitly, so this direct approach is not practical. Luckily, the 2d perspective of the Schwarzian theory as a limit of 2D Virasoro CFT theory gives another way to solve the problem. 
In this section we will combine the calculation presented above with the ideas of \cite{Jackson:2014nla} to compute the exact OTO four-point function.  Technical details are delegated to Appendix \ref{App:FM}.

Most recent studies of OTO correlation functions in (putative) chaotic systems have focussed on the time-dependence. However, as originally pointed out in \cite{Jackson:2014nla}, to exhibit the dynamical mechanism that underlies the Lyapunov behavior, it is equally informative to study the four-point function in Fourier space, in which one fixes the energies of the intermediate states. The latter approach is also more naturally incorporated into our construction of the Schwarzian amplitudes in terms of the momentum space amplitudes.

Before we turn to the derivation, let us first write out the explicit form for the OTO four point function, as follows from the application of the Feynman rules presented in section \ref{sect:overview} to the diagram on the right-hand side of Figure \ref{fig:fourptfiga}:
\bea
\label{oto4pt}
G^{\rm OTO}_{\Deltal_1\Deltal_2}\; \is\;  \int\!\! dk_1^2dk_4^2dk_s^2 dk_t^2\, \sinh 2 \pi k_1 \sinh 2 \pi k_4\sinh 2 \pi k_s \sinh2\pi k_t~ \nonumber \\[2.5mm]
\! & \! \! & \! \!\!\!\!\!\!\!\!\!\!\!\!\!\!\!\! \times\; 
\frac{\bigl| \Gamma(\Deltal_2 + ik_4\pm ik_s)\spc \Gamma(\Deltal_1 + ik_1\pm i k_s)\spc \Gamma(\Deltal_2 + ik_1\pm ik_t)\Gamma(\Deltal_1 + ik_4\pm i k_t)\bigr|}{\Gamma(2\Deltal_1)\, \Gamma(2\Deltal_2)}\quad\\[3.5mm]
\! & \!\! & \times \; R_{k_sk_t}\! \left[\; {}^{k_4}_{k_1} \;{}^{\Deltal_2}_{\Deltal_1} \right] \; \times\; \nonumber   e^{ -   k_1^2(\tau_3-\tau_1) - k_t^2(\tau_3 -\tau_2) -k_4^2(\tau_4-\tau_2)- k_s^2(\beta-\tau_4 +\tau_1) } {}_{\strut{}}.
\eea
The essential new ingredient in this expression is the R-matrix $R_{k_sk_t}\! \left[\; {}^{k_4}_{k_1} \;{}^{\Deltal_2}_{\Deltal_1} \right]$. Its explicit form is given in Appendix \ref{App:FM}.
By gauge/gravity duality this quantity describes the S-matrix corresponding to scattering of particles close to the horizon of a black hole. In other words, the integrand is already capturing the information we need to relate the out-of-time-ordered correlation function to gravitational properties of the event horizon. In the remainder of this section, we will explain how the above result arises from the correspondence with 2D CFT, how to extract the time dependence and Lyapunov behavior, and how it matches with $2\to 2$ scattering via gravitational shockwaves.

\subsection{The R-matrix}\label{sec:otoc}

The OTO four-point function in the Schwarzian theory corresponds to a two-point function of two bulk Liouville vertex operators that are spacelike separated, so that the past lightcones of the two operators cross each other as indicated on the right in Figure \ref{figZZhyp}. From the point of view of the 2D CFT, this means that one of the chiral conformal blocks has been analytically continued to an OTO conformal block
\bea
\label{otozz}
G_{\Deltal_1\Deltal_2}^{\rm OTO} \is \lb ZZ| V_{\Deltal_1} (z_1,\bar{z}_1)V_{\Deltal_2} (z_2,\bar{z}_2) | ZZ\rb_{\rm OTO}, \\[3mm]
 \is \int dP dQ~\Psi^\dagger_{\rm ZZ}(P)\Psi_{\rm ZZ}(Q) ~\lb P | V_{\Deltal_1} (z_1,\bar{z}_1)V_{\Deltal_2} (z_2,\bar{z}_2)| Q\rb_{\rm OTO}.\nonumber
\eea
where the integrand factorizes in terms of CFT kinematic data as 
\bea\label{eq:4ptL}
\lb P | V_{\Deltal_1} (z_1,\bar{z}_1)V_{\Deltal_2} (z_2,\bar{z}_2)| Q\rb_{\rm OTO} \is\\[3mm]
\! & \!\! & \hspace{-4.5cm} \int dP_s~C(-P,\Deltal_1,P_s)\, C(-P_s,\Deltal_2, Q) \; \mathcal{F}^{\rm\spc OTO}_{P_s}\left[\;{}_{P}^{\Deltal_1} \;{}_{Q}^{\Deltal_2}\right](z_1,z_2)\; \mathcal{F}_{P_s}\left[\;{}_{P}^{\Deltal_1} \;{}_{Q}^{\Deltal_2}\right](\bar{z}_1, \bar{z}_2)\nonumber
\ea
Here the OTO label indicates that we have  applied a specific monodromy transformation to the 2D conformal block. The effect of this monodromy transformation in the Schwarzian limit can be found in the following way. 

The argument of the s-channel conformal block is $z=z_1/z_2$, which goes to zero in the time-ordered case. Inserting the two operators in opposite order gives $z'=1/z=z_2/z_1\to\infty$. The 2D conformal block behaves non-trivially in the limit where the cross ratio becomes infinite. Even though we do not know the explicit expression for the full conformal block, we can use the R-matrix transformation of Ponsot and Teschner \cite{PT}
\bea
\label{pstrafo}
\mathcal{F}_{P_s}[\,{}_1^2\;{}_4^3\,](z') \is \int dP_t~ R_{P_sP_t}\left[\, {}^{2}_{1} \;{}^{3}_{4} \, \right]~ \mathcal{F}_{P_t}[\, {}_1^3\;{}_4^2\,](1/z') 
\eea
to extract its exact behavior in the large cross-ratio regime $z' \to \infty$ by using the fact that
the conformal block inside the integral in (\ref{pstrafo}) becomes trivial for $z = 1/z'\to 0$. 
Inserting the transformed conformal block into (\ref{otozz}) and (\ref{eq:4ptL}), we obtain the momentum integral representation of the out-of-time-ordered four-point function. The total calculation procedure can be graphically represented~as
\bea
G_{\Deltal_1\Deltal_2}^{\rm OTO}&=&\int dP dQ~\Psi_{\rm ZZ}^\dagger(P)\Psi_{\rm ZZ}(Q)\;\times\, \int\! dP_s ~\;
 \begin{tikzpicture}[scale=0.3, baseline={([yshift=-0.1cm]current bounding box.center)}]
 \draw[color=black,thick](0,-3.5) -- (0,3.5);
\draw (0.6,-3.5) node {\footnotesize $P$};
\draw (0.6,3.5) node {\footnotesize$Q$};
\draw (0.8,0) node {\footnotesize $P_s$};
\draw (2.4,2.2) node {\small $\Deltal_1$};
\draw[color=black,thick] (0,-1.5) -- (1.9,-1.5) to [bend right=40] (3.5,0) to [bend left=30] (7-1.9,1.5) -- (7,1.5) ;
\draw[color=black,thick] (0,1.5) --(1.9,1.5) to [bend left=40] (3.5-0.3,0.4);
\draw[color=black,thick] (3.5+0.3,-0.5) to [bend right=40] (7-1.9,-1.5) -- (7,-1.5);
\draw (2.4,-2.2) node {\small $\Deltal_2$};
 \draw[color=black,thick](7+0,-3.5) -- (7+0,3.5);
\draw (7-0.6,-3.5) node {\footnotesize $P$};
\draw (7-0.6,3.5) node {\footnotesize $Q$};
\draw (7-0.8,0) node {\footnotesize $P_s$};
\end{tikzpicture} \\[3mm]
& & \hspace{-2cm} = \ \int dP dQ~\Psi_{\rm ZZ}^\dagger(P)\Psi_{\rm ZZ}(Q)\;\times\, \int\! dP_s dP_t~~{R}_{P_sP_t} \ \;
 \begin{tikzpicture}[scale=0.3, baseline={([yshift=-0.1cm]current bounding box.center)}]
 \draw[color=black,thick](0,-3.5) -- (0,3.5);
\draw (0.6,-3.5) node {\footnotesize $P$};
\draw (0.6,3.5) node {\footnotesize $Q$};
\draw (0.9,0) node {\footnotesize $P_s$};
\draw (3.5,2.3) node {\small $\Deltal_1$};
\draw (3.5,-2.3) node {\small $\Deltal_2$};
\draw[color=black,thick] (0,-1.5) -- (7,-1.5) ;
\draw[color=black,thick] (0,1.5) -- (7,1.5);
 \draw[color=black,thick](7+0,-3.5) -- (7+0,3.5);
\draw (7-0.6,-3.5) node {\footnotesize $P$};
\draw (7-0.6,3.5) node {\footnotesize $Q$};
\draw (7-0.9,0) node {\footnotesize $P_t$};
\end{tikzpicture}\nonumber
\ea

Our remaining task is to compute the appropriate large $c$ limit of the crossing kernel of 2D CFT conformal blocks. This calculation is performed in Appendix \ref{App:FM}. The 2d crossing kernels, i.e. the $F$-matrix and $R$-matrix, are explicitly known and expressed in terms of $U_q(\mathfrak{sl}(2,\mathbb{R}))$ $6j$-symbols \cite{PT}.  Perhaps unsurprisingly, we will find that in the Schwarzian limit the Ponsot-Teschner result for the quantum 6j-symbols reduces to known expressions for the classical 6j-symbols of $SU(1,1)$.

The R-matrix and fusion matrix of 2D Virasoro conformal blocks are related via
\bea
\label{rf}
R_{\alpha_s\alpha_t}\left[\; {}^{\alpha_3}_{\alpha_4} \;{}^{\alpha_2}_{\alpha_1} \right] \is e^{ 2 \pi i(\Delta_2+\Delta_4-\Delta_s-\Delta_t)} 
F_{\alpha_s\alpha_t}\left[\; {}^{\alpha_3}_{\alpha_4} \;{}^{\alpha_2}_{\alpha_1} \right].
\eea
The F-matrix, in turn, is expressed in terms of the quantum $6j$-symbol via \cite{Teschner:2012em}
\bea
F_{\alpha_s\alpha_t}\left[\; {}^{\alpha_3}_{\alpha_4} \;{}^{\alpha_2}_{\alpha_1} \right] \is |S_b(2\alpha_t)S_b(2\alpha_s)| \sqrt{\frac{C(\alpha_4,\alpha_t,\alpha_1)C(\bar{\alpha}_t,\alpha_3,\alpha_2)}{C(\alpha_4,\alpha_3,\alpha_s)C(\bar{\alpha}_s,\alpha_2,\alpha_1)}} ~\sixj{\alpha_1}{\alpha_3}{\alpha_2}{\alpha_4}{\alpha_s}{\alpha_{t}}_b^{},
\eea
where $S_b(x)$ denotes the double Sine function and $C(\alpha_3,\alpha_2,\alpha_1)$ is the DOZZ  three point function \cite{DOZZ}. In the 1D limit, we need to take two $\alpha$'s to be real and proportional to $\alpha_i = b \Deltal_i$ with $\ell_i$ finite and the other four of the form $\alpha_j = \frac{Q}{2} + i b k_j$ with $k_j$ finite. Specifically, we will choose
\bea\label{eq:Schlimit}
\alpha_1\! \is \! \Deltal_1 b, ~~~~~~~~~~~\alpha_2=\frac{Q}{2}+i bk_2,~~~~~~~~~~~\alpha_s = \frac{Q}{2}+i b k_s, \nn
\alpha_3\! \is \! \Deltal_3 b, ~~~~~~~~~~~\alpha_4=\frac{Q}{2}+i bk_4,~~~~~~~~~~~\alpha_t=\frac{Q}{2}+i b k_t.\nonumber
\ea
For the application to the Schwarzian theory, we must further take the classical limit of the quantum 6j-symbols
\bea
\label{sixjlimit}
\sixj{\Deltal_1}{\Deltal_3}{k_2}{k_4}{k_s}{k_{t}} \! & \equiv & \! \lim_{b\to 0} \; 2 \pi b^3 ~\sixj{\alpha_1}{\alpha_3}{\alpha_2}{\alpha_4}{\alpha_s}{\alpha_{t}}_b^{}.
\eea

Returning to equation (\ref{pstrafo}) and using the above formulas, we are now obtain an explicit expression for the large $z$ limit of . The conformal block in the right hand side again becomes trivial since $z'=1/z\to0$. The answer can be written as  (see Appendix \ref{App:FM})
\bea
\label{substitute}
\mathcal{F}_{P_s}^{\rm OTO}[\, {}_1^2\;{}_4^3\, ](\tau) \is \int\! dP_t~e^{- \tau P_t^2/b^2} \; R_{P_sP_t}\left[\, {}^{2}_{1} \;{}^{3}_{4} \, \right] \nonumber \\[-1.5mm]\\[-1.5mm]
= \; \int \! dk^2_t \sinh 2\pi k_t  \hspace{-6mm} & & \hspace{-2mm} e^{- \spc k_t^2\tau}\; 
\left| \frac{ \Gamma({\ell_1}+ ik_4 \pm i k_t)\spc \Gamma(\ell_3 +i k_2\pm i k_t)}{\Gamma({\ell_1}+ ik_2 \pm i k_s)\spc \Gamma(\ell_3 +i k_4\pm i k_s)}\right|\; \sixj{\Deltal_1}{\Deltal_3}{k_2}{k_4}{k_s}{k_{t}} 
\nonumber
\eea
with the 6j-symbol as defined via (\ref{sixjlimit}). Note that in the 1D limit the dimensions of the operators that appear in the phase factor in equation (\ref{rf}) are all equal to $\frac{c}{24}+\mathcal{O}(1/c)$. The phase factor thus becomes trivial.

To obtain the out-of-time-ordered four point function we make the above substitution inside of the integral expression (\ref{eq:4ptL}).
This leads to the final expression given in equation (\ref{oto4pt}), where we define the Schwarzian $R$-matrix via
\beq
\label{rmatrix}
{R}_{k_sk_t}\left[\; {}^{k_4}_{k_1} \;{}^{\Deltal_2}_{\Deltal_1} \right] = \sixj{\Deltal_1}{\Deltal_2}{k_1}{k_4}{k_s}{k_{t}}.
\eeq
With this definition, the R-matrix is naturally a unitary operator relative to the spectral measure $d\mu(k)$.

\subsection{Schwarzian $6j$-symbols}\label{sec:6j}

In this section we present the explicit expression for the Schwarzian limit of the $6j$-symbols of the Virasoro CFT. A general expression for this quantity at finite $c$, and its relation with the monodromy of the 2D conformal blocks, was found by B.~Ponsot and J.~Teschner in \cite{PT}. For our purpose, we need to take the large $c$ limit outlined above. Details of the calculation are given in Appendix \ref{App:FM}. After some straightforward algebra, one arrives at the somewhat daunting looking integral expression (\ref{appfinal}). 
The integral can be done by the method of residues. The final result can be organized in the following symmetric expression\\[-1mm]
\bea\label{eq:6jSch}
\sixj{\Deltal_1}{\Deltal_3}{k_2}{k_4}{k_s}{k_{t}} \is \sqrt{\Gamma(\Deltal_1\pm i k_2 \pm ik_s)\Gamma(\Deltal_3 \pm ik_2\pm ik_t) \Gamma(\Deltal_1\pm i k_4 \pm i k_t)\Gamma(\Deltal_3\pm i k_4 \pm i k_s)}\nonumber\\[2mm]
\! & \!\! & \qquad \qquad \times\; \mathbb{W}(k_s, k_t ; \Deltal_1 + i k_4,\Deltal_1 - i k_4, \Deltal_3 - i k_2,\Deltal_3 + i k_2),_{\strut{}}
\ea
where we define $\Gamma(x\pm y\pm z)$ as a shorthand for the product of the gamma function with four combinations of signs. The function that appears in the right hand side is a rescaled version of the Wilson function introduced by  W.~Groenevelt \cite{groenevelt}. The original function introduced in \cite{groenevelt} is denoted by $\mathbb{W}(\alpha,\beta;a,b,c,d) = \phi_{\alpha}(\beta;a,b,c,1-d)$ and it is proportional to a generalized hypergeometric function ${}_7F_6$ evaluated at $z=1$ whose coefficients depend on $\alpha$, $\beta$, $a$, $b$, $c$ and $d$.

Given that the above expression was obtained as a limit of the quantum 6j-symbol, it is natural to suspect that the result can be interpreted as a classical $6j$-symbol.  The above  indeed matches with the $6j$-symbol associated to the Lie algebra $\mathfrak{su}(1,1)$ found by  W.~Groenevelt \cite{groenevelt}. The heavy operators with label $k_i$ correspond to the principal unitary series representations of $\mathfrak{su}(1,1)$, while the light operators $\Deltal_i$ correspond to the discrete series.\footnote{Note that even though $SU(1,1)$ and $SL(2,\mathbb{R})$ are isomorphic, their tensor categories are different and they have different $6j$-symbols.} 
The expression \eqref{rmatrix} enjoys tetrahedral symmetry that acts by permutations on the six spin labels.\footnote{The classical 6j-symbol of any Lie group can indeed be written as the expectation value of six Wilson lines, glued together into a tetrahedron, of the corresponding 2D BF-gauge theory. }  In addition, the 6j-symbols satisfy the unitarity condition 
\bea
\int \! dk_s\spc \rho(k_s)\spc \sixj{\Deltal_1}{\Deltal_3}{k_2}{k_4}{k_s}{k_{t}}^{\dagger} \spc \sixj{\Deltal_1}{\Deltal_3}{k_2}{k_4}{k_s}{k_{t'}} \is \frac{1}{\rho(k_t)}\spc\delta(k_t\! -k_{t'}), \quad \quad \rho(k) = 2k \sinh(2\pi k)\quad
\eea
which underscores the proposed holographic interpetation of the R-matrix (\ref{rmatrix}) as describing a gravitational scattering amplitude in the bulk. This unitarity condition is also responsible for the crossing symmetry of the 2D Liouville four point function \cite{PT}. 

Wilson functions also appeared recently as a fusion matrix of conformal blocks in a toy-model CFT with $SL(2,\mathbb{R})$ symmetry \cite{Hogervorst:2017sfd}\cite{Hogervorst:2017kbj}. It would be interesting to understand how these two approaches are related.

\def\mM{\mbox{\small {\rm M}}}
\def\Ez{\nspc\smpc\alpha\nspc\smpc}
\def\Eo{\nspc\smpc\epsilon_0\nspc\smpc}
\def\Et{\nspc\smpc\omega\nspc\smpc}
\def\Ef{\nspc\smpc\beta\nspc\smpc}

\subsection{Gravitational Scattering and Chaos}
\label{sect53}
We have shown that the characteristic behavior of the OTO correlation function is governed by an R-matrix in the form of a 6j-symbol. This R-matrix is a unitary matrix, that incorporates the gravitational bulk scattering amplitude in momentum space. 
 In this section we summarize how one can extract the characteristic Lyapunov exponent from the R-matrix. Our discussion here closely follows the derivation given in section 1 
 of \cite{Jackson:2014nla} for case of  AdS${}_3$/CFT${}_2$. 

The R-matrix depends on six parameters, $\ell_1,\ell_3, k_2,k_4, k_s$ and $k_t$. Following \cite{Jackson:2014nla}, let us label the four momenta as follows
\bea
\label{massp}
k_2\spc =\, \mM\smpc , \ \ \  \qquad \quad \ \;  & & \quad k_s\spc = \spc \mM+\alpha,\nonumber \\[-3mm]\\[-3mm]
k_{4}\spc  =\spc  \mM\nspc + \omega, \quad \qquad  & & \quad k_t \spc =\spc \mM+ \beta\spc .\nonumber
\eea
We will assume that we are in the regime $\mM\gg \alpha,\beta,\omega \gg \ell_1, \ell_3$. So to isolate the leading order behavior, we will set $\ell_1=\ell_3=0$. In this notation, the explicit integral formula for the R-matrix  takes the form
\bea
\label{rmatrixab}
R_{\alpha\beta}\!\! \is \!\!\! \int_{\cal C} \frac{du}{2\pi} \, \frac{\Gamma (u)\Gamma (i \alpha\nspc - \nspc u ) \Gamma (i (\alpha\nspc -\nspc\omega )\nspc - \nspc u)\Gamma (u\nspc-\nspc i (\alpha \nspc +\nspc \beta \nspc -\nspc \omega ))}{\Gamma (u\nspc -\nspc i \alpha ) \Gamma (u\nspc -\nspc i \alpha\nspc +\nspc i \omega )}\Gamma (u\nspc -\nspc 2 i \mM_{2\alpha} )  \Gamma (u\nspc +\nspc 2i\mM_{\beta + \omega-\alpha} )\nonumber\\[2mm]
& & \mbox{${}$}\hspace{-1.1cm}\times\, \sqrt{\frac{\Gamma (-i  \alpha ) \Gamma (i \beta )\Gamma (i ( \omega\nspc-\nspc\alpha  )) \Gamma (-\nspc i ( \omega\nspc -\nspc\beta ))}{\Gamma (i \alpha ) \Gamma (-i  \beta )\Gamma (-\nspc i (\omega \nspc -\nspc \alpha))\Gamma (i ( \omega \nspc -\nspc \beta ))}}\;
\sqrt{\frac{ \Gamma (2i\mM_\alpha )\Gamma (-2i\mM_\beta )\Gamma (2i\mM_{\alpha +\omega} ) \Gamma (-2i\mM_{\beta+\omega} )}
{\Gamma (-2i \mM_\alpha ) \Gamma (2i\mM_\beta) \Gamma (-2i \mM_{\alpha +\omega})\Gamma (2i\mM_{\beta +\omega })}} \nonumber\\
\eea
\vspace{-11mm}

\noindent
where $2i\mM_{\alpha} = 2i\mM + i \alpha$, etc.

How do we extract physical information from this expression for $R_{\alpha\beta}$? Since in the Schwarzian QM, it represents an exchange property of the momentum space amplitude, it is useful to label the operators by means of their momentum, or rather, by means of the amount they shift the momentum of the state on which they act. Concretely,  we can define the action of a momentum space operator $A_\omega$ via the algebraic rule
\bea
A_\omega\, |\mM\rangle = \gamma_A (\omega) \, \bigl| \mM+\omega\ra,
\eea
where $|\mM\ra$ denotes an energy eigenstate with $SL(2,\mathbb{R})$ spin $j= -\frac 12 + i\mM$, with $\gamma_A(\omega) = \gamma_{\ell_A}(\mM,\mM+\omega)$ the vertex function. This algebraic rule allows us to multiply operators and keep track of the time dependence through the usual Schr\"odinger evolution. This prescription works as long as the operators are time-ordered. 

The R-matrix (\ref{rmatrixab}) prescribes what happens if we exchange two operators and place them in out-of-time order. Schematically,
 \bea
{B}_{\omega-\alpha}\, A_\alpha\, | \mM \rangle \is \sum_\beta {R}_{\alpha\beta} \; A_{\omega-\beta}\, B_\beta\, | \mM\rangle.
  \eea 
Here $\sum_\beta$ is short-hand for $\int\! d\beta \rho(\mM\!+\!\beta)$ with $\rho(\mM\!+\!\beta) = (\mM\!+\!\beta)\sinh(2\pi(\mM\!+\!\beta))$ the spectral density of the intermediate state $|\mM+\beta\rangle$.

\begin{figure}[t]
\centering
\includegraphics[width=.38\textwidth]{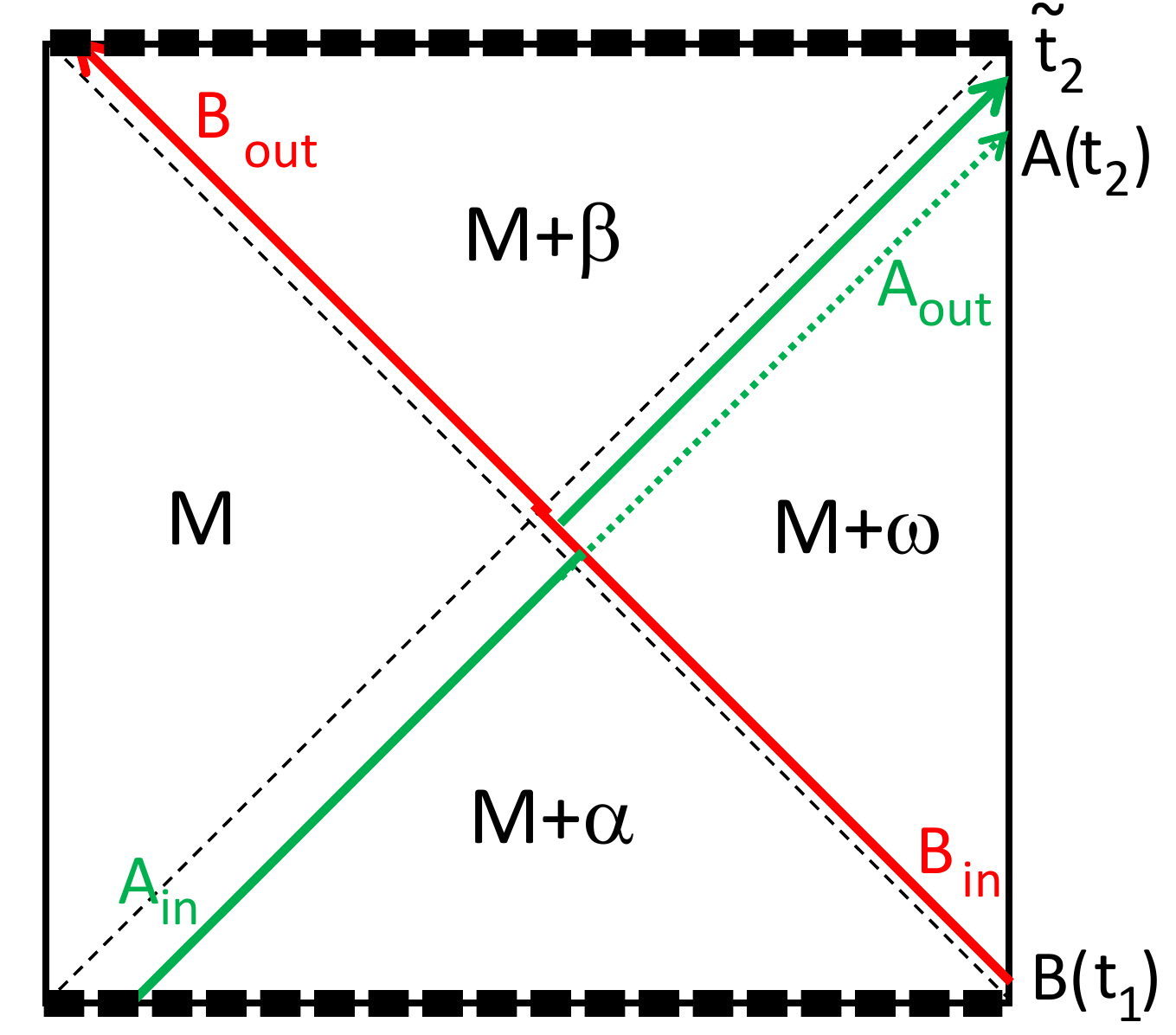}
\caption{\small The R-matrix describes the gravitational shockwave interaction between an infalling and outgoing matter perturbation near a black hole horizon. The particle trajectories divide the space-time into four regions.}
\label{shockshift}
\end{figure}

From the bulk perspective, this exchange algebra expresses the physical effect of an ingoing perturbation created by $B_{\omega-\alpha}$ (the `butterfly') on an outgoing signal $A$, as indicated in Figure \ref{shockshift}. To see the associated Lyapunov behavior, we need to translate the scattering phase to the time domain. This is done via the standard rules of geometric optics, which is justified since in the regime of interest, the phase of the R-matrix is rapidly changing with frequency. Focussing on this phase factor, we write
\bea
\label{rexps}
 {R}_{\alpha\beta} \is  
 e^{iI_{\alpha\beta}}. 
\eea
Next, we localize the operators $A$ and $B$ in time by considering them as wave-packets with a given approximate frequency.  In the leading order stationary phase approximation the exchange relation then takes the form
\bea
\label{xch}
B_{{\omega-\alpha}}(t_1) \, A_{\alpha}(t_2) \is \spc e^{i I_{\alpha\beta}} \, A_{\omega-\beta}(\smpc \tilde{t}_2\smpc) \,  B_{\beta}(\tilde{t}_1),
\eea
where the value of $\beta$, $\tilde{t_2}$ and $\tilde{t_1}$ on the right-hand side are fixed by the stationary phase criterion. Let us introduce the time differences
\bea
t_\alpha = t_2-t_1, \qquad \qquad t_\beta = \tilde{t}_2 - \tilde{t}_1.
\eea
These time differences are linked through the frequency dependence of the scattering phase ${R}_{\alpha\beta} = e^{iI_{\alpha\beta}}$ via the Hamilton-Jacobi (geometric optics) equations
\bea
\label{hj}
t_\alpha =-\frac{\partial I_{\alpha\beta}}{\partial E_\alpha} = -\frac{1}{2 \mM} \frac{\partial I_{\alpha\beta}}{\partial \alpha},
\qquad \qquad
t_\beta \!\is \! \frac{\partial I_{\alpha\beta}}{\partial E_\beta} = \frac{1}{2 \mM} \frac{\partial I_{\alpha\beta}}{\partial \beta}, 
\eea
which follow from the fact that both sides of the exchange relation (\ref{xch}) have the same dependence on $\alpha$ and $\beta$.\footnote{The phase $I_{\alpha\beta}$ is the generating function of the canonical transformation between the initial and final canonical variables $(E_\alpha,t_\alpha)$ and $(E_\beta,t_\beta)$.}

The prediction from bulk gravity is that the time delay $\tilde{t}_2 -t_2$ of the outgoing signal $A$ due to the perturbation $B$ grows exponentially with the time separation $t_2-t_1$
\bea
\tilde{t}_2 - t_2 \sim e^{\lambda_{\rm M} (t_2-t_1)}, \qquad \qquad \lambda_{\rm M} \, =\, \frac{2\pi}{\beta_{\rm M}},
\eea
with $\beta_{\rm M}$ the temperature of the black hole dual to the state $| \mM\ra$. We have 
\bea
S(\mM) = 2\pi \mM, \qquad \ E(\mM) = \mM^2, \qquad \ \beta_{\rm M} = \frac{\pi}{ \mM} , \qquad\
\lambda_{\rm M} \is \frac{2\pi}{\beta_{\rm M}}\,= \, 2 \mM\, .
\eea
Can we extract this from the exact expression \eqref{rmatrixab} for the R-matrix?

We have thus far not been able to find a precise enough way to evaluate the integral \ref{rmatrixab}. So we will proceed by making a plausible assumption, in the form of the following
\smallskip

\noindent
{\it Ansatz: in the semi-classical regime, the integral \eqref{rmatrixab} is dominated by the residue at $u=0$.}

\smallskip

\noindent
The pole at $u=0$ appears due to the $\Gamma(u)$ factor in the integrand. The above hypothesis is supported by several pieces of evidence. First, a naive application of Stirling and the stationary phase approximation indeed points to the existence of a saddle point near $u=0$. Secondly, as we will see shortly, via this Ansatz we can make contact with the semi-classical analysis of \cite{Jackson:2014nla}, which applies in the regime of large $c$ and large conformal dimensions $\Delta - \frac{c-1}{24}$ of order $c$. We leave the further justification of the above simplifying Ansatz for future study.

Evaluating the residue and approximating $\log\bigl(\frac{\Gamma(i(2{\rm M} + \alpha))}{\Gamma(-i(2{\rm M} + \alpha))}\bigr) \sim 2 i \alpha \log(2\mM)$, we find that 
\bea  
R_{\alpha\beta}^{(u=0)} \is e^{-i (\alpha +\beta-\omega)t_{\rm M} }\; \Gamma (i (\omega\nspc-\nspc \alpha \nspc -\nspc \beta ))\;
\sqrt{\frac{\Gamma (i  \alpha ) \Gamma (i \beta )\Gamma (-i ( \omega\nspc-\nspc\alpha  )) \Gamma (-\nspc i ( \omega\nspc -\nspc\beta ))}{\Gamma (-i \alpha ) \Gamma (-i  \beta )\Gamma ( i (\omega \nspc -\nspc \alpha))\Gamma (i ( \omega \nspc -\nspc \beta ))}}\;\nonumber
\eea
with $t_{\rm M} \simeq \log( 2 \mM)$. Now using Stirling gives
\bea\label{iabshock}
I_{\alpha\beta} \! &\! \simeq \! &\!
\spc  \alpha \log \alpha \spc + \spc \beta\log \beta
- ( \omega\nspc -\nspc \alpha) \log(\omega\nspc -\nspc \alpha) - \spc (\Et\nspc -\nspc\Ef) \log(\Et\nspc -\nspc \Ef) \nonumber\\[2mm] & &  \qquad  -\;  (\Ez + \beta - \Et\smpc ) \log\bigl(\Ez+ \Ef  - \Et \bigr)- (\Ez + \beta - \Et\smpc )  t_R ,
\eea 
which is identical to the formula for $I_{\alpha\beta}$ derived in \cite{Jackson:2014nla} from both 2+1D gravity and from 2D Virasoro CFT. Using this expression for $I_{\alpha\beta}$ and the geometric optics relations (\ref{hj}) we can now compute the time difference $t_\beta$ as functions of $\alpha$, $\omega$, and the time difference  $t_\alpha$ 
\bea t_\alpha - t_R\! \is\!  \frac 1 {\lambda_{\rm M}} \,\log \Bigl(\spc\frac{\alpha +  \beta-\omega}{2\smpc \alpha \smpc (\omega - \alpha)}\spc \Bigr), \qquad \qquad
t_\beta -t_R = \frac 1 {\lambda_{\rm M}} \,\log \Bigl(\spc \frac{\alpha +  \beta-\omega }{2 \smpc \beta \smpc (\omega - \beta)\spc }\spc\Bigr),
\eea 
or equivalently \cite{SS, Jackson:2014nla}
\bea
\label{relno}
\beta \is \omega - \alpha\spc +\spc  2 \spc\alpha \smpc (\omega-\alpha)
\spc e^{\lambda_{\rm M} (t_\alpha-t_R)},\\[2mm]
\label{relnt}
\alpha \is \omega - \beta + \spc 2\spc \beta \smpc (\omega -\beta) \spc e^{\lambda_{\rm M} (t_\beta-t_R)}.
\eea
These two relations are identical to the ones derived for the shockwave scattering process near a black hole in 2+1D AdS space-time \cite{SS}, and also match with the expected behavior in the AdS${}_2$ Jackiw-Teitelboim model. 
Equation (\ref{relno}) determines $\beta$ and $t_\beta$ as a function of $\alpha$ and the time difference $t_\alpha = t_2-t_1$. One finds
\bea
\label{timeshift}
\tilde{t_2}-t_2 \is  - \frac 1{ \lambda_{\rm M}}\,\log \Bigl(\frac{\omega-\alpha}{\beta}\Bigr) \, \simeq \, \frac{2 \alpha}{\lambda_{\rm M}}\, e^{\lambda_{\rm M}(t_2-t_1-t_R)},
\eea
which exhibits the expected maximal Lyapunov growth.

\section{SUSY generalizations}
 In the previous sections we presented a picture where the Schwarzian theory is completely solved in terms of known objects of 2D Liouville theory. In this section we will generalize these arguments to the supersymmetric cases. These theories can likewise be realized as the low energy effective theory of certain supersymmetrized SYK models \cite{Fu:2016vas}. For the $\mathcal{N}=1$ Schwarzian theory it is also known how it arises from a Jackiw-Teitelboim supergravity approach \cite{Forste:2017kwy}.

\subsection{ $\mathcal{N}=1$ Schwarzian theory}\label{sec:N1}

The super-Schwarzian is defined in $\mathcal{N}=1$ ($\tau,\theta$) superspace by:
\begin{equation}
\text{Sch}(\tau) \equiv \text{Sch}_f(\tau) + \theta \text{Sch}_b(\tau) = \frac{D^4\theta'}{D\theta} - 2 \frac{D^3\theta'D^2\theta'}{(D\theta')^2},
\end{equation}
with $D= \partial_\theta + \theta \partial_\tau$ the superderivative and $\theta'= \sqrt{\partial_\tau f}\left(\theta + \eta + \frac{1}{2}\theta\eta\partial_\tau \eta\right)$ as defined in \cite{Friedan:1986rx, Cohn:1986wn} for the reparametrization $f$ and its superpartner $\eta$.
Via the same arguments as for the bosonic theory, we can view the super-Schwarzian theory as the $c\to \infty$ limit of $\mathcal{N}=1$ super-Liouville theory between a pair of ZZ-branes. As usual, one has different sectors depending on the fermionic boundary conditions both between the ZZ-branes (open channel) and along the small circle (closed channel).
This gives four possibilities $NS$, $R$, $\widetilde{NS}$ or $\widetilde{R}$, where the tilde means that we insert a $(-1)^F$ in the partition function for the corresponding sector.

From the 2D Liouville perspective, different choices of brane configurations and different sectors correspond to the quantization of different coadjoint orbits. The one relevant for the application of the $\mathcal{N}=1$ Schwarzian as a low energy theory is 
\beq
{\rm Diff} \big(S^{1|1}\big)/{\rm OSp}(1|2).
\eeq
The path integral over this space was formulated and studied in \cite{Fu:2016vas} and \cite{wittenstanford}. This space is parametrized by a bosonic mode $f(\tau)$ associated to reparametrizations and a fermionic mode $\eta(\tau)$. From the 2D perspective it will turn out the relevant sector is $\widetilde{NS}$. Following the same procedure as in the construction of bosonic branes in Liouville one can solve the modular bootstrap to find the exact partition function associated to ZZ-branes in this section \cite{Fukuda:2002bv, Ahn:2002ev}. For the $\mathcal{N}=1$ case it is still given by the identity character 
\beq\label{eq:N1ch}
Z_{\mathcal{N}=1}= \chi_0^{\widetilde{NS}}(q) = {\rm Tr}_{\rm NS} (-1)^F q^{L_0 - \frac{c}{24}} = q^{-\frac{c-1}{24}}~(1+q)\sqrt{\frac{\theta_4(\tau)}{\eta(\tau)^3}}.
\eeq
Taking a parametrization similar to the bosonic case $q=e^{-\frac{48\pi^2}{\beta c}}$ and the limit $c\to\infty$ we obtain
\beq
Z_{\mathcal{N}=1}= e^{S_0^{\mathcal{N}=1}} ~\left(\frac{\pi}{\beta}\right)^{1/2} \exp\Bigl(\spc \frac{\pi^2}{\beta}\Bigr),
\eeq
where $S_0^{\mathcal{N}=1}$ denotes the zero-point entropy of the system. It is possible to see explicitly from \eqref{eq:N1ch} that because of supersymmetry, we obtain a vanishing zero-point energy $E_0^{\mathcal{N}=1}=0$, although the zero-point entropy is still divergent $S_0^{\mathcal{N}=1}\sim \log b$. The modular transfomation of this character automatically gives the exact density of states of the theory 
\bea
\label{N1mod}
\chi_0^{\widetilde{{\rm NS}}} \left(q\right)\! \is\! \int_0^\infty\!\!\! dP~S^P_0~\chi^{\rm R}_P(\tilde{q})
, \qquad \  \tilde{q} \spc = \spc e^{-\frac{\beta c}{12}},\qquad \ 
\chi_P(\tilde{q}) \spc = \spc \sqrt{\frac{\theta_4(\tilde{\tau})}{\eta(\tilde{\tau})}}\frac{ \tilde{q}^{P^2}}{\eta(\tilde\tau)},
\eea
 where the modular S-matrix corresponding to the $\mathcal{N}=1$ extension of the Virasoro algebra is given by 
\bea
S^P_0 \is 4 \cosh \bigl( 2 \pi b P\bigr) \cosh\Bigl( \mbox{\Large $\frac{\mbox{\footnotesize $2$} \pi P}{b}$} \Bigr).
\eea
Notice that the modular transformation turns the $\widetilde{NS}$ sector into the $R$-sector. Taking the appropriate limit to recover the Schwarzian theory gives a density of states 
\bea
Z_{\mathcal{N}=1}(\beta) \is  \int_0^\infty\! \!\! d\mu(k) \, e^{-\beta E(k)}, \qquad \quad d\mu(k) = dk \cosh(2\pi k),
\eea
which matches the result found in \cite{wittenstanford}.

The modular bootstrap of $\mathcal{N}=1$ super-Liouville also provides an expression for the ZZ-brane wavefunction. Moreover, a generalization of the DOZZ formula which gives the OPE coefficients of local operators is also known. Combining these two pieces of information, in the same way as was done for the bosonic case, we can obtain correaltion functions of local operators between the branes \footnote{The only interesting correlator is when a $NS$ vertex operator is inserted, as inserting one $R$ vertex operator yields zero, by (spacetime) fermion number conservation.}. The details and outcomes of these calculations are left for appendix \ref{sec:AppN1}, where we also explain the connection between primary operators in $\mathcal{N}=1$ Liouville and the natural super-Schwarzian $OSp(1|2)$-invariant bilocal operators.   
 
The main observable is a $\mathcal{N}=1$ generalization of the one studied in the bosonic case, which we denote by 
\bea
\label{N1bilocal}
\mathcal{O}_\Deltal(\tau_1,\tau_2) \equiv 
\left( \frac{\sqrt{f'(\tau_1)f'(\tau_2)}}{\frac{\beta}{\pi}\sin \frac{\pi}{\beta}[f(\tau_1)-f(\tau_2)]} \right)^{2\Deltal}+ ~({\rm fermion ~bilinears}).
\eea
The explicit form of the extra fermionic terms is given in appendix \ref{sec:AppN1}. The exact expectation value of this operator $G^\beta_\Deltal(\tau_1,\tau_2)= \lb \GO^{\mathcal{N}=1}_{\Deltal}(\tau_1,\tau_2) \rb$ is given by
\bea\label{eq:n12ptexact}
~~ &&\hspace{-1.1cm} G^\beta_\Deltal(\tau_1,\tau_2)\, = \,    \frac{2 \spc e^{-\frac{\pi^2}{\beta}}}{ \pi^{5/2} \beta^{-1/2}}\int^{\strut{}}_{\strut{}} \!\! dk_1 dk_2\spc \cosh(2 \pi k_1) \cosh( 2 \pi k_2)\, e^{-\tau k_1^2 - (\beta-\tau)k_2^2}\, \nonumber\\
&&\hspace{1.2cm}\times \spc \bigg(\frac{\Gamma\bigl({\small \frac{1}{2}}+\Deltal\pm i(k_1- k_2)\bigr)\Gamma\bigl(\Deltal\pm i(k_1+ k_2)\bigr)}{\Gamma(2\Deltal)} + (k_2 \to -k_2 )\bigg)
\ea
where $\tau=\tau_{12}$. The two-point function of its superpartner can also be computed, and we refer to appendix \ref{sec:AppN1} for the result.

Additionally, stress tensor insertions in Liouville theory lead to (the bosonic piece of) super-Schwarzian insertions that are dealt with completely analogously as in the bosonic case, and leads to the constant energy:
\bea
\la \text{Sch}_{b}(\tau)\ra \is \frac{1}{\beta} + \frac{2\pi^2}{\beta^2}.
\eea
The fermionic piece of the super-Schwarzian $\text{Sch}_f(\tau)$ analogously arises from the Liouville supercurrent $T_F$ (with $R$- boundary conditions along the circle) and has a zero one-point function due to worldsheet fermion conservation. Its two-point function does not vanish, and is just as the bosonic stress tensor two-point function constant up to contact terms. The constant piece is readily seen to be the square root of the corresponding bosonic piece, due to $G_0^2 = L_0$ in the parent 2d theory. 
Importantly, it requires the same $1/b^2$ rescaling to define a finite quantity: $T_F(w) \to \frac{1}{2b^2}\text{Sch}_{f}(\tau)$, consistent with 1d supersymmetry: $T_F(w) + \theta T(w) \to \text{Sch}_f(\tau) + \theta \text{Sch}_b(\tau)$.

Going beyond the two-point function, one finds again a Feynman diagram decomposition which is structurally identical to the bosonic case. The spectral measure now takes the form
\begin{equation}
d\mu(k) = dk \cosh(2\pi k),
\end{equation}
while the vertices for bosonic and superpartner insertions are respectively given by
\begin{align}
\gamma_\ell(k_1,k_2)^2 &= {\frac{\Gamma\bigl(\frac{1}{2}+\ell \pm i(k_1-k_2)\bigr) \Gamma\bigl(\ell \pm i(k_1+k_2)\bigr) + (k_2 \to -k_2)}{2\Gamma(2\ell)}}, \\[1.5mm]
\gamma_\ell^{\Psi}(k_1,k_2)^2 &= {\frac{(k_1+k_2)^2\Gamma\bigl(\frac{1}{2}+\ell \pm i(k_1-k_2)\bigr) \Gamma\bigl(\ell \pm i(k_1+k_2)\bigr) + (k_2 \to -k_2)}{2\Gamma(2\ell)}}.
\end{align}
The R-matrix should be computed by taking the Schwarzian limit of the $U_q(\mathfrak{osp}(1|2))$ quantum group 6j-symbols for four Ramond continuous near-parabolic insertions and two light NS insertions. While several results are known on this object, the fusion matrix with this specific configuration is not yet available \cite{Hadasz:2013bwa, Pawelkiewicz:2013wga}.

Finally, we briefly comment on the possibility of considering other fermionic boundary conditions. Using either the characters or the known wavefunctions, one immediately deduces that the spectral density for both the $R$- and the $NS$-sector is given by
\bea
\rho(E) \is \sinh(2\pi \sqrt{E}).
\eea
The characters and their 1d Schwarzian limit are summarized as follows.
\begin{equation}
\!\begin{array}{c|c|c}
\text{Sector} \! & \! 2d \! & \! 1d \\
\hline
NS \! & \! \text{Tr}_{NS}q^{L_0-c/24} \! & \! \text{bosonic} \\
\widetilde{NS} \! & \! \text{Tr}_{NS}(-)^F q^{L_0-c/24} \! & \! Z \\
R \! & \! \text{Tr}_{R}q^{L_0-c/24} \! & \! \text{bosonic} \\
\widetilde{R} \! & \! \text{Tr}_{R}(-)^F q^{L_0-c/24} = \text{Witten index} = 0 \! & \! \text{Witten index} = 0
\end{array}\! \nonumber
\end{equation}
Only one interesting supersymmetric sector remains in the 1D limit, and that is the one of the $\mathcal{N}=1$ Schwarzian theory introduced above.
For $NS$- and $R$-sectors, no fermionic zero-mode along the circle survives and these sectors then give non-supersymmetric 1D thermal models, identical to the bosonic theory. 
The partition function of the $\widetilde{R}$-sector contains periodic zero-modes along the circle and periodic fermionic boundary conditions along the Schwarzian thermal circle, identifying it as the Witten index both in 2D and in 1D.

\subsection{ $\mathcal{N}=2$ Schwarzian theory}\label{sec:N2}
In this section we want to identify which sector of $\mathcal{N}=2$ super-Liouville generates the path integral over the orbit 
\beq
{\rm Diff}\big(S^{1|2}\big)/{\rm OSp}(2|2),
\eeq
relevant for the $\mathcal{N}=2$ super-Schwarzian theory \cite{Cohn:1986wn, Fu:2016vas}. The character of the identity representation is given by 
\beq
{\rm ch}_{\widetilde{NS}} (q,y) ={\rm Tr}_{\rm NS} (-1)^F q^{L_0 - \frac{c}{24}}y^{\Deltal_0} = \frac{e^{\pi i \hat{c} \frac{z^2}{\tau}}~q^{-\frac{1}{4b^2}}(1-q)}{(1-y q^{\frac{1}{2}})(1-y^{-1} q^{\frac{1}{2}})} \frac{\theta_3(q,-y)}{\eta(\tau)^3},
\eeq
where $\hat{c}=\frac{c}{3}=1+\frac{2}{b^2}$. A special feature of the $\mathcal{N}=2$ case is that this is not equal to the partition function of a pair of ZZ-branes anymore. In \cite{Eguchi:2003ik} T.~Eguchi and Y.~Sugawara solved the modular bootstrap for $\mathcal{N}=2$ super-Liouville, see also \cite{Ahn:2003tt}. They found that, in order to do this, one is forced to take a sum over spectral flow. Moreover, the construction only works for rational central charge $\hat{c}=1+\frac{2K}{N}$ for any $K,N\in \mathbb{Z}$. The partition function of the pair of ZZ-brane is equal to the extended character which is defined as 
\beq\label{eq:N2exch}
Z_{\mathcal{N}=2}=\chi_0^{\widetilde{NS}}(q,y) = \sum_{n\in N\mathbf{Z}} q^{\frac{\hat{c}}{2}n^2}y^{\hat{c} n} {\rm ch}_{\widetilde{NS}} (q,y).
\eeq
We will take $N$ to be finite and $K\to \infty$ to take the $c\to\infty$ limit. Taking the limit $\tau \to \frac{2 \pi }{\beta} b^2$ and $z=\alpha \tau$ with $\alpha$ fixed, equation (\ref{eq:N2exch}) becomes 
\beq
Z_{\mathcal{N}=2}=\frac{4}{\pi} \sum_{n\in N\mathbf{Z}} \frac{\cos \pi (\alpha+n)}{1-4(\alpha+n)^2} ~e^{\frac{\pi^2}{\beta}\bigl(1-4(\alpha+n)^2\bigr)}.
\eeq
This expression coincides with the exact partition function found in \cite{wittenstanford} if we identify $\alpha$ to be proportional to the chemical potential. The parameter $N$ corresponds to the size of the compact boson present in the $\mathcal{N}=2$ multiplet. From a path integral point of view (as opposed to invoking the modular bootstrap) the same happens in $\mathcal{N}=2$ super-Liouville, since besides the field $\phi$ we have another boson which is compact, usually denoted $Y$, with a coupling similar to bosonic sine-Liouville theory. Finally, this parameter also corresponds to the R-charge of the fermions in the SYK model.

 As anticipated, we find that all divergences in this limit disappear. We find $E_0^{\mathcal{N}=2}=0$ due to supersymmetry as in the $\mathcal{N}=1$ case, but we also find a finite zero-point entropy $S_0^{\mathcal{N}=2}=\log \frac{4}{\pi}$. This seems to indicate that even though bosonic and $\mathcal{N}=1$ Schwarzian theory should be interpreted as a low energy effective theory of a QM system, the $\mathcal{N}=2$ super-Schwarzian might be a well-defined theory by itself.

Following the procedure applied to the bosonic case, we can read off the density of states of the theory from the modular properties of the identity character. The modular transformation of the identity character in the $\widetilde{NS}$ sector is given by 
\bea\label{eq:modN2ch}
{\rm ch}_{\widetilde{NS}}\Big(\frac{-1}{\tau},\frac{z}{\tau}\Big) \! & \!=\! & \! \int_{-\infty}^\infty d\omega\int_0^\infty dp \frac{\sinh(\pi \mathcal{Q} p)\sinh(2\pi\frac{p}{\mathcal{Q}})}{\mathcal{Q} \big| \sinh \pi \left( \frac{p}{\mathcal{Q}} + i \frac{\omega}{\mathcal{Q}^2}\right)\big|^2} {\rm ch}_{\rm cont}^{\rm R}(p,\omega;\tau,z)\nn
\! & \!\! & \!+2\sum_{n\in\mathbf{Z}} \int_{-\frac{1}{2}}^{\frac{1}{2}} d\omega~ \cos \pi \omega ~{\rm ch}_{\rm BPS}^{\rm R}(\omega,n;\tau,z),
\ea
where $\mathcal{Q}^2=2K/N$. The integral is over the continuous representation with Liouville momenta $p$ and R-charge $\omega$ in the R-sector. The second line corresponds to a sum over BPS states in the R-sector. These can have arbitrary charge $\omega$ but the $\mathcal{N}=2$ super-Virasoro algebra implies that they have a fixed dimension $\Delta_{\rm BPS}^{\rm R} = \frac{c}{24}$ independent of the charge. We give some more details of these representations and their characters in Appendix \ref{App:N2}. A similar formula also exists for the modular transformation of the extended characters 
\bea\label{eq:exchmod}
\chi_{\widetilde{NS}}\Big(\frac{-1}{\tau},\frac{z}{\tau}\Big) \! & \!=\! & \! \frac{1}{N} \sum_{m\in\mathbf{Z}_{2 NK}}\int_0^\infty dp \frac{\sinh(\pi \mathcal{Q} p)\sinh(2\pi\frac{p}{\mathcal{Q}})}{\mathcal{Q} \big| \sinh \pi \left( \frac{p}{\mathcal{Q}} + i \frac{\omega}{\mathcal{Q}^2}\right)\big|^2}~ \chi_{\rm cont}^{\rm R}\Big(p,\frac{m}{N};\tau,z \Big)\nn
\! & \!\! & \!+\frac{2}{N}\sum_{n\in\mathbf{Z}} \sum_{m=1}^{N-1}~ \sin \pi \frac{m}{N} ~\chi_{\rm BPS}^{\rm R}\Big(\frac{m}{N}-\frac{1}{2},n;\tau,z\Big).
\ea
 At finite $N$ the expressions look similar but now instead of having an integral over charges we have a sum over discrete charges. This was one of the original motivation to take spectral flow into account, since having a continuum of charges does not seem physical.

Before writing down the density of states as a function of energy and charge we will restrict to the case $\alpha=0$ to match \eqref{eq:exchmod} with the expression found in \cite{wittenstanford}. If we take  \eqref{eq:exchmod} in the appropriate limit, and perform the integrals over $\omega$, using some results left for the appendix, we directly obtain 
\beq
\rho_{\alpha=0}(E)=\sum_{n\in\mathbf{Z}} \frac{2 \cos \pi n}{1-4 n^2} \left(\frac{\sqrt{1-4n^2}I_1\bigl(2\pi\sqrt{(1-4 n^2)E}\bigr)}{\sqrt{E}} + \delta(E) \right),
\eeq
where the integral over the Bessel function comes from the integral over non-BPS states and the delta function comes from the BPS states. While this is obtained in \cite{wittenstanford} by performing an inverse Laplace transform, in our approach both terms have a physical origin. For arbitrary $\alpha$, \eqref{eq:exchmod} gives the following density of states as a function of both energy and charge
\bea
Z(\beta,\alpha)\! & \!=\! & \!\frac{1}{N}\sum_{m\in \mathbf{Z}} \int_{\frac{m^2}{4N^2}}^\infty \frac{dE}{8E}\; \textstyle {\sinh\Bigl(\spc 2 \pi \sqrt{E-\frac{m^2}{4N^2}}\spc \Bigr)}  \; e^{-\beta E} \; \bigl( y^{\frac{m}{N}+\frac{1}{2}} +  y^{\frac{m}{N}-\frac{1}{2}} \bigr) \nn
\! & \!\! & \qquad \qquad +\;  \frac{2}{N} \sum_{m=1}^{N-1}~\int_0^\infty dE~ \sin \pi \frac{m}{N} ~\delta(E)~e^{-\beta E}~y^{\frac{m}{N}-\frac{1}{2}}.
\ea

From this expression we can obtain the density of states $\rho(E,Q)$. If we redefine the chemical potential such that charge is either integer or half-integer $y\to y^{N}$ and shift $m$ in order to get a dependence $y^Q$ then we obtain 
\bea
\rho_{\rm cont}(E,Q)&=&\frac{1}{8N}\frac{\sinh(2 \pi \sqrt{E-E_0^+(Q)})}{E}~ \Theta(E-E_0^+(Q))~+~\big(+\leftrightarrow-\big),\\
\rho_{\rm disc}(E,Q) &=& \delta(E) ~\frac{2}{N}\cos \pi \frac{Q}{N}~~ \Theta(2|Q|-N),
\ea
where $\Theta(x)$ corresponds to the Heaviside step function and we defined the two charge-dependent threshold energies as following $E_0^\pm (Q) \equiv \left( \frac{Q}{2N} \pm \frac{1}{4} \right)^2$. 

The modular transformation gives us explicit expressions for these densities which would be hard to find starting from the expression with Bessel functions. From these, we can point out some features which agree with the DH calculation \cite{wittenstanford}. The zero energy contribution from BPS states appear only for a finite charge range $2|Q|<\frac{N}{2}$. The continuous part of the spectrum consists of two terms that start contributing at different energies. The lowest one gives the minimum energy possible for the continuous spectrum with a fixed charge, which is  
\beq
E_{\rm min}(Q)=\left(\frac{|Q|}{2 N} - \frac{1}{4}\right)^2.
\eeq
Finally, depending on whether $N$ is even or odd, the sum over charges is either over integer $Q$ or half-integer $Q$ respectively.

\section{Concluding Remarks}
We have presented a connection between the Schwarzian theory and a 2D CFT with Virasoro symmetry. In particular, the relevant 2D theory is Liouville between ZZ-branes. Using this connection and the output of the conformal bootstrap of Liouville \cite{PT}, we have obtained exact correlation functions of the Schwarzian theory. The final answers are summarized in section \ref{sect:overview}.  

Recently, there has been considerable progress towards understanding the simplest version of holography between quantum gravity on nearly AdS$_2$ and a nearly conformal quantum mechanics CFT$_1$. The original motivation for studying the Schwarzian theory is to understand  this duality in more detail. 
 In section \ref{sec:2ptZZ} we described how to perform an analytic continuation that takes our finite temperature two-point function to the thermofield double state. Moreover, as explained in section \ref{sect5}, the OTO correlators are built up from insertions of the R-matrix, which have a holographic interpretation in terms of gravitational scattering. In section \ref{sect53} we give some evidence of the Lyapunov behavior of the OTO correlator when $t \gtrsim \beta$. For even larger times $t \gg C$, starting from equation (\ref{oto4pt}) we can deduce a power-law decay as $\sim t^{-6}$ consistent with the results of \cite{altland}. We leave the detailed verification of this proposed interpretation and the study of its implications to future work.

Throughout this work, we have only discussed $SL(2,\mathbb{R})$ invariant observables. It is however also possible to construct $SL(2,\mathbb{R})$ covariant and local operators and only demand invariance of the entire correlator. Correlators of such operators are heavily constrained by the equality of the Casimir and the Hamiltonian; it is possible to construct an analogue of the Knizhnik-Zamolodchikov equation in this context. As this is slightly orthogonal to our main results, we present this and a related bootstrap analysis of such correlators in Appendix \ref{sect:KZ}.

We can mention some connections between our calculations and black hole physics by studying the two-point function. The exact answer was given in equation (\ref{realt2pt}). After labeling the integration variables as $k^2_2= E$ and $k_1^2=E+\omega$ the real-time two-point function can be written in the form 
\beq
G_\ell^{\pm}(t) \sim \int dE\spc \rho(E) \spc e^{-\beta E} \int d\omega\spc e^{i \omega t} \spc |\mathcal{A}_E (\omega, \ell) |^2,
\eeq
where we include in the amplitude $\mathcal{A}_E$ the density of states $\rho(E+\omega)$ and the gamma functions coming from the OPE coefficients in equation (\ref{eq:2ptexact}). If we take the expression integrated over $\omega$, but not yet over $E$, we can interpret it as an amplitude in gravity involving a black hole of fixed mass $E$. Then $ |\mathcal{A}_E (\omega, \ell) |^2$ corresponds to the amplitude associated to the process $E\to E+\omega \to E$ with the absorption/emission of a particle of mass $\ell$ and energy $\omega$. Following for example \cite{Krasnov:2004ki, Turiaci:2016cvo}, we can obtain the quasi-normal mode frequencies of the black hole as the location of the poles of $ |\mathcal{A}_E (\omega, \ell) |^2$ as a function of $\omega$. For example, if we take $\omega \ll E$ then $|\mathcal{A}_E (\omega, \ell) |^2 \sim \Gamma\big(\ell \pm i \frac{\omega}{2\sqrt{E}} \big)$. Choosing the required contour for the retarded Green function, we pick up poles in $\omega$ on the lower half plane and we recover $\omega_n = - i \frac{2\pi}{\beta} (\ell + n)$, with $\beta = \frac{dS}{dE} = \frac{\pi}{\sqrt{E}}$ and $n\in \mathbb{N}$. These are the quasi-normal frequencies of the 2D black hole or, equivalently, the dimensional reduction of BTZ. We note that compared to that case, no angular momentum parameter nor Virasoro descendant modes appear. Of course, we could have obtained this directly by taking the Fourier transform of the classical answer \eqref{sp2pt}. For arbitrary values of $\omega$ and $E$, the resonances in the amplitude give a non-perturbative definition of the quasi-normal modes.  

To conclude, we mention some other possible generalizations of the formalism presented in this paper. It would be interesting to extend our analysis of correlation functions to the supersymmetric cases beyond $\mathcal{N}=1$. The $\mathcal{N}=2$ theory has an abelian R-symmetry making it a non-trivial extension. $\mathcal{N}=2$ Liouville theory is also related to 2D black holes through an FZZ-like duality. The $\mathcal{N}=4$ case would also be instructive, as it would correspond to an SYK-like model with non-abelian symmetry.
Another natural generalization is to study the generalized Schwarzian theory that arises by taking the 1D limit of 2D Toda theory (see for example \cite{Floch:2015hwo}), which has an extended symmetry algebra,
and to construct the corresponding generalized SYK model and Jackiw-Teitelboim theory that would have this extended Schwarzian model as its low energy description. 
Likewise, it would be interesting to understand whether the methods of this paper can be applied to the 2D Schwarzian theory proposed in \cite{Turiaci:2017zwd}.

\begin{center}
{\bf Acknowledgements}
\end{center}

\vspace{-2mm}

We thank N. Callebaut, R. Dijkgraaf, H. Lam, B. Le Floch, J. Maldacena, D. Stanford, E. Witten and Z. Yang for useful discussions. T.M. acknowledges financial support from the Research Foundation-Flanders (FWO Vlaanderen). The research of H.V. is supported by NSF grant PHY-1620059.

\begin{appendix}

\def\WW{\mbox{\small W}}

\section{Stress tensor correlators}
\label{sect:stress}
The simplest correlation functions are those of the Schwarzian derivative itself:
\beq
{\rm Sch}(f,\tau) \equiv \bigl\{ \tan \frac{\pi f(\tau)}{\beta} , \tau\bigr\},
\eeq
which correspond to insertions of the Hamiltonian itself. From the 2D perspective, we can obtain these correlation functions via the dictionary
\bea
T(w) \; \leftrightarrow \; \frac{c}{12} \, {\rm Sch}(f,\tau).
\eea
These correlation functions are fixed by Virasoro Ward identities. The case of a single insertion corresponds to taking the derivative of the partition function in the open channel with respect to the modulus $q$:
\bea
\lb T (z) \rb \is \frac{1}{Z} \frac{\partial Z}{\partial \log q}.
\eea
It is instructive to evaluate both sides via the representation of the partition function in terms of the ZZ boundary states
\bea
\lb ZZ | T | ZZ\rb \is \frac{1}{Z} \int dk^2~e^{-\beta k^2} \sinh(2 \pi k) ~\lb k| T | k\rb.
\eea
The stress-tensor one-point function is constant on the cylinder: upon mapping the plane to the cylinder via $z=e^{-\frac{w}{b}}$, using the standard anomalous transformation law for the stress tensor and that $h_k = \frac{Q^2}{4} + b^2k^2$, one finds
\bea
\lb k| T(w) | k\rb \is \frac{k^2}{b^2} - \frac{1}{24b^4}  \quad \rightarrow \qquad
\la T(z) \ra \, = \, \frac{c}{12}\Bigl(\, \frac{2\pi^2}{\beta^2} +\frac{3}{\beta}\,\Bigr),
\eea
where we dropped a constant ground state energy contribution $E_0$. So we deduce that
\bea
\la{\rm Sch}(f,\tau) \ra \is  \frac{2\pi^2}{\beta^2} +\frac{3}{\beta},
\eea
which reproduces the result in \cite{wittenstanford}.

 Correlation functions of more insertions can be deduced in the same way. The two-point function of the stress-tensor on the cylinder reads
\bea
\left\langle k\right|T(w_1)T(w_2)\left|k\right\rangle \is \frac{c}{2b^8\left(2\sinh\frac{w_{12}}{2b^2}\right)^4} + \frac{2h_k}{b^8\left(2\sinh\frac{w_{12}}{2b^2}\right)^2} + \frac{h^2}{b^4}.
\eea
Next we take the 1D limit. If $w_{1} \neq w_{2}$, the pole terms vanish in the $b\to0$ limit, only the $h^2$-part remains. One obtains
\bea
\left\langle ZZ\right|{\rm Sch}(f,\tau_1)\, {\rm Sch}(f,\tau_2)\left|ZZ\right\rangle  \is  \frac{4\pi^4}{\beta^4} + \frac{20\pi^2}{\beta^3} + \frac{15}{\beta^2}.
\eea
When $w_{1} = w_{2}$, the pole terms in the above expression become genuine contact terms. Using the standard $TT$ OPE, and defining $\lb\!\lb A B \rb \! \rb = \lb A B \rb - \lb A \rb \lb B \rb $, one finds that
\bea
\la\!\la {\rm Sch}(f,\tau_1)\, {\rm Sch}(f,\tau_2)\ra\!\ra \is - 2\delta''(\tau_{12}) - 4 \la {\rm Sch}(f,\tau_1)\ra \delta(\tau_{12}) + \frac{8 \pi^2}{\beta^3} + \frac{6}{ \beta^2}.
\eea
This coincides with the result of \cite{wittenstanford}.

Similarly, a 1D version of the Ward identity can be derived:
\bea
&&\left\lb {\rm Sch}(f,\tau)   \prod_i \mathcal{O}_{\Deltal}(\tau_{2i-1},\tau_{2i})\right\rb- \lb {\rm Sch}(f,\tau) \rb \left\lb\prod_i \mathcal{O}_{\Deltal}(\tau_{2i-1},\tau_{2i})\right\rb=\nn
&&= \left(  \sum_{j=1}^{2n} 2\Deltal\delta(\tau-\tau_j) -2 \, {\rm sgn}(\tau-\tau_j) \partial_{\tau_j}\right) \left\lb\prod_i \mathcal{O}_{\Deltal}(\tau_{2i-1},\tau_{2i})\right\rb.
\ea
Generalizing to non-equal $\ell$'s is straightforward. It would be interesting to interpret this formula from the point of view of AdS$_2$ gravity as a soft-mode theorem of boundary gravitons. 

A similar analysis can be done for the supersymmetric extensions of the Schwarzian theory. 

\section{Fusion matrix in the Schwarzian limit}\label{app:ZZ}

\subsection{Special functions}
In this section we will present some definitions and properties of special functions appearing recurrently in Liouville theory, both in the DOZZ formula for the OPE coefficients and in the fusion matrix. 

All special function are built upon a deformed version of the Gamma function 
\beq
\Gamma_b(x) \equiv \frac{\Gamma_2(x | b, b^{-1})}{\Gamma_2 (Q/2|b,b^{-1})},
\eeq
where $\Gamma_2(z|\epsilon_1,\epsilon_2)$ is the Barnes double gamma function. This function is uniquely defined by the properties under a shift in $b$ or $b^{-1}$ of its argument. 
In this paper we are interested in the $b\to0$ limit. In this regime one can approximate this function by 
\bea
\Gamma_b(b x) &\to& (2 \pi b^3)^{\frac{1}{2}(x-\frac{1}{2})}\Gamma(x),\\
\Gamma_b(Q-bx) &\to& (2\pi b)^{-\frac{1}{2}\left(x-\frac{1}{2}\right)}.
\ea

In the DOZZ formula this function appears in a specific combination $\Upsilon_b(x) \equiv \frac{1}{\Gamma_b(x)\Gamma_b(Q-x)}$. Using the expressions above in the small $b$ limit we can approximate this function by 
\beq
\Upsilon_b(b x) \to \frac{b^{\frac{1}{2}-x}}{\Gamma(x)} .
\eeq
This result was used when the Schwarzian limit of the DOZZ formula was taken. 

Finally, in the integral formula for the fusion matrix the relevant combination of $\Gamma_b$ is called the double-sine function and defined by $S_b(x)\equiv \frac{\Gamma_b(x)}{\Gamma_b(Q-x)}$. It satisfies
\bea
S_b(x+b)&=& 2 \sin (\pi b x) S_b(x),\\
S_b\Bigl(x+\frac{1}{b}\Bigr) &=& 2 \sin \Bigl(\frac{\pi x}{b}\Bigr) S_b(x).
\ea
If we keep $x,y$ fixed and take $b\to 0$ the following limits are 
\bea
S_b(b x) &\to&(2\pi b^2)^{x-\frac{1}{2}}\Gamma(x)\\
S_b\Bigl(\frac{1}{2b}+b x\Bigr) &\to& 2^{x-\frac{1}{2}}\\
S_b\Bigl(\frac{1}{2b}+b x+\frac{y}{b}\Bigr) &\to& \Bigl(\frac{\cos \pi y}{2}\Bigr)^{\frac{1}{2}-x} e^{-\frac{1}{b^2}\int_0^\infty \frac{dt}{t} \left[\frac{\sinh 2 y t}{2 t \sinh t}-\frac{y}{t} \right]},~~|{\rm Re} y|<\frac{1}{2}
\ea
These and more results can be found in \cite{Ribault:2007td}. In the next section of this appendix we will use these approximate expressions to obtain the Schwarzian limit of the fusion matrix.

\subsection{Fusion matrix}\label{App:FM}
A formula for the $6j$-symbols of $U_q(\mathfrak{sl}(2,\mathbb{R}))$, derived in \cite{Teschner:2012em}, is given by 
\bea\label{app:6j3}
\sixj{\alpha_1}{\alpha_3}{\alpha_2}{\alpha_4}{\alpha_s}{\alpha_{t}}_b^{}
&=&\Delta(\alpha_s,\alpha_2,\alpha_1)\Delta(\alpha_4,\alpha_3,\alpha_s)\Delta(\alpha_t,\alpha_3,\alpha_2)
\Delta(\alpha_4,\alpha_t,\alpha_1)\nn
&&\qquad \times\int_{\mathcal{C}}du\;
S_b(u-\alpha_{12s}) S_b(u-\alpha_{s34}) S_b(u -\alpha_{23t}) \\
&&\!\!\!\!\!
S_b(u-\alpha_{1t4})
S_b( \alpha_{1234}-u) S_b(\alpha_{st13}-u) 
S_b(\alpha_{st24}-u) S_b(2Q-u)\nonumber
\ea
where, following the notations of \cite{Teschner:2012em} we defined the following normalization factors 
\beq
\Delta(\alpha_3,\alpha_2,\alpha_1)\equiv \left(\frac{S_b(\alpha_1+\alpha_2+\alpha_3-Q)}{S_b(\alpha_1+\alpha_2-\alpha_s)S_b(\alpha_1+\alpha_s-\alpha_2)S_b(\alpha_2+\alpha_s-\alpha_1)}\right)^{1/2}.
\eeq
 The integral is defined for the cases in which all $\alpha_k\in \frac{Q}{2} + i \mathbb{R}$. The contour $\mathcal{C}$ approaches $2Q+i\mathbb{R}$ at infinity and passes the real axis in the interval $(3Q/2,2Q)$. In order to take the Schwarzian limit we need an analytic continuation of this formula. Unfortunately, although this representation makes the symmetries of the $6j$-symbol manifest, it does not allow for a natural evaluation in the Schwarzian limit. Instead we will start a representation that does not have the symmetry manifest but makes the analytic continuation straightforward
\beq
\sixj{\alpha_1}{\alpha_3}{\alpha_2}{\alpha_4}{\alpha_s}{\alpha_{t}}_b^{}
 = \frac{M(\alpha_s,\alpha_2,\alpha_1)M(\bar{\alpha}_4,\alpha_3,\alpha_s)}{M(\alpha_t,\alpha_3,\alpha_2)M(\bar{\alpha}_4,\alpha_t,\alpha_1)} ~\sixjAN{\alpha_1}{\alpha_3}{\alpha_2}{\bar{\alpha}_4}{\alpha_s}{\alpha_t}
\eeq
with $\bar{\alpha}_4\equiv Q-\alpha_4$. The prefactors are defined as 
\beq
M(\alpha_3,\alpha_2,\alpha_1)=\left( \frac{S_b(\alpha_1+\alpha_2-\alpha_3)S_b(\alpha_1+\alpha_2+\alpha_3-Q)}{S_b(\alpha_1+\alpha_3-\alpha_2)S_b(\alpha_2+\alpha_3-\alpha_1)}\right)^{1/2},
\eeq
and following \cite{Teschner:2001rv} we use the following integral representation of the asymmetric $6j$-symbol
\bea
\sixjAN{\alpha_1}{\alpha_3}{\alpha_2}{\bar{\alpha}_4}{\alpha_s}{\alpha_t}\equiv \frac{S_b(\alpha_2+\alpha_s-\alpha_1)S_b(\alpha_t+\alpha_1-\alpha_4)}{S_b(\alpha_2+\alpha_t-\alpha_3)S_b(\alpha_s+\alpha_3-\alpha_4)}\int_{-i\infty}^{i\infty} ds~\prod_{i=1}^4 \frac{S_b(U_i+s)}{S_b(V_i+s)}
\ea
where the $U_i$ and $V_i$ factors are defined as 
\bea
&\!\!\!\!\!\!\!\!\!\!\!\!U_1= \alpha_s+\alpha_1-\alpha_2 ~~~&V_1=2Q+\alpha_s-\alpha_t-\alpha_2-\alpha_4\nn
&U_2=Q+\alpha_s-\alpha_1-\alpha_2 ~~~&V_2=Q+\alpha_s+\alpha_t-\alpha_2-\alpha_4\nn
&\!\!\!\!\!\!\!\!\!\!\!\!U_3= \alpha_s+\alpha_3-\alpha_4 ~~~&V_3=2\alpha_s\nn
&U_4=Q+\alpha_s-\alpha_3-\alpha_4~~~&V_4=Q.
\ea
In the limit we are interested in, which we refer to as the Schwarzian limit, we choose the following set of parameters 
\bea\label{eq:Schlimit}
\alpha_1\! \is \! \Deltal_1 b, ~~~~~~~~~~~\alpha_2=\frac{Q}{2}+i bk_2,~~~~~~~~~~~\alpha_s = \frac{Q}{2}+i b k_s\nn
\alpha_3\! \is \! \Deltal_3 b, ~~~~~~~~~~~\alpha_4=\frac{Q}{2}+i bk_4,~~~~~~~~~~~\alpha_t=\frac{Q}{2}+i b k_t\nonumber
\ea
It is important to make this identification in this order since otherwise the $b\to0$ limit will be ill-defined. One can check that all the pre-factors have a well-defined $b\to0$ limit, by using the identities of the previous appendix involving double sine functions. Having done this, the only non-trivial aspect of the calculation is the integral appearing in the definition, which we denote by
\bea
I(j_1,k_2,j_3,k_4;k_2,k_t) \equiv \int_{-i\infty}^{i \infty} \frac{ds}{2\pi i}~4 \pi^2 b^3~\prod_{i=1}^4 \frac{S_b(U_i+s)}{S_b(V_i+s)}.
\ea
In the Schwarzian limit the integrand becomes
\bea\label{app:integrand}
4 \pi^2 b^{4}~\prod_{i=1}^4 \frac{S_b(U_i+s)}{S_b(V_i+s)} &=&\frac{\Gamma(s+i[k_t-k_s+k_2+k_4])\Gamma(s-i[k_s+k_t-k_2-k_4])}{\Gamma(s+j_1-i(k_s-k_2))\Gamma(s+j_3-i(k_s-k_4))}\nn
&&\Gamma(s-2ik_s)\Gamma(s)\Gamma(j_1+i (k_s-k_2)-s)\Gamma(j_3+i(k_s-k_4)-s)\nonumber
\ea

Before writing down the answer for this integral, let's consider the most general case and solve the following problem 
\beq
I=\int_{-i \infty}^{+i\infty} \frac{ds}{2\pi i}~\frac{\Gamma(a_1+s)\Gamma(a_2+s)\Gamma(a_3+s)\Gamma(a_4+s)}{\Gamma(b_1+s)\Gamma(b_2+s)} ~\Gamma(A-s)\Gamma(B-s).
\eeq
This can be computed by the method of residues. The integral is done over the imaginary axis, and we take a contour that leaves the poles of $\Gamma(A-s)$ and $\Gamma(B-s)$ to the right, and all the other poles to the left. For the integral relevant for the computation of the $6j$-symbols this is the proper contour to take. If we close the contour to the right and pick up only the poles of $\Gamma(A-s)$ and $\Gamma(B-s)$, this integral is given by 
\beq\label{app:integral}
 I=\frac{\Gamma(B-A)\prod_{i=1}^4\Gamma(A+a_i)}{\Gamma(A+b_1)\Gamma(A+b_2)} ~_4F_3\Big[\mbox{\small$\begin{array}{cccc}\! A+a_1 \!\! & \!\! A+a_2 \!\! &\!\! A+a_3 &\!\! A+a_4 \!\nspc \\[-1mm] \!\! A+b_1 \!\! &\!\! A+b_2 \!\! &\!\! 1+A-B \!\nspc \end{array}$} ; 1 \Big] + (A\leftrightarrow B),
\eeq
where the generalized hypergeometric function is defined as 
\beq
~_4F_3\Big[\mbox{\small$\begin{array}{cccc}\! a_1 \!\! & \!\! a_2 \!\! &\!\! a_3 &\!\! a_4 \!\nspc \\[-1mm] \!\! b_1 \!\! &\!\! b_2 \!\! &\!\! b_3 \!\nspc \end{array}$} ; z \Big] \equiv \sum_{n=0}^\infty \frac{ (a_1)_n(a_2)_n(a_3)_n(a_4)_n}{(b_1)_n(b_2)_n(b_3)_n}\frac{z^n}{n!}.
\eeq
For the particular choice of parameters that appear in the $6j$-symbol integral it is instructive to recognize this sum of hypergeometric functions as a Wilson function. This function was introduced in \cite{groenevelt} and is defined as
\beq
\mathbb{W}(\alpha,\beta;a,b,c,d) \equiv \frac{\Gamma(d-a)~_4F_3\Big[\mbox{\small$\begin{array}{cccc}\! a+i\beta \!\! & \!\! a-i\beta \!\! &\!\! \tilde{a}+i \alpha &\!\! \tilde{a}-i\alpha \!\nspc \\[-1mm] \!\!\!\! a+b \!\! &\!\!\!\!\! a+c \!\! &\!\! 1+a-d \!\nspc \end{array}$} ; 1 \Big]}{\Gamma(a+b)\Gamma(a+c)\Gamma(d\pm i \beta) \Gamma(\tilde{d}\pm i\alpha)}  + (a\leftrightarrow d),
\eeq
where $\tilde{d}=(b+c+d-a)/2$ and $\tilde{a}=(a+b+c-d)/2$. As seen from the definition, this is not the most general sum of hypergeometric functions and it requires a specific relation between its parameters. It is explained in \cite{groenevelt} that the Wilson function can also be written as a single ${}_7F_6$ hypergeometric function evaluated at $z=1$, which makes some non-trivial symmetries of its parameters more transparent. 

If we use the general integral result given in equation (\ref{app:integral}), and evaluate for the particular parameters of equation (\ref{app:integrand}), to apply it to the Schwarzian limit of the $6j$-symbol, then we obtain the final answer for the integral as 
\bea
I &=& \frac{\Gamma(j_3-j_1+ik_2-ik_4)\Gamma(j_1 - ik_2 \pm ik_s) \Gamma(j_1+ik_4\pm ik_t)}{\Gamma(2 j_1)\Gamma(j_1+j_3-ik_2+ik_4)}\nn
&&~\nn
&&\times{}_4F_3\Big[\mbox{\small$\begin{array}{cccc}\! j_1 + ik_4+ ik_t \!\! & \!\! j_1 + ik_4- ik_t \!\! &\!\! j_1 - ik_2- ik_s &\!\! j_1 - ik_2+ ik_s \!\nspc \\[-1mm] \!\! 2j_1 \!\! &\!\! j_1+j_3-i k_2 +ik_4 \!\! &\!\! 1+j_1-j_3-ik_2 +ik_4 \!\nspc \end{array}$} ; 1 \Big]\nn
&&~\nn
&&+~ \frac{\Gamma(j_1-j_3+ik_4-ik_2)\Gamma(j_3 - ik_4 \pm ik_s) \Gamma(j_3+ik_2\pm ik_t)}{\Gamma(2 j_3)\Gamma(j_1+j_3+ik_2-ik_4)}\\
&&~\nn
&&\times{}_4F_3\Big[\mbox{\small$\begin{array}{cccc}\! j_3 + ik_2+ ik_t \!\! & \!\! j_3 + ik_2- ik_t \!\! &\!\! j_3 - ik_4- ik_s &\!\! j_3 - ik_4+ ik_s \!\nspc \\[-1mm] \!\! 2j_3 \!\! &\!\! j_1+j_3-i k_4 +ik_2 \!\! &\!\! 1+j_3-j_1-ik_4 +ik_2 \!\nspc \end{array}$} ; 1 \Big]\nonumber
\ea 
It is straightforward to notice that this sum of hypergeometric functions is indeed of the Wilson function type if we make the identification of $\alpha \to k_s$, $\beta \to k_t$ and 
\bea
a\is  j_1 + i k_4,\qquad \quad
b \, =\,  j_1 - i k_4, \qquad \quad 
c \, = \,  j_3 - i k_2,\\[2mm]
d \is j_3 + i k_2,\qquad \quad 
\tilde{a} \, =\, j_1 -ik_2,\qquad \quad 
\tilde{d}\, =\, j_3-ik_4.\nonumber
\ea
In terms of these parameters that appear in the definition of the Wilson function, the integral we computed can be rewritten simply in terms of the Wilson function
\beq
I = \Gamma(d\pm i \beta)\Gamma(\tilde{d}\pm i \alpha)\Gamma(a\pm i \beta )\Gamma(\tilde{a}\pm i \alpha)~\mathbb{W}(\alpha, \beta ; a,b,c,d).
\eeq
After including the prefactors we can use this formula for the integral to write the final expression of the $6j$-symbols of Liouville theory in the Schwarzian limit: 
\bea\label{app:6jSch}
\sixj{\alpha_1}{\alpha_3}{\alpha_2}{\alpha_4}{\alpha_s}{\alpha_{t}} \is \sqrt{\Gamma(j_1\pm i k_2 \pm ik_s)\Gamma(j_3 \pm ik_2\pm ik_t)\Gamma(j_1\pm i k_4 \pm i k_t)\Gamma(j_3\pm i k_4 \pm i k_s)}\nn
&&\qquad \  \ \times\ \ \mathbb{W}(k_s, k_t ; j_1 + i k_4,j_1 - i k_4, j_3 - i k_2,j_3 + i k_2),
\ea
where the notation $\Gamma(x\pm y \pm z)$ means to take the product of all four sign combinations. 
Using the final expression, let's see how the symmetries of the $6j$-symbols are recovered. 
\begin{itemize}
\item First of all, the Wilson function is symmetric in all the last four arguments. This means there is a symmetry $k_4 \to -k_4$ and $k_2 \to -k_2$. This symmetry is preserved by the prefactor. 
\item Another symmetry is to exchange $j_1\leftrightarrow j_3$ together with $k_2 \leftrightarrow k_4$, which is also preserved by the prefactor. This is equivalent to the relation 
\beq
\sixj{\alpha_1}{\alpha_3}{\alpha_2}{\alpha_4}{\alpha_s}{\alpha_{t}} = \sixj{\alpha_3}{\alpha_1}{\alpha_4}{\alpha_2}{\alpha_s}{\alpha_{t}}.
\eeq
\item In \cite{groenevelt} another relation is proven, referred to as duality of the Wilson function, 
\bea
\mathbb{W}(k_s, k_t ; j_1\! +\! i k_4,j_1\! -\! i k_4, j_3 \! -\!  i k_2,j_3\! +\! i k_2) \is \mathbb{W}(k_2, k_4 ; j_1\! +\! i k_t,j_1\! -\! i k_t, j_3\! - \! i k_s,j_3 \! +\!  i k_s). \nonumber
\ea 
This is also preserved by the prefactor and implies 
\beq
\sixj{\alpha_1}{\alpha_3}{\alpha_2}{\alpha_4}{\alpha_s}{\alpha_{t}} = \sixj{\alpha_1}{\alpha_3}{\alpha_s}{\alpha_t}{\alpha_2}{\alpha_4}.
\eeq
\item Finally, one can also exchange in both the Wilson function and the prefactor $k_s\leftrightarrow k_t$ together with $k_2 \leftrightarrow k_4$, namely 
\bea
\mathbb{W}(k_s, k_t ; j_1\! +\! i k_4,j_1\! -\! i k_4, j_3\! -\!  i k_2,j_3\!+\! i k_2)\is \mathbb{W}(k_t, k_s ; j_1\! +\! i k_2,j_1 \! -\! i k_2, j_3 \! -\!  i k_4,j_3\! + \! i k_4),\nonumber
\ea
which implies 
\beq
\sixj{\alpha_1}{\alpha_3}{\alpha_2}{\alpha_4}{\alpha_s}{\alpha_{t}} = \sixj{\alpha_1}{\alpha_3}{\alpha_4}{\alpha_2}{\alpha_t}{\alpha_s}.
\eeq
\end{itemize}
At the level of the Wilson function, the unitarity of the $6j$-symbols has already been proven by Groenevelt. After including the right prefactors, the $6j$-symbol generates an integral transformation equivalent to what he denotes as a Wilson transform of type 1. 

To conclude this appendix, we give an integral expression of the $6j$-symbol which will be useful in the main text (here $k_{a\pm b} = k_a \pm k_b$, etc) 
\bea
~~\sixj{\Deltal_1}{\Deltal_3}{k_2}{k_4}{k_s}{k_{t}} \is \sqrt{\frac{\Gamma(\Deltal_1+ik_2 \pm ik_s)\Gamma(\Deltal_3-ik_2\pm ik_t)\Gamma(\Deltal_1-i k_4 \pm ik_t) \Gamma(\Deltal_3+ik_4\pm ik_s)}{\Gamma(\Deltal_1-ik_2 \pm ik_s)\Gamma(\Deltal_3+ik_2\pm ik_t)\Gamma(\Deltal_1+i k_4 \pm ik_t) \Gamma(\Deltal_3-ik_4\pm ik_s)}}\nonumber \\[2mm]
\! & \!\! & \!\hspace{-2.6cm}\times \int\limits_{-i\infty}^{i\infty}\!\! \frac{du}{2\pi i} \, \frac{\Gamma(u)\Gamma(u\! -\! 2ik_s)\Gamma(u\! +\!  i k_{2+4-s+t})\Gamma(u\! -\! i k_{s+t-2-4})\Gamma(\Deltal_1\!  + \! i k_{s-2}\! -\! u)\Gamma(\Deltal_3\! +\! i k_{s-4}\! -\! u)}{\Gamma(u\! +\! \Deltal_1\!  - \! i k_{s-2})\Gamma(u\! +\! \Deltal_3\! -\! ik_{s-4})} .\nonumber \\[-4mm]
\label{appfinal}
\ea

\section{Details on supersymmetric cases}\label{sec:AppN1}

\subsection{$\mathcal{N}=1$ two-point functions}
To describe correlation functions, we again propose to insert the super-Liouville primaries $V_\alpha = e^{\alpha \phi}$ in the theory to describe Schwarzian correlators in the appropriate limit. The resulting bilocal Schwarzian operators correspond to an arbitrary reparametrization of the superconformal two-point function, written in superspace as:
\begin{equation}
\frac{1}{\tau_1-\tau_2-\theta_1\theta_2}\,  \to \,  \frac{D_1 \theta'_1 D_2 \theta'_2}{\tau_1'-\tau_2'-\theta_1'\theta_2'},
\end{equation}
where $\tau' = f(\tau+\theta \eta(\tau))$ and $\theta'$ is defined in the main text. Just as in the bosonic case, two approaches can be followed. When using the minisuperspace limit, we need the $\mathcal{N}=1$ generalization of the relevant momentum-eigenstate wavefunction. As the circular dimension has periodic (Ramond) boundary conditions, the fermions have zero-modes as well, and the minisuperspace limit is modified compared to the bosonic case. The momentum-eigenstate wavefunction is given by ($z= e^{b\phi}$) \cite{Douglas:2003up}\cite{Seiberg:2003nm}
\bea
\psi_{P+}(\phi) \is \frac{2}{\Gamma\left(-\frac{iP}{b}+\frac{1}{2}\right)}\sqrt{z}\left(K_{\frac{iP}{b}+\frac{1}{2}}(z) + K_{\frac{iP}{b}-\frac{1}{2}}(z)\right). 
\eea
The relevant matrix element is then
\bea
\left\langle Q\right| V_{jb}\left|P\right\rangle \is \frac{4}{\Gamma\left(\frac{iP}{b}+\frac{1}{2}\right)\Gamma\left(-\frac{iQ}{b}+\frac{1}{2}\right)}\nonumber \\
\times \! && \!\int dz z^j\, \Big(K_{\frac{iQ}{b}+\frac{1}{2}}(z) + K_{\frac{iQ}{b}-\frac{1}{2}}(z)\Big)\Big(K_{-\frac{iP}{b}+\frac{1}{2}}(z) + K_{-\frac{iP}{b}-\frac{1}{2}}(z)\Big),
\eea
which can be evaluated explicitly.
Alternatively, one can take the limit directly in the $\mathcal{N}=1$ DOZZ formula. To derive this result, we use the expansion of the ZZ-brane state in Ishibashi states 
\beq
\label{ishi}
|ZZ\rb = \int dP~\Psi(P) |P,R,+\rb = \int dP~\Psi(P)\left( | \Theta^{++}_{Q/2+iP}\right\rangle + \left|\Theta^{--}_{Q/2+iP} \rb\right) + (\text{descendants}),
\eeq
where $\Theta^{\pm\pm}_{\alpha} = \sigma^{\pm\pm}V_{\alpha}$ is a Ramond primary with $\sigma^{\pm\pm}$ the spin field in the R-sector. In the Schwarzian limit, descendants in (\ref{ishi}) are dropped again and two primaries remain, requiring four different terms to compute in the 3-point function, which are two by two equal. \\
Following \cite{Fukuda:2002bv}, we define
\bea
\tilde{C}_1 \is \la V_{\alpha_1}\Theta^{\pm\pm}_{\alpha_2}\Theta^{\mp\mp}_{\alpha_3}\ra, \qquad \qquad 
\tilde{C}_2 \, = \, \la V_{\alpha_1}\Theta^{\pm\pm}_{\alpha_2}\Theta^{\pm\pm}_{\alpha_3}\ra.
\eea
The $\mathcal{N}=1$ DOZZ formula is given by\footnote{$\Upsilon_{NS}(x)= \Upsilon\left(\frac{x}{2}\right)\Upsilon\left(\frac{x+Q}{2}\right)$, $\Upsilon_{R}(x)= \Upsilon\left(\frac{x+b}{2}\right)\Upsilon\left(\frac{x+b^{-1}}{2}\right)$.}
\bea
\tilde{C}_1 \is \frac{\Upsilon_{NS}'(0)\Upsilon_{NS}(2\alpha_1)\Upsilon_{R}(2\alpha_2)\Upsilon_{R}(2\alpha_3)}{\Upsilon_{R}(\alpha_{1+2+3}-Q)\Upsilon_{NS}(\alpha_{1+2-3})\Upsilon_{R}(\alpha_{2+3-1})\Upsilon_{NS}(\alpha_{3+1-2})}, \\[2mm]
\tilde{C}_2 \is \frac{\Upsilon_{NS}'(0)\Upsilon_{NS}(2\alpha_1)\Upsilon_{R}(2\alpha_2)\Upsilon_{R}(2\alpha_3)}{\Upsilon_{NS}(\alpha_{1+2+3}-Q)\Upsilon_{R}(\alpha_{1+2-3})\Upsilon_{NS}(\alpha_{2+3-1})\Upsilon_{R}(\alpha_{3+1-2})}.
\eea
Setting $\alpha_1=jb$, $\alpha_2 = Q/2 + ibk_1$ and $\alpha_3 = Q/2 + ibk_2$ and taking the small $b$-limit, one obtains
\bea
\tilde{C}_1 \is\frac{1}{2b}\frac{\left|\Gamma\left(\frac{1}{2}(j+i(k_1-k_2))\right)\Gamma\left(\frac{1}{2}(1+j+i(k_1+k_2))\right)\right|^2}{\Gamma(j)\Gamma(1/2-ik_1)\Gamma(1/2-ik_2)}, \\
\tilde{C}_2 \is \frac{1}{2b}\frac{\left|\Gamma\left(\frac{1}{2}(j+i(k_1+k_2))\right) \Gamma\left(\frac{1}{2}(1+j+i(k_1-k_2))\right)\right|^2}{\Gamma(j)\Gamma(1/2-ik_1)\Gamma(1/2-ik_2)}.
\eea

Comparing $2(\tilde{C}_1+\tilde{C}_2)$ to the minisuperspace computation above, one finds agreement. 
Setting $j=2\ell$ and rescaling $k_i$, the final two-point function is given by:
\begin{align}
\label{super2pt}
G_\ell(\tau) \,\, \! & \!= \frac{e^{-\frac{\pi^2}{\beta}}}{\pi^{\frac{5}{2}}\Gamma(2\ell) \beta^{-\frac{1}{2}}}\int dk_1 dk_2 e^{-\tau k_1^2 - (\beta-\tau)k_2^2} \cosh \left(2\pi k_1\right) \cosh \left(2\pi k_2\right)  \nonumber \\[2mm]
\! & \!\times \left(\Gamma\bigl(\textstyle \frac{1}{2}+\ell \pm i(k_1-k_2)\bigr)\, \Gamma\bigl(\ell \pm i(k_1+k_2)\bigr) + (k_2 \to -k_2)\right).
\end{align}
The normalization of these functions can be fixed/checked by noting that for $\ell\to 0$, one has $G_\ell \to 1$ and that the small $\tau$-expansion starts with $G_\ell = \frac{1}{\tau^{2\ell}}$ with unit prefactor.

Next we consider the two-point correlator of the superpartner $\Lambda_\alpha =- \alpha^2 \psi\bar{\psi}V_\alpha$ of $V_\alpha$, obtained by taking the superpartner in both left- and right-moving sectors simultaneously. Going back to the original (S)CFT, we are computing the Schwarzian correlation function of the bilocal observable:
\begin{equation}
\Psi_{\ell}(\tau_{1},\tau_2) = \left\langle \Psi(\tau_1)\Psi(\tau_2)\right\rangle_{\text{SCFT}},
\end{equation}
with $\Psi(\tau)$ the superpartner of the bosonic operator $\mathcal{O}(\tau)$. 
Using the superconformal Ward identity (see e.g. \cite{Fukuda:2002bv}), one finds for the three-point function on the sphere:
\bea
\left\langle \Lambda_{\alpha_1}(z,\bar{z}) \Theta^{\pm\pm}_{\alpha_2}(0)\Theta^{\pm\pm}_{\alpha_3}(\infty)\right\rangle \is -\frac{b^2}{2\left|z\right|}\left(k_1-k_2\right)^2 \left\langle V_{\alpha_1}(z,\bar{z}) \Theta^{\pm\pm}_{\alpha_2}(0)\Theta^{\mp\mp}_{\alpha_3}(\infty)\right\rangle, \\[2mm]
\left\langle \Lambda_{\alpha_1}(z,\bar{z}) \Theta^{\pm\pm}_{\alpha_2}(0)\Theta^{\mp\mp}_{\alpha_3}(\infty)\right\rangle \is - \frac{b^2}{2\left|z\right|}\left(k_1+k_2\right)^2 \left\langle V_{\alpha_1}(z,\bar{z}) \Theta^{\pm\pm}_{\alpha_2}(0)\Theta^{\pm\pm}_{\alpha_3}(\infty)\right\rangle.
\eea
These are the analogues of the DOZZ formula for $\Lambda_\alpha$. As $\alpha_1 =j b$, the factors of $b$ in the above expression can be cancelled by extracting them from the superpartner as $\tilde{\Lambda}_{\alpha_1} = j^2 \psi\bar{\psi}V_{\alpha_1}$, a consistency check that the $b\to0$ limit gives something non-trivial. 

Transforming to the cylinder $z=e^{-w/b^2}$ gives an additional factor of $-\frac{1}{b^2}\left|z\right|$ (compared to the underlying $V_\alpha$) and finally leads in the Schwarzian limit to the fermionic two-point function in the Schwarzian theory:\footnote{A check on some relative signs is that it is invariant under $\left\{\tau \to \beta-\tau, \, k_1 \leftrightarrow k_2\right\}$, and that it is manifestly invariant under $k_1 \to -k_1$ and $k_2 \to -k_2$. }
\begin{align}
G^\Psi_\ell(\tau) &= \frac{e^{-\frac{\pi^2}{\beta}}}{\pi^{\frac{5}{2}}\Gamma(2\ell) \beta^{-\frac{1}{2}}}\int dk_1 dk_2 e^{-\tau k_1^2 - (\beta-\tau)k_2^2} \cosh \left(2\pi k_1\right) \cosh \left(2\pi k_2\right) \\[2mm]
&\times \Bigl(\left(k_1+k_2\right)^2\Gamma\bigl(\textstyle\frac{1}{2}+\ell \pm i(k_1-k_2)\bigr)\spc\Gamma\bigl(\ell \pm i(k_1+k_2)\bigr) + (k_2 \to -k_2)\Bigr). \nonumber
\end{align}
Its small time decay starts at $G^\Psi_\ell = \frac{2\ell}{\tau^{2\ell+1}}$, identifying it with the superpartner in the Schwarzian theory. Numerically, the qualitative profile in time is very similar to all earlier cases discussed.

Note that this expression cannot be combined with its counterpart (\ref{super2pt}) in superspace in terms of the superdistance $\mathcal{T} = \tau_{12} - \theta_1\theta_2$, corresponding to the spontaneous breaking of supersymmetry at finite temperature. At zero temperature, supersymmetry is restored, and the superspace expression is written as:
\begin{align}
G_\ell(\mathcal{T})_{\beta \to +\infty} &= \frac{1}{\pi^{2}}\int dk \spc e^{-\mathcal{T} k^2} \cosh (2\pi k) \spc \frac{\Gamma\bigl(\textstyle\frac{1}{2}+\ell \pm ik\bigr) \Gamma\bigl(\ell \pm ik\bigr)}{\Gamma(2\ell)}.
\end{align}

\subsection{$\mathcal{N}=2$ density of states}\label{App:N2}
In this section we will outline the details of the derivation of the charge-integrated density of states obtained in \cite{wittenstanford} by starting from the character modular transformation. We will write the partition function as a sum over spectral flow parameter $n$ and take the modular transformation of each individual character. Therefore we will apply equation (\ref{eq:modN2ch}) for $y=e^{2 \pi i \alpha_n}$ and sum over $n$, where $\alpha_n=\alpha+n$. Lets start by considering the integral over the non-BPS continuous spectrum 
\beq
 \int_{-\infty}^\infty d\omega\int_0^\infty dp \frac{\sinh(\pi \mathcal{Q} p)\sinh(2\pi\frac{p}{\mathcal{Q}})}{\mathcal{Q} \big| \sinh \pi \left( \frac{p}{\mathcal{Q}} + i \frac{\omega}{\mathcal{Q}^2}\right)\big|^2} {\rm ch}_{\rm cont}^{\rm R}(p,\omega;\tau,z).
\eeq
The character of the continuous representation in the Ramond sector has the limit 
\beq
{\rm ch}_{\rm cont}^{\rm R}(p,\omega;\tau,z) = q^{P^2+\frac{b^2\omega^2}{4}} y^\omega \frac{\theta_{10}(q,y)}{\eta(q)^3} \to e^{-\beta \left( \rho^2+\frac{\omega^2}{4}\right)} y^\omega 2 \cos \pi \alpha_n.
\eeq
where we call $P=b \rho$, $q=e^{-\beta b^{-2}}$ and take $b\to0$. The integrand in the modular transformation has the following $b\to0$ limit 
\beq
\frac{\sinh(\pi \mathcal{Q} p)\sinh(2\pi\frac{p}{\mathcal{Q}})}{\mathcal{Q} \big| \sinh \pi \left( \frac{p}{\mathcal{Q}} + i \frac{\omega}{\mathcal{Q}^2}\right)\big|^2}  \to \frac{1}{\pi b} \frac{\rho \sinh 2 \pi \rho}{\rho^2+\frac{\omega^2}{4}}.
\eeq
Then the contribution from non-BPS states is given by 
\beq
Z_n^{\rm non-BPS} = 2 \cos \pi \alpha_n \int d\omega \int_0^\infty \rho d \rho~\frac{\sinh 2 \pi \rho}{\rho^2 + \frac{\omega^2}{4} } e^{-\beta  \left( \rho^2+\frac{\omega^2}{4}\right)} e^{2 \pi i \alpha_n \omega}.
\eeq
Now we will change variables from $\rho$ to $E\equiv \rho^2+\frac{\omega^2}{4}$ and change the order of integration such that the $\omega$ integral is done first. This gives 
\beq
Z_n^{\rm non-BPS} = 2 \cos \pi \alpha_n \int dE~  e^{-\beta E}  \int_{-2\sqrt{E}}^{2\sqrt{E}} d\omega~\frac{\sinh 2 \pi \sqrt{E-(\omega/2)^2}}{2 \pi E }e^{2 \pi i \alpha_n \omega}.
\eeq
The integral over $\omega$ can be done exactly and gives 
\beq
\int_{-2\sqrt{E}}^{2\sqrt{E}} d\omega~\frac{\sinh 2 \pi \sqrt{E-(\omega/2)^2}}{2 \pi E }e^{2 \pi i \alpha_n \omega} = \frac{I_1\bigl(2 \pi \sqrt{(1-4 \alpha_n^2)E}\bigr)}{\sqrt{(1-4 \alpha_n^2)E}},
\eeq
which matches the continuous part of the density found in \cite{wittenstanford}. For the BPS contribution the story is easier. The character of the BPS in the Ramond sector is 
\beq
\frac{y^\omega}{1+y} \frac{\theta_2(q,y)}{\eta(q)^3},
\eeq
and the integral over $\omega$ is easily perform. Since for BPS R-states $\Delta=c/24$, or Liouville momentum $P=0$, we have to introduce the integral over energy by hand with a delta function at $E=0$. This can be seen from the super-Virasoro algebra 
\beq
\{ G_r^+, G_s^-\} = 2 L_{s+r} + (r-s)J_{r+s} + \frac{c}{12}( 4r^2-1) \delta_{r+s}.
\eeq
If we look for BPS states in the NS-sector the fact that $G_{-1/2}$ annihilates the states is equivalent to the usual condition $h=|Q|/2$. If on the other hand we are in the Ramond sector, $G_0$ is the operator that annihilates the state. From the algebra one deduces that this implies the unusual BPS relation $\Delta^R_{\rm BPS} = \frac{c}{24}$ independent of the charge. Finally, since $q\to0$ the spectral flowed contributions can be neglected.

\section{Knizhnik-Zamolodchikov equation}
\label{sect:KZ}
It is possible to consider operators in the Schwarzian theory that depend only on a single (Lorentzian) time variable:
\begin{equation}
\phi^{j}(x,t) = \left(\frac{\dot{f}(t)}{(f(t)-x)^2}\right)^{j}, \quad x\in\mathbb{R},
\end{equation}
transforms under $SL(2,\mathbb{R})$ in the spin $j$ representation:\footnote{We will suppress the $x$ and $t$ labels of these operators from here on.}
\begin{equation}
\phi^{j}(x,t) \to \phi^{j}(x',t)(Cx'+D)^{2j},
\end{equation}
where $x = \frac{Ax'+D}{Cx'+D}$. As they are not $SL(2,\mathbb{R})$-invariant, this goes beyond the interpretation of $SL(2,\mathbb{R})$ as a gauge symmetry. Nonetheless, it is interesting to pursue this line of thought a bit further.
We are not aware of any natural origin of such operators in either the SYK model or in the Jackiw-Teitelboim model. 
 
All of the realizations of the Schwarzian theory discussed in section \ref{sect:schro} exhibit a $SL(2,\mathbb{R})$ symmetry whose Casimir equals the Hamiltonian operator. This heavily constrains correlation functions.

We will now present a short bootstrap analysis of correlation functions in theories with this property.
The discussion is similar to that given by Teschner in \cite{Teschner:1997fv} when discussing the minisuperspace limit of the $H_3^+$ WZW model using the ``baby conformal bootstrap". Even though the $H_3^+$ model is related to the Schwarzian at the level of the partition function (as shown earlier), it seems less useful when considering correlation functions. In fact, we will study the $SL(2,\mathbb{R})$ version of Teschner's $SL(2,\mathbb{C})$ story, studied recently in \cite{Hogervorst:2017sfd}. 

Classifying field insertions according to irreps of this algebra, the generators act on the quantum fields as
\begin{equation}
\left[\ell_a, \phi^{j}\right] = - t^{(j)}_a \phi^{j},
\end{equation}
with Casimir $H$ itself:
\begin{equation}
\left[H, \phi_{j}\right] = -2t^{(j)}_n\phi_j \ell_n + C_j \phi_{j} = -2t^{(j)}_n \ell_n \phi_j - C_j \phi_{j}.
\end{equation}

Without loss of generality, we assume we are working with an $SL(2,\mathbb{R})$ invariant vacuum.\footnote{Via the state-operator correspondence we can rewrite non-trivial in- and out-states as operator insertions.}
$SL(2,\mathbb{R})$ invariance of correlators requires
\begin{equation}
\sum_{k=1}^{n}t^{(k)}_a\la\phi_1 \hdots \phi_n\ra =  0, \quad a=-1,0,+1.
\end{equation}
In analogy with 2d CFT, we can write a Knizhnik-Zamolodchikov equation that links the Hamiltonian operator with the Casimir $ g^{ab}\ell_a \ell_b = \ell_{0}^2 - \frac{1}{2}\left\{\ell_1, \ell_{-1}\right\}$. Inserting a ``null" field 
\begin{equation}
\bigl[H - \bigl(\ell_0^2 -\textstyle \frac{1}{2}\left\{\ell_1, \ell_{-1}\right\}\bigr), \phi_i \bigr] = 0
\end{equation}
in the correlator, one finds
\begin{align}
\Bigl( i \partial_{t_i} + g^{ab}\sum_{\substack{k=1 \\ k\neq i}}^{n} \text{sgn}(i-k)t^{(k)}_a \otimes t^{(k)}_b \Bigr)\left\langle \phi_1 \hdots \phi_n\right\rangle= 0, \quad i=1, ..., n.
\end{align}
Compared to 2d CFT, note the replacement of the single pole by a sign-function. 
Summing this expression over $i$, one finds time translation invariance:
\begin{equation}
\label{timet}
i\sum_i \partial_{t_i}\left\langle \phi_1 \hdots \phi_n\right\rangle = 0.
\end{equation}

$SL(2,\mathbb{R})$ invariance combined with the KZ condition then constrains the two- and three-point function to be of the form (with the Casimir $\omega_k = j_k(j_k+1)$):\footnote{E.g. for the three-point function, one finds
\beq
\left(i\partial_{t_1} - \omega_1 \right)\left\langle \phi_1 \phi_2 \phi_3\right\rangle = 0,  \qquad
\left(i\partial_{t_2} + (\omega_1-\omega_3) \right)\left\langle \phi_1 \phi_2 \phi_3\right\rangle = 0, \qquad
\left(i\partial_{t_3} +\omega_3 \right)\left\langle \phi_1 \phi_2 \phi_3\right\rangle = 0.
\eeq}
\begin{align}
\left\langle \phi_1^{j_1}\phi_2^{j_2}\right\rangle &= \delta_{j_1, j_2}\frac{e^{-i\omega_1 (t_1-t_2)}}{(x_1-x_2)^{-4j_1}}, \\
\left\langle \phi_1^{j_1}\phi_2^{j_2}\phi_3^{j_3}\right\rangle
&= C_{123} \frac{e^{-i\omega_1 t_1 + i \omega_3 t_3 + i (\omega_1-\omega_3)t_2}}{(x_1-x_2)^{-2j_1-2j_2+2j_3}(x_1-x_3)^{-2j_1-2j_3+2j_2}(x_2-x_3)^{-2j_2-2j_3+2j_1}}.
\end{align}
 
The four-point function can be solved explicitly, just like in 2d CFT. The diagram is 
\vspace{0.2cm}
\begin{center}
\includegraphics[width=0.36\textwidth]{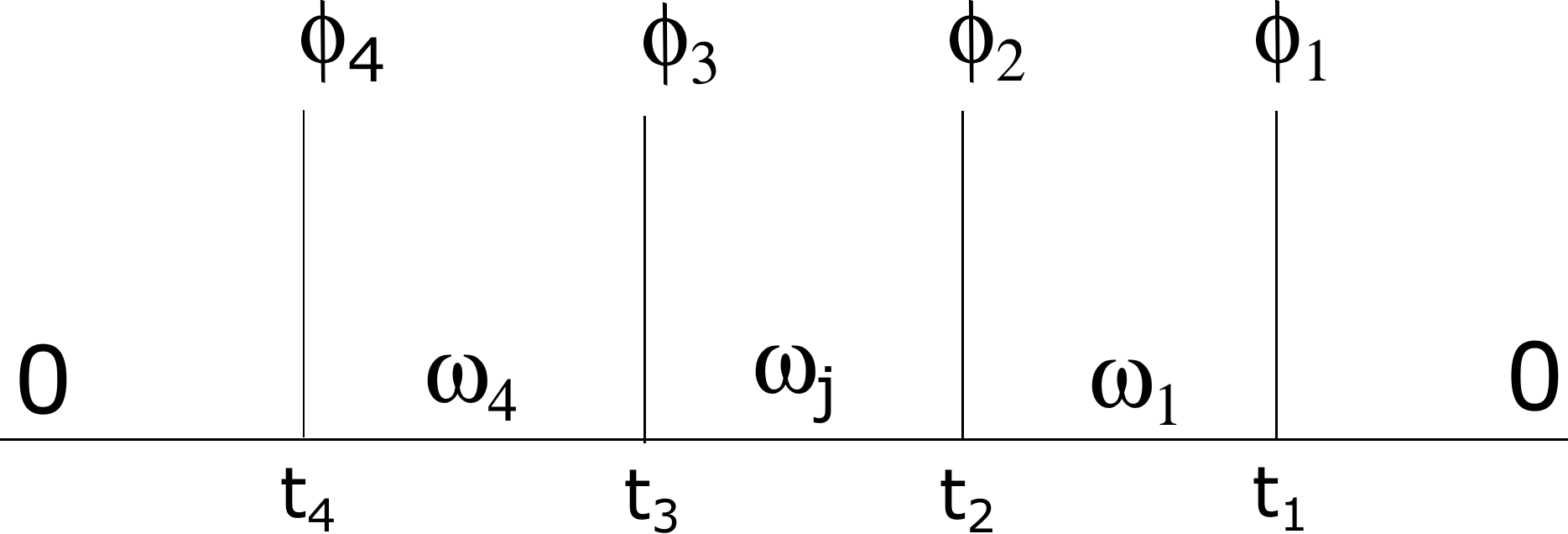}
\end{center}
\vspace{0.2cm}
Upon writing the $SL(2,\mathbb{R})$ currents as differential operators
\begin{align}
J_{-1} = \partial_x, \quad J_{0} = -x \partial_x + j, \quad J_{+1} = x^2 \partial_x - 2j x,
\end{align}
and defining the cross-ratio $x= \frac{x_{21}x_{43}}{x_{31}x_{42}}$ and time difference $t=t_3-t_2$, one rewrites the four-point function in terms of a single function $F(x,t)$:
\begin{equation}
\left\langle \phi_4^{j_4}\phi_3^{j_3}\phi_2^{j_2}\phi_1^{j_1}\right\rangle = x_{43}^{j_4+j_3-j_2-j_1}x_{42}^{2j_2}x_{41}^{j_4+j_1-j_3-j_2}x_{31}^{j_3+j_2+j_1-j_4}e^{-i\omega_4 t_{43}}e^{-i\omega_1 t_{21}}F(x,t).
\end{equation}
For the time-dependence, this is the 1d analogue of the general conformal structure in 2d CFT. The time difference $t$ plays the role of the cross-ratio in this case. 

The KZ equation for $t_1$ and $t_4$ are identically satisfied, the one for $t_3+t_2$ is also trivial. 
Finally, the sum of all four equations lead to time translation invariance (\ref{timet}). 
The only remaining non-trivial equation, for $t=t_3-t_2$, is
\bea
& \! & \!\bigg[ i\partial_t + x^2(1-x)\partial_x^2 + \left((j_1+j_3-j_4+3j_2-1)x^2-2(j_1+j_2)x\right)\partial_x \nonumber \\
& \! & \!+  2(j_2j_4-j_2j_3-j_2j_1-2j_2^2)x+2j_1j_2+j_2+j_2^2+j_1+j_1^2 \bigg] F(x,t) = 0.
\eea

Projecting onto a fixed intermediate channel, we set $\partial_t = -i\omega_j$ and find the solution:
\begin{align}
F^j(x,t) = e^{-i\omega_j t} x^{j_1+j_2-j}\bigg[ & {}_2F_1(j_1-j_2-j,j_4-j_3-j;-2j;x) \nonumber \\
\! & \!+ \lambda\, x^{1+2j}{}_2F_1(j_1-j_2+j+1,j_4-j_3+j+1;2+2j;x) \bigg],
\end{align}
in terms of the coefficient $\lambda$ which can be fixed by monodromy considerations. 
As usual, the second solution is just the reflected one ($j\to-j-1$) and can be taken care of by incorporating this into the region of integration. 

This determines the 4-point function in the form:
\bea
\left\langle \phi_4^{j_4}\phi_3^{j_3}\phi_2^{j_2}\phi_1^{j_1}\right\rangle \is \!x_{43}^{j_4+j_3-j_2-j_1}x_{42}^{2j_2}x_{41}^{j_4+j_1-j_3-j_2}x_{31}^{j_3+j_2+j_1-j_4} e^{-i\omega_4 t_{43}}e^{-i\omega_1 t_{21}} \nonumber \\
&& \times \int dj e^{-i\omega_{j}t_{32}} C(j,j_3,j_4)C(-j-1,j_1,j_2)\, \mathcal{F}^{s}_{j}\left[\!\begin{array}{cc}
j_3 \! & \! j_2 \\
j_4 \! & \! j_1 \end{array}\! \right](x),
\eea
with the $s$-channel blocks:
\bea
\mathcal{F}^{s}_{j}\left[\!\begin{array}{cc}
j_3 \! & \! j_2 \\
j_4 \! & \! j_1 \end{array}\! \right](x) = x^{j_1+j_2-j}{}_2F_1(j_1-j_2-j,j_4-j_3-j;-2j;x).
\eea
The OPE coefficients $C(j,j_3,j_4)$ and range of integration/summation over intermediate states cannot be fixed without additional dynamical information about the theory. 

The above discussion treats $SL(2,\mathbb{R})$ as a global symmetry. Gauging it requires making each correlator as a whole $SL(2,\mathbb{R})$-invariant. This could be achieved by a suitable integration over the $x$-variable, but we will not go into this here. The main message is that gauge-invariant correlators could be constructed using gauge non-invariant operators. 

When specifying to the Schwarzian theory, the Schwarzian itself is $SL(2,\mathbb{R})$-invariant and satisfies the above equations in a trivial way. Its equation of motion dictates that  $\partial_t\text{Sch}(f,t) = 0$, which means that in particular its Casimir eigenvalue vanishes.  
When restricting to local gauge-invariant operators, this would be the only surviving operator (up to trivial modifications) for which the above discussion applies.

\end{appendix}


\begin{thebibliography}{99}


\bibitem{SS} S.~H.~Shenker and D.~Stanford, ``Black holes and the butterfly effect,'' JHEP {\bf 1403}, 067 (2014)
  \href{http://arxiv.org/abs/1306.0622}{{\ttfamily arXiv:1306.0622 [hep-th]}}; ``Multiple Shocks,'' JHEP {\bf 1412}, 046 (2014) \href{http://arxiv.org/abs/1312.3296}{{\ttfamily arXiv:1312.3296 [hep-th]}}.

\bibitem{KitaevTalks}
A. Kitaev, Talk given at the Fundamental Physics Prize Symposium, \href{https://www.youtube.com/watch?v=OQ9qN8j7EZI}{Nov. 10, 2014}; A. Kitaev, KITP
seminar, \href{http://online.kitp.ucsb.edu/online/joint98/kitaev/}{Feb. 12, 2015}; ``A simple model of quantum holography,'' talks at KITP,  \href{http://online.kitp.ucsb.edu/online/entangled15/kitaev/}{April 7, 2015} and \href{http://online.kitp.ucsb.edu/online/entangled15/kitaev2/}{May 27, 2015}.
  
  \bibitem{MSS}J.~Maldacena, S.~H.~Shenker and D.~Stanford, ``A bound on chaos,''  \href{http://arxiv.org/abs/1503.01409}{{\ttfamily arXiv:1503.01409 [hep-th]}}
  
       \bibitem{Polchinski:2015cea} 
  J.~Polchinski,
  ``Chaos in the black hole S-matrix,''
   \href{http://arxiv.org/abs/1505.08108}{{\ttfamily arXiv:1505.08108 [hep-th]}}.
  
  \bibitem{Jensen:2016pah}
  K.~Jensen,
  ``Chaos and hydrodynamics near AdS$_2$,''
  \href{http://arxiv.org/abs/1605.06098}{{\ttfamily arXiv:1605.06098 [hep-th].}}

  
\bibitem{Sachdev:1992fk}
S.~Sachdev and J.-w. Ye, ``{Gapless spin fluid ground state in a random,
  quantum Heisenberg magnet},''
 Phys. Rev. Lett.
  {\bfseries 70} (1993) 3339,
\href{http://arxiv.org/abs/cond-mat/9212030}{{\ttfamily arXiv:cond-mat/9212030
  [cond-mat]}}.
  
\bibitem{Polchinski:2016xgd} 
  J.~Polchinski and V.~Rosenhaus,
  ``The Spectrum in the Sachdev-Ye-Kitaev Model,''
  JHEP {\bf 1604}, 001 (2016)
 \href{http://arxiv.org/abs/1601.06768}{{\ttfamily arXiv:1601.06768 [hep-th]}}.

\bibitem{Jevicki:2016bwu} 
  A.~Jevicki, K.~Suzuki and J.~Yoon,
  ``Bi-Local Holography in the SYK Model,''
 JHEP {\bf 1607}, 007 (2016)
  \href{http://arxiv.org/abs/1603.06246}{{\ttfamily arXiv:1603.06246 [hep-th]}}.
  
\bibitem{Maldacena:2016hyu} 
  J.~Maldacena and D.~Stanford,
  ``Remarks on the Sachdev-Ye-Kitaev model,''
Phys.\ Rev.\ D {\bf 94}, no. 10, 106002 (2016)
\href{http://arxiv.org/abs/1604.07818}{{\ttfamily arXiv:1604.07818 [hep-th]}}. 


\bibitem{Jevicki:2016ito} 
  A.~Jevicki and K.~Suzuki,
  ``Bi-Local Holography in the SYK Model: Perturbations,''
  JHEP {\bf 1611}, 046 (2016)
  \href{http://arxiv.org/abs/1608.07567}{{\ttfamily arXiv:1608.07567 [hep-th]}}.

\bibitem{Witten:2016iux} 
  E.~Witten,
  ``An SYK-Like Model Without Disorder,''
  \href{http://arxiv.org/abs/1610.09758}{{\ttfamily arXiv:1610.09758 [hep-th]}}; I.~R.~Klebanov and G.~Tarnopolsky,
  ``Uncolored random tensors, melon diagrams, and the Sachdev-Ye-Kitaev models,''
  Phys.\ Rev.\ D {\bf 95}, no. 4, 046004 (2017)
  \href{http://arxiv.org/abs/1611.08915}{{\ttfamily arXiv:1611.08915 [hep-th]}}.
  
\bibitem{Jackiw:1984je} 
  R.~Jackiw,
  ``Lower Dimensional Gravity,''
  Nucl.\ Phys.\ B {\bf 252}, 343 (1985);  C.~Teitelboim,
  ``Gravitation and Hamiltonian Structure in Two Space-Time Dimensions,''
  Phys.\ Lett.\  {\bf 126B}, 41 (1983).
  
\bibitem{Almheiri:2014cka} 
  A.~Almheiri and J.~Polchinski,
  ``Models of AdS$_{2}$ backreaction and holography,''
  JHEP {\bf 1511}, 014 (2015)
  \href{http://arxiv.org/abs/1402.6334}{{\ttfamily arXiv:1402.6334 [hep-th]}}.
  

\bibitem{Maldacena:2016upp} 
  J.~Maldacena, D.~Stanford and Z.~Yang,
  ``Conformal symmetry and its breaking in two dimensional Nearly Anti-de-Sitter space,''
  PTEP {\bf 2016}, no. 12, 12C104 (2016)
 \href{http://arxiv.org/abs/1606.01857}{{\ttfamily arXiv:1606.01857 [hep-th]}}.
 

\bibitem{Engelsoy:2016xyb} 
  J.~Engelsoy, T.~G.~Mertens and H.~Verlinde,
  ``An investigation of AdS$_{2}$ backreaction and holography,''
 JHEP {\bf 1607}, 139 (2016)
 \href{http://arxiv.org/abs/1606.03438}{{\ttfamily arXiv:1606.03438 [hep-th]}}.
 
\bibitem{Cvetic:2016eiv} 
  M.~Cvetic and I.~Papadimitriou,
  ``AdS$_{2}$ holographic dictionary,''
  JHEP {\bf 1612}, 008 (2016)
  Erratum: [JHEP {\bf 1701}, 120 (2017)]
   \href{http://arxiv.org/abs/1608.07018}{{\ttfamily arXiv:1608.07018 [hep-th]}}.
  
  \bibitem{Kirillov}
A.~A.~Kirillov, ``Orbits of the Group of Diffeomorphisms of a Circle and Local Lie
Superalgebras," Functional Analysis and Its Applications {\bf 15} no. 2, (1981) 135-137.

 \bibitem{LazutkinSegal}
  V.F.~Lazutkin and T.F.~Pankratova, Funkts. Anal. Prilozh. 9 (1975); S.~Segal, ``Unitary representations of some infinite dimensional groups," Commun. Math. Phys. 80, 307 (1981).
  
\bibitem{Witten:1987ty} 
  E.~Witten,
  ``Coadjoint Orbits of the Virasoro Group,''
  Commun.\ Math.\ Phys.\  {\bf 114}, 1 (1988).
  
\bibitem{AS}
  A.~Alekseev and S.~L.~Shatashvili,
  ``Path Integral Quantization of the Coadjoint Orbits of the Virasoro Group and 2D Gravity,''
  Nucl.\ Phys.\ B {\bf 323}, 719 (1989); ``From geometric quantization to conformal field theory,''
  Commun.\ Math.\ Phys.\  {\bf 128}, 197 (1990).


\bibitem{altland} 
   D.~Bagrets, A.~Altland and A.~Kamenev, ``Sachdev-Ye-Kitaev model as Liouville quantum mechanics,''
  Nucl.\ Phys.\ B {\bf 911}, 191 (2016)
   \href{http://arxiv.org/abs/1607.00694}{{\ttfamily arXiv:1607.00694 [cond-mat.str-el]}};
  ``Power-law out of time order correlation functions in the SYK model,''
  \href{http://arxiv.org/abs/1702.08902}{{\ttfamily arXiv:1702.08902 [cond-mat.str-el]}}.
  

\bibitem{Mandal:2017thl} 
  G.~Mandal, P.~Nayak and S.~R.~Wadia,
  ``Virasoro coadjoint orbits of SYK/tensor-models and emergent two-dimensional quantum gravity,''
  \href{http://arxiv.org/abs/1702.04266}{{\ttfamily arXiv:1702.04266 [hep-th]}}.
  
\bibitem{wittenstanford}
  D.~Stanford and E.~Witten,
  ``Fermionic Localization of the Schwarzian Theory,''
    \href{http://arxiv.org/abs/1703.04612}{{\ttfamily arXiv:1703.04612 [hep-th]}}.
  
  \bibitem{PT}
B.~Ponsot and J.~Teschner,
  ``Liouville bootstrap via harmonic analysis on a noncompact quantum group,''
   \href{https://arxiv.org/abs/hep-th/9911110}{{\ttfamily hep-th/9911110}}; 

\bibitem{Teschner:2001rv} 
  J.~Teschner,
  ``Liouville theory revisited,''
  Class.\ Quant.\ Grav.\  {\bf 18}, R153 (2001)
   \href{https://arxiv.org/abs/hep-th/0104158}{{\ttfamily [hep-th/0104158]}}.
  
\bibitem{Teschner:2012em} 
  J.~Teschner and G.~Vartanov,
  ``6j symbols for the modular double, quantum hyperbolic geometry, and supersymmetric gauge theories,''
  Lett.\ Math.\ Phys.\  {\bf 104}, 527 (2014)
   \href{https://arxiv.org/abs/1202.4698}{{\ttfamily [arXiv:1202.4698 [hep-th]]}}.
  
   \bibitem{groenevelt}
  W.~Groenevelt, ``The Wilson function transform," \href{https://arxiv.org/abs/math/0306424}{{\ttfamily arXiv:0306424 [math.CA]}}; ``Wilson function transforms related to Racah coefficients," 	\href{https://arxiv.org/abs/math/0501511}{{\ttfamily arXiv:math/0501511 [math.CA]}}.

\bibitem{Kitaev16} 
 A.~Kitaev. Talk given IAS chaos workshop, Oct. 18, 2016.

\bibitem{Comtet} 
  A.~Comtet and P.~J.~Houston,
  ``Effective Action on the Hyperbolic Plane in a Constant External Field,''
  J.\ Math.\ Phys.\  {\bf 26}, 185 (1985); A.~Comtet,
  ``On the Landau Levels on the Hyperbolic Plane,''
  Annals Phys.\  {\bf 173}, 185 (1987).
  
\bibitem{Duistermaat:1982vw} 
  J.~J.~Duistermaat and G.~J.~Heckman,
  ``On the Variation in the cohomology of the symplectic form of the reduced phase space,''
  Invent.\ Math.\  {\bf 69}, 259 (1982).
  
\bibitem{Cotler:2016fpe} 
  J.~S.~Cotler {\it et al.},
  ``Black Holes and Random Matrices,''
    \href{http://arxiv.org/abs/1611.04650}{{\ttfamily arXiv:1611.04650 [hep-th]}}.
  
  \bibitem{LaurenHV}
L.~McGough and H.~Verlinde,
  ``Bekenstein-Hawking Entropy as Topological Entanglement Entropy,''
  JHEP {\bf 1311}, 208 (2013)
     \href{http://arxiv.org/abs/1308.2342}{{\ttfamily arXiv:1308.2342 [hep-th]}}.

\bibitem{Cardy} 
  J.~L.~Cardy,
  ``Boundary Conditions, Fusion Rules and the Verlinde Formula,''
  Nucl.\ Phys.\ B {\bf 324}, 581 (1989).
  
\bibitem{Fateev:2000ik} 
  V.~Fateev, A.~B.~Zamolodchikov and A.~B.~Zamolodchikov,
  ``Boundary Liouville field theory. 1. Boundary state and boundary two point function,''
  \href{https://arxiv.org/abs/hep-th/0001012}{{\ttfamily hep-th/0001012}};
  A.~B.~Zamolodchikov and A.~B.~Zamolodchikov,
  ``Liouville field theory on a pseudosphere,''
   \href{https://arxiv.org/abs/hep-th/0101152}{{\ttfamily hep-th/0101152}}.

\bibitem{Gukov:2008ve} 
  S.~Gukov and E.~Witten,
  ``Branes and Quantization,''
  Adv.\ Theor.\ Math.\ Phys.\  {\bf 13}, no. 5, 1445 (2009)
  \href{https://arxiv.org/abs/0809.0305}{{\ttfamily [arXiv:0809.0305 [hep-th]]}}.
  
\bibitem{Czech:2015qta} 
  B.~Czech, L.~Lamprou, S.~McCandlish and J.~Sully,
  ``Integral Geometry and Holography,''
  JHEP {\bf 1510}, 175 (2015)
       \href{http://arxiv.org/abs/1505.05515}{{\ttfamily arXiv:1505.05515 [hep-th]}}.
  
\bibitem{Czech:2016xec} 
  B.~Czech, L.~Lamprou, S.~McCandlish, B.~Mosk and J.~Sully,
  ``A Stereoscopic Look into the Bulk,''
  JHEP {\bf 1607}, 129 (2016)
         \href{http://arxiv.org/abs/1604.03110}{{\ttfamily arXiv:1604.03110 [hep-th]}}.
  
\bibitem{Dorn:2006ys} 
  H.~Dorn and G.~Jorjadze,
  ``Boundary Liouville theory: Hamiltonian description and quantization,''
  SIGMA {\bf 3}, 012 (2007)
   \href{https://arxiv.org/abs/hep-th/0610197}{{\ttfamily [hep-th/0610197]}}; ``Operator Approach to Boundary Liouville Theory,''
  Annals Phys.\  {\bf 323}, 2799 (2008)
   \href{https://arxiv.org/abs/0801.3206}{{\ttfamily [arXiv:0801.3206 [hep-th]]}}.
  
\bibitem{Balog:1997zz} 
  J.~Balog, L.~Feher and L.~Palla,
  ``Coadjoint orbits of the Virasoro algebra and the global Liouville equation,''
  Int.\ J.\ Mod.\ Phys.\ A {\bf 13}, 315 (1998)
  \href{https://arxiv.org/abs/hep-th/9703045}{{\ttfamily [hep-th/9703045]}}.
  
  \bibitem{DOZZ} 
    H.~Dorn and H.~J.~Otto,
  ``Two and three point functions in Liouville theory,''
  Nucl.\ Phys.\ B {\bf 429}, 375 (1994)
   \href{https://arxiv.org/abs/hep-th/9403141}{{\ttfamily [hep-th/9403141]}}; A.~B.~Zamolodchikov and A.~B.~Zamolodchikov,
  ``Structure constants and conformal bootstrap in Liouville field theory,''
  Nucl.\ Phys.\ B {\bf 477}, 577 (1996)
  \href{https://arxiv.org/abs/hep-th/9506136}{{\ttfamily [hep-th/9506136]}}; J.~Teschner,
  ``On the Liouville three point function,''
  Phys.\ Lett.\ B {\bf 363}, 65 (1995)
   \href{https://arxiv.org/abs/hep-th/9507109}{{\ttfamily [hep-th/9507109]}}.
  
  
  
  		\bibitem{Maldacena:2001kr}
  J.~M.~Maldacena,
  ``Eternal black holes in anti-de Sitter,''
  JHEP {\bf 0304} (2003) 021
  \href{https://arxiv.org/abs/hep-th/0106112}{{\ttfamily [hep-th/0106112]}}.

\bibitem{Gao:2016bin}
  P.~Gao, D.~L.~Jafferis and A.~Wall,
  ``Traversable Wormholes via a Double Trace Deformation,''
  \href{https://arxiv.org/abs/1608.05687}{{\ttfamily arXiv:1608.05687 [hep-th]}}.

\bibitem{Maldacena:2017axo}
  J.~Maldacena, D.~Stanford and Z.~Yang,
  ``Diving into traversable wormholes,''
  Fortsch.\ Phys.\  {\bf 65} (2017) no.5,  1700034
  \href{https://arxiv.org/abs/1704.05333}{{\ttfamily[arXiv:1704.05333 [hep-th]]}}.
  
  
   \bibitem{Jackson:2014nla} 
S.~Jackson, L.~McGough, and H.~Verlinde, ``{Conformal Bootstrap, Universality
  and Gravitational Scattering},''
   Nucl. Phys.
  {\bfseries B901} (2015) 382--429,
\href{http://arxiv.org/abs/1412.5205}{{\ttfamily arXiv:1412.5205 [hep-th]}}.

\bibitem{Hogervorst:2017sfd} 
  M.~Hogervorst and B.~C.~van Rees,
  ``Crossing Symmetry in Alpha Space,''
   \href{http://arxiv.org/abs/1702.08471}{{\ttfamily arXiv:1702.08471  [hep-th]}}.

\bibitem{Hogervorst:2017kbj} 
  M.~Hogervorst,
  ``Crossing Kernels for Boundary and Crosscap CFTs,''
     \href{http://arxiv.org/abs/1703.08159}{{\ttfamily arXiv:1703.08159  [hep-th]}}.
     
\bibitem{Fu:2016vas}
  W.~Fu, D.~Gaiotto, J.~Maldacena and S.~Sachdev,
  ``Supersymmetric Sachdev-Ye-Kitaev models,''
 Phys.\ Rev.\ D {\bf 95} (2017) no.2,  026009
 Addendum: [Phys.\ Rev.\ D {\bf 95} (2017) no.6,  069904]
          \href{http://arxiv.org/abs/1610.08917}{{\ttfamily arXiv:1610.08917 [hep-th]}}.
          
\bibitem{Forste:2017kwy}
  S.~Forste and I.~Golla,
 ``Nearly AdS$_2$ Sugra and the Super-Schwarzian,''
  \href{http://arxiv.org/abs/1703.10969}{{\ttfamily arXiv:1703.10969 [hep-th]}}.

 
\bibitem{Friedan:1986rx} 
  D.~Friedan,
  ``Notes On String Theory And Two-dimensional Conformal Field Theory,''
  In Santa Barbara 1985, Proceedings, Unified String Theories, 162-213 and Chicago Univ. - EFI 85-99 (85,REC.APR.86) 60p.
  
\bibitem{Cohn:1986wn} 
  J.~D.~Cohn,
  ``N=2 Super-Riemann Surfaces,''
  Nucl.\ Phys.\ B {\bf 284}, 349 (1987).
  
\bibitem{Fukuda:2002bv}
  T.~Fukuda and K.~Hosomichi,
  ``Super Liouville theory with boundary,''
  Nucl.\ Phys.\ B {\bf 635} (2002) 215
 \href{https://arxiv.org/abs/hep-th/0202032}{{\ttfamily [hep-th/0202032]}}.
  
\bibitem{Ahn:2002ev} 
  C.~Ahn, C.~Rim and M.~Stanishkov,
  ``Exact one point function of N=1 superLiouville theory with boundary,''
  Nucl.\ Phys.\ B {\bf 636}, 497 (2002)
    \href{https://arxiv.org/abs/hep-th/0202043}{{\ttfamily [hep-th/0202043]}}.
    
 
\bibitem{Hadasz:2013bwa} 
  L.~Hadasz, M.~Pawelkiewicz and V.~Schomerus,
  ``Self-dual Continuous Series of Representations for U$_q$(sl(2)) and U$_q$(osp(1$|$2)),''
  JHEP {\bf 1410}, 91 (2014)
    \href{http://arxiv.org/abs/1305.4596 }{{\ttfamily arXiv:1305.4596  [hep-th]}}.
  
\bibitem{Pawelkiewicz:2013wga} 
  M.~Pawelkiewicz, V.~Schomerus and P.~Suchanek,
  ``The universal Racah-Wigner symbol for U$_q$(osp(1$|$2)),''
  JHEP {\bf 1404}, 079 (2014)
      \href{http://arxiv.org/abs/1307.6866 }{{\ttfamily arXiv:1307.6866  [hep-th]}}.
			
\bibitem{Eguchi:2003ik} 
  T.~Eguchi and Y.~Sugawara,
  ``Modular bootstrap for boundary N = 2 Liouville theory,''
  JHEP {\bf 0401}, 025 (2004)
    \href{https://arxiv.org/abs/hep-th/0311141}{{\ttfamily [hep-th/0311141]}}.
		
\bibitem{Ahn:2003tt} 
  C.~Ahn, M.~Stanishkov and M.~Yamamoto,
  ``One point functions of N = 2 superLiouville theory with boundary,''
  Nucl.\ Phys.\ B {\bf 683}, 177 (2004)
   \href{https://arxiv.org/abs/hep-th/0311169}{{\ttfamily [hep-th/0311169]}}.
  
\bibitem{Krasnov:2004ki} 
  K.~Krasnov and S.~N.~Solodukhin,
  ``Effective stringy description of Schwarzschild black holes,''
  Adv.\ Theor.\ Math.\ Phys.\  {\bf 8}, no. 3, 421 (2004)
  \href{https://arxiv.org/abs/hep-th/0403046}{{\ttfamily [hep-th/0403046]}}.
  
\bibitem{Turiaci:2016cvo} 
  G.~Turiaci and H.~Verlinde,
  ``On CFT and Quantum Chaos,''
  JHEP {\bf 1612}, 110 (2016)
      \href{http://arxiv.org/abs/1603.03020 }{{\ttfamily arXiv:1603.03020  [hep-th]}}.
   
   
\bibitem{Floch:2015hwo} 
  B.~Le Floch,
  ``S-duality wall of SQCD from Toda braiding,''
  \href{https://arxiv.org/abs/1512.09128}{{\ttfamily arXiv:1512.09128 [hep-th]}}.
  
  
\bibitem{Turiaci:2017zwd} 
  G.~Turiaci and H.~Verlinde,
  ``Towards a 2d QFT Analog of the SYK Model,''
     \href{http://arxiv.org/abs/1701.00528 }{{\ttfamily arXiv:1701.00528  [hep-th]}}. 

  
\bibitem{Ribault:2007td} 
  S.~Ribault,
  ``Boundary three-point function on AdS2 D-branes,''
  JHEP {\bf 0801}, 004 (2008)
   \href{https://arxiv.org/abs/0708.3028}{{\ttfamily [arXiv:0708.3028 [hep-th]]}}.
    
  
 
  
  
\bibitem{Douglas:2003up}
  M.~R.~Douglas, I.~R.~Klebanov, D.~Kutasov, J.~M.~Maldacena, E.~J.~Martinec and N.~Seiberg,
  ``A New hat for the c=1 matrix model,''
  In *Shifman, M. (ed.) et al.: From fields to strings, vol. 3* 1758-1827
  \href{https://arxiv.org/abs/hep-th/0307195}{{\ttfamily [hep-th/0307195]}}.

\bibitem{Seiberg:2003nm}
  N.~Seiberg and D.~Shih,
  ``Branes, rings and matrix models in minimal (super)string theory,''
  JHEP {\bf 0402} (2004) 021
  \href{https://arxiv.org/abs/hep-th/0312170}{{\ttfamily [hep-th/0312170]}}.
	
\bibitem{Teschner:1997fv} 
  J.~Teschner,
  ``The Minisuperspace limit of the sl(2,C) / SU(2) WZNW model,''
  Nucl.\ Phys.\ B {\bf 546}, 369 (1999)
  \href{https://arxiv.org/abs/hep-th/9712258}{{\ttfamily [hep-th/9712258]}}.


\end{thebibliography}
\end{document}